\documentclass[aps,prx,twocolumn,superscriptaddress,floatfix,nofootinbib]{revtex4-1}
\usepackage{graphicx,amsfonts,amssymb,amsmath,hyperref,hypcap,enumerate}
\usepackage{xcolor}
\usepackage{tabls}

\newcommand{\bit}{\begin{itemize}}
\newcommand{\eit}{\end{itemize}}
\newcommand{\f}{\frac}
\renewcommand{\>}{\right\rangle}
\newcommand{\<}{\left\langle}
\newcommand{\ba}{\begin{align}}
\newcommand{\ea}{\end{align}}
\newcommand{\be}{\begin{equation}}
\newcommand{\ee}{\end{equation}}
\newcommand{\bi}{\begin{itemize}}
\newcommand{\ei}{\end{itemize}}
\newcommand{\lf}{\left(}
\newcommand{\ri}{\right)}
\newcommand{\dd}{\mathrm{d}}

\newcommand{\Tr}{\operatorname{Tr}}

\newcommand{\bra}[1]{\< #1 \right|}
\newcommand{\ket}[1]{\left| #1 \>}

\newcommand{\bigsection}{\section}
\newcommand{\smallsection}{\subsection}

\newcommand{\appendixsection}{\subsection}

\graphicspath{{Figures/}}

\begin{document}
\title{Valence Bonds in Random Quantum Magnets: \\ Theory and Application to YbMgGaO$_4$}
\author{Itamar Kimchi}
\affiliation{Department of Physics, Massachusetts Institute of
Technology, Cambridge, MA 02139, USA}
\author{Adam Nahum}
\affiliation{Department of Physics, Massachusetts Institute of
Technology, Cambridge, MA 02139, USA}
\affiliation{
Theoretical Physics, Oxford University, 1 Keble Road, Oxford OX1 3NP, United Kingdom}
\author{T. Senthil}
\affiliation{Department of Physics, Massachusetts Institute of
Technology, Cambridge, MA 02139, USA}
\begin{abstract}
We analyze the effect of quenched disorder on spin-1/2 quantum magnets in which magnetic frustration promotes the formation of local singlets. Our results include a theory for 2d valence-bond solids subject to weak bond randomness, as well as extensions to stronger disorder regimes where we make connections with quantum spin liquids. We find, on various lattices, that the destruction of a valence-bond solid phase by weak quenched disorder leads inevitably to the nucleation of topological defects carrying spin--1/2 moments. This renormalizes the lattice into a strongly random spin network with interesting low-energy excitations. Similarly when short-ranged valence bonds would be pinned by stronger disorder, we find that this putative glass is unstable to defects that carry spin--1/2 magnetic moments, and whose residual interactions decide the ultimate low energy fate. Motivated by these results we conjecture Lieb-Schultz-Mattis-like restrictions on ground states for disordered magnets with spin--1/2 per statistical unit cell. These conjectures are supported by an argument for 1d spin chains. We apply insights from this study to  the phenomenology of YbMgGaO$_4$, a recently discovered triangular lattice spin--1/2 insulator which was proposed to be a quantum spin liquid. We instead explore a description based on the present theory. Experimental signatures, including unusual specific heat, thermal conductivity, and dynamical structure factor, and their behavior in a magnetic field, are predicted from the theory, and compare favorably with existing measurements on YbMgGaO$_4$ and related materials.  
\end{abstract}
\maketitle

\tableofcontents



\begin{figure}[b]
\includegraphics[width=\columnwidth]{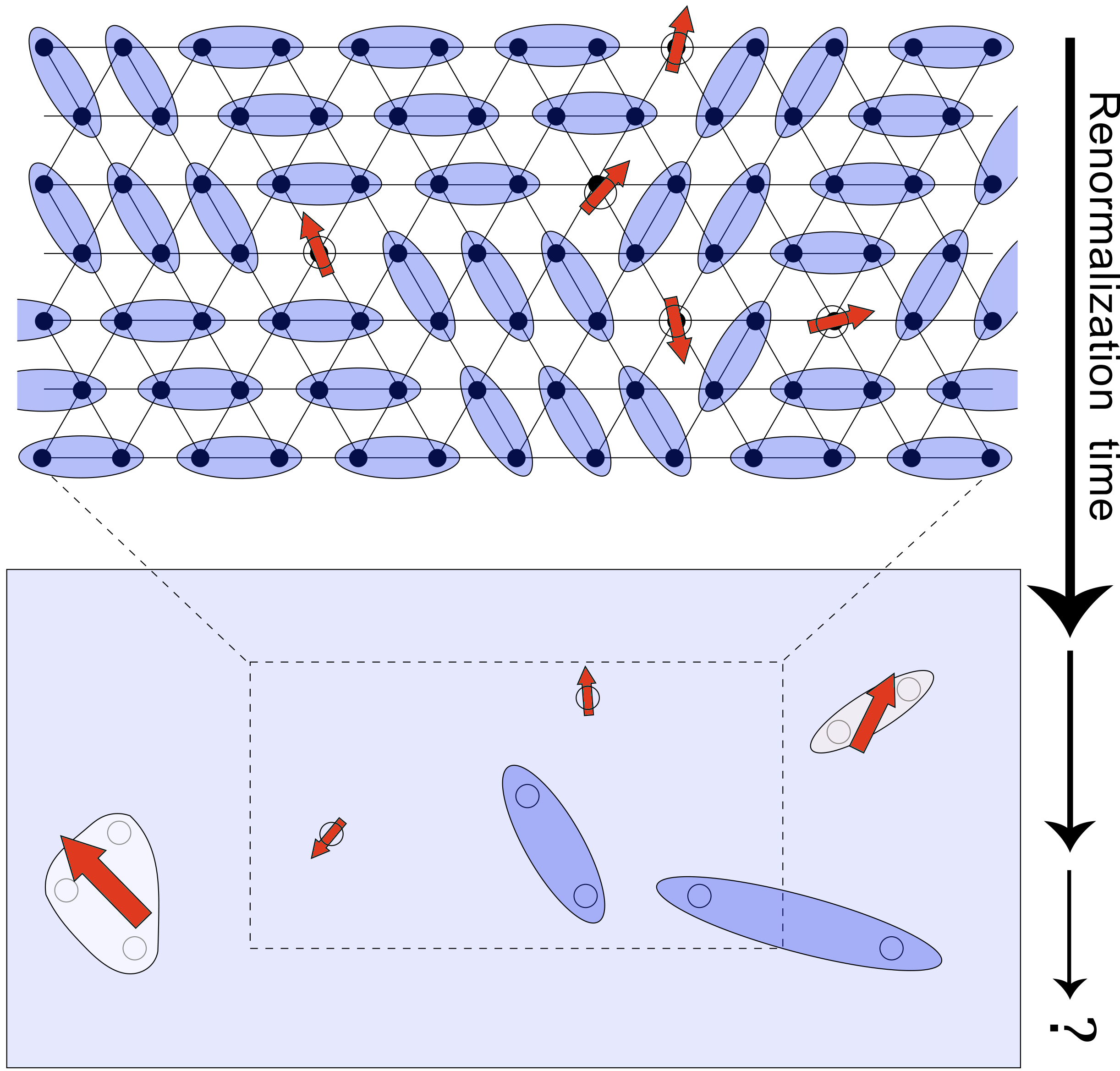}
\caption[]{ \textbf{Spin--1/2 defects in a valence bond solid.} 
The Valence Bond Solid (VBS) patterns on the triangular lattice shown here [top] admit two types of defects: point defects (vortices) and line defects (domain walls). Each vortex hosts a protected spin--1/2 in its core. A particular type of domain wall, here shown separating the sets of domains on the left and the right, also carries dangling spin--1/2. 
When bond randomness destroys long-range VBS lattice-symmetry-breaking order it also nucleates  a random network of these defects [bottom]: RG flows at low energies can produce longer-range random singlets, larger-spin clusters and spin glass freezing of the nucleated spins.
} \label{fig:VBS}
\end{figure}

\bigsection{Introduction}
Magnetic insulators often exhibit quenched disorder from material defects, 
including randomness in the
 strengths of 
 magnetic exchanges. How this bond randomness interacts with geometrical frustration and quantum fluctuations of the moments  is an interesting and largely open question,  especially urgent in view of  perplexing  experiments that seem related to both features.

%
%

\begin{table}[]
\centering
\begin{tabular}{l|l l}
\hline\hline
                    & \multicolumn{2}{c}{Disorder strength} \\
Clean-limit phase  & Weak            & Intermediate         \\ \hline
Quantum spin liquid \ \  & Stable          & Sec.~\ref{strong_disorder_section} A           \\
Valence bond solid  & Sec.~\ref{VBSdis} \ \ \ \          & Sec.~\ref{strong_disorder_section} A,B         \\ \hline\hline
\end{tabular}
\caption{Fate of quantum-paramagnet phases in a spin--1/2 2d lattice system upon adding bond-randomness disorder.
In both regimes (defined in Sec.~\ref{sec:introsetting}) we find a sparse strongly-random spin network must necessarily emerge (Figs.~\ref{fig:VBS},\ref{fig:vbginstability}), giving rise to interesting low energy spin excitations.}
\label{tab:disorder}
\end{table}


For concreteness, let us focus on the prominent recently-discovered material YbMgGaO$_4$.\cite{srep,Zhang2015,muSR,Li2016, Mourigal2016,Zhao2016,Gegenwart2017,Rueegg2017,Zhang2017,Zhao2017,Mourigal2017,Wen2017} 
This is a layered insulator in which the Yb sites yield  effective $S=1/2$ magnetic moments (arising from the strong spin-orbit coupling of the heavy 
Yb$^{3+}$ ions). These moments form a stack of two-dimensional triangular lattices.
Magnetic frustration arises from the triangular geometry and potentially from spin-orbit coupling (SOC). 
Disorder is intrinsic for YbMgGaO$_4$: its magnetic Yb$^{3+}$ layers  are separated by  layers of non-magnetic Mg$^{2+}$ and Ga$^{3+}$ ions which randomly occupy a single $R\bar{3}m$ crystallographic position\cite{srep,Zhang2015,Zhang2017}, forming a triangular lattice with Mg/Ga occupancy described by a disordered Ising variable. 
The random electric field from the Mg/Ga site (which sits directly above or below a {Yb-O-Yb} oxygen site) can modify the {Yb-O-Yb} magnetic superexchange as well as the Yb effective spin--1/2 $g$-factor; 
experimentally, the  distribution of $g$-factors\cite{Zhang2017} shows direct evidence for Hamiltonian randomness, though  exchange randomness\cite{Wen2017} has been difficult to quantify.

In YbMgGaO$_4$, neutron scattering\cite{Mourigal2016,Zhao2016,Gegenwart2017,Rueegg2017} and $\mu$SR\cite{muSR} studies at temperatures down to around 50 mK (less than two percent of the Curie-Weiss temperature $\theta_{CW}$)  found no signs of magnetic ordering or of frozen moments. Even more unusually, careful comparisons with the nonmagnetic analogue LuMgGaO$_4$ exposed a peculiar power-law $C(T)\sim T^{0.7}$ heat capacity of the magnetic moments,\footnote{Note that the identical oxidation states of Yb$^{3+}$ and Lu$^{3+}$ imply that the magnetic compound and its non-magnetic analogue  have the same charge disorder in the Mg/Ga layers.}
 extending over a decade in temperature from 1 K down to 60 mK\cite{srep}. This fractional power law inspired theoretical work\cite{Chen2016,Zhao2016,Wang2016,Chen2016b,Chen2017}  that interpreted these observations as evidence for a  spin-liquid phase with a spinon Fermi surface. However a comparison of inelastic neutron scattering\cite{Mourigal2016,Zhao2016,Gegenwart2017,Rueegg2017,Zhao2017} at low and high temperatures shows little direct evidence\footnote{A sharp $2 k_f$ edge will be a direct signature of such a spinon Fermi surface and this has not been demonstrated.} of a Fermi surface, and even more strikingly measurements of thermal conductivity\cite{Li2016} found that $\kappa/T$  vanishes with temperature, substantially complicating any interpretation in terms of itinerant spinons. 

This complex phenomenology brings to mind doped semiconductors\cite{Kamimura1982,Ramakrishnan1985} such as Si:P, where a broad 
distribution of couplings between magnetic moments, at randomly located dopant sites, was argued to generate a \textit{random singlet} phase\cite{Lee1981,Lee1982, Sachdev1988, Fisher1994} --- in which   each spin $i$ forms a singlet with another spin $j$, 
with weak singlets forming across  arbitrarily large distances.
 The random singlet phase exists in 1d, where it is tractable using the strong disorder renormalization group, which iteratively integrates out the strongest couplings.   In higher dimensions the above cartoon is not accurate for  disordered spin networks: numerical strong disorder renormalization group calculations show that there is formation of higher-spin clusters as well as long-range singlets, and likely spin glass freezing at the lowest energies (where strong disorder RG breaks down).\cite{Lee1982, Igloi2003,Fisher2000,Troyer2012} However, {if} the initial coupling distribution is parametrically broad, strong disorder RG will capture the physics over a parametrically large range of lengthscales. Thus this is a mechanism for power law heat capacity without frozen moments over a large energy range, as is indeed experimentally observed in doped semiconductors. 

Motivated by these developments, in this paper we study the effects of disorder on quantum paramagnetic phases that arise in frustrated quantum magnets. Our main concern is the interplay between (1) valence bond physics and (2) randomness in the exchange couplings.

\smallsection{Setting}
\label{sec:introsetting}
Consider a magnet which, in the clean limit, is in a spin-gapped paramagnetic phase. It is useful to have a heuristic picture in mind for the various energy scales. The spin  gap $\Delta_S$, which is the energy scale for creating spin-carrying excitations, may be loosely associated with the energy scale of formation of singlet valence bonds between pairs of local moments. At a second energy scale $\Delta_{VB}$, which we may loosely identify with the energy scale for valence-bond-type excitations that do not carry spin, these valence bonds may freeze into a Valence Bond Solid (VBS) phase that breaks lattice symmetries, or they may form a
quantum liquid. The latter phase is a gapped quantum spin liquid known as a short-ranged Resonating Valence Bond (RVB) state.    The ratio ${\Delta_{VB}}/{\Delta_S}$ is typically of order one,  but it will sometimes be instructive  to consider the limit where the scales become well separated, ${\Delta_{VB}}/{\Delta_S} \ll 1$.  

What is the fate of such a system when the exchange couplings are random? We will restrict attention to the situation in which the width $\Delta$ of the distribution of the random exchanges is smaller than the spin gap $\Delta_S$ of the clean magnet, but will consider both the case where randomness is parametrically weak ($\Delta \ll \Delta_{VB},\, \Delta_S$), and the case where randomness is weak compared to the spin gap, but has a strong pinning effect on the valence bonds (loosely speaking, $\Delta_{VB} \lesssim \Delta \ll \Delta_S$).

\begin{figure}[b]
\includegraphics[width=0.88\columnwidth]{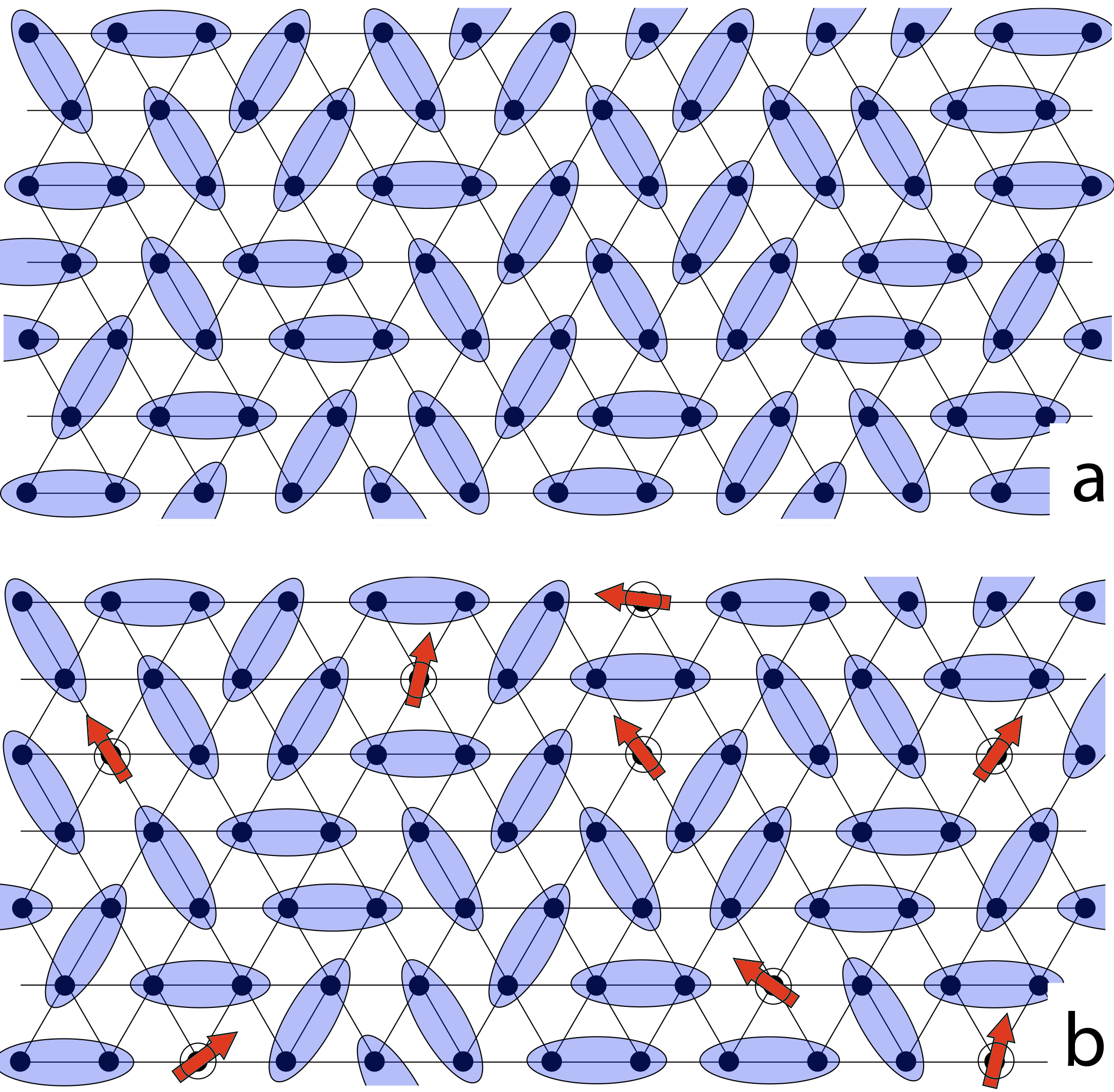}
\caption[]{ \textbf{Pinned singlets instability.} 
\textbf{(a)} The putative 2d valence bond glass phase consists of pinned short-ranged singlets.  \textbf{(b)} We find that in the thermodynamic limit such a valence bond glass is unstable to the nucleation of defect monomers that carry spin--1/2. The RG flow from the resulting random spin network then determines the low energy physics, as in Fig.~\ref{fig:VBS}. 
} \label{fig:vbginstability}
\end{figure}

If the randomness  $\Delta$ is the weakest energy scale in the problem ({\em i.e}  $\Delta \ll  \Delta_{VB}, \,\Delta_S$) then it has strikingly different effects on VBS and RVB phases.   RVB phases are stable. However VBS phases in $d \leq 2$  are unstable, as we review below, because disorder couples linearly to the VBS order parameter.  We will study the eventual fate of the disordered VBS state in various examples in both 1d and 2d.

In the weak disorder limit, we introduce a theoretical mechanism by which
a regime of strong randomness --- involving broad coupling distributions and random connectivities --- 
can arise in   2d quantum magnets even when disorder is weak at the lattice scale.  
The low-energy physics is determined by   a multi-stage RG flow that begins with the nucleation of  topological defects. These  carry spin, despite the fact that the disorder scale  is small compared to the spin gap of the clean system, $\Delta \ll \Delta_S$. This flow has several well-defined regimes when the initial disorder is weak. We characterize these regimes  on the triangular and square lattices, showing that gapless spinful excitations necessarily emerge.  At the very largest scales, the defect spins can break spin symmetry, for example via spin-glass order.

We will also study the case $\Delta_{VB} \lesssim \Delta \ll \Delta_S$, where  randomness strongly pins the valence bonds.  
 This limit is well modeled by a dimer model with random energies. We will show that broadly similar results obtain in both the weakly disordered VBS model and the random dimer model. At first sight one might have guessed that strong pinning would allow a distinct paramagnetic phase made only of randomly pinned singlets (a `valence bond glass'), but we show that this phase is unstable to the nucleation of spinful defects in 2d. 

These analyses raise interesting questions about what kinds of states can exist, even in principle, for disordered magnets on various lattices.  Our results in the two limits of disorder strength motivate conjectures in the spirit of the Lieb-Schultz-Mattis theorem of clean magnets. Our conjectures rule out certain kinds of spin-gapped states for disordered Hamiltonians that preserve spin symmetry and preserve lattice  symmetry on average, with spin-1/2 per unit cell of the statistical translation symmetry. We describe some constraints that we expect in physical terms, without aspiring to mathematical rigour.  We propose that a useful perspective on Lieb-Schultz-Mattis-like constraints in disordered systems is from the standpoint of the entanglement structure of ground states.

The picture that emerges for the pinning of valence bonds suggests an application of the above ideas to YbMgGaO$_4$ (which we discuss in detail) and other frustrated materials. 
Indeed an isostructural compound, YbZnGaO$_4$, was recently studied\cite{Wen2017} and shown to exhibit phenomenology closely related to that of YbMgGaO$_4$. The YbZnGaO$_4$ dynamical structure factor $S(q,\omega)$ shows similar features to YbMgGaO$_4$; thermal conductivity $\kappa(T)$ is  close to indistinguishable; and, remarkably, YbZnGaO$_4$ was also found to show a power law magnetic specific heat with an anomalous exponent, here $C(T) \sim T^{0.59}$.  Our random-singlet-inspired picture, in which the anomalous $C(T)$ exponent is an effective exponent over a range of temperature scales,
and can take different values depending on the disorder distribution,  
appears to be a good candidate for the phenomenological description. 

The importance of disorder in the YbMgGaO$_4$ family of compounds was discussed for $g$-factor randomness\cite{Zhang2017} as well as for exchange randomness: 
in particular the study of YbZnGaO$_4$  also reported  new measurements of AC susceptibility for both compounds\cite{Wen2017}, which find broad excitation spectra suggested to be associated with spin freezing at ultralow temperatures $T_f < 0.1$ K. This suggests that the interplay of disorder with frustration should be part of the description of these magnets. Here we consider this interplay together with the additional ingredient of spin--1/2 quantum fluctuations and singlet formation. 
 Interestingly, the signatures of spin freezing in the two Yb magnets were reported to be distinct from conventional spin glasses: $T_f$  in both compounds is $\sim 20$ times smaller than the  temperature of the peak in specific heat, in contrast to conventional spin glasses, and the spin freezing entropy was also reported to be anomalously small\cite{Wen2017}. As elaborated on below, ultralow temperature spin freezing of remnant defect spins is indeed the likely RG fate for the defect spin network that arises on the triangular lattice, suggesting the above ideas may be on the right track for these materials.

In the literature, short-ranged random singlet type phases have been proposed\cite{sandvik1995disorder,Sakai2014,Shimokawa2014,Kawamura2017} for lattice magnets in numerical studies that found an apparent suppression of ordered phases under strong bond randomness.
The suppression occurs when disorder strength becomes of order unity, i.e. couplings distributed across $\{J{\pm} \Delta\}$ seem to require\footnote{ 
For a clean system in a VBS phase, a smaller $2 \Delta {\gtrsim} J/2$ was recently reported\cite{Kawamura2017}, though the Imry-Ma mechanism discussed below implies that this critical disorder strength would vanish if the system size were taken to the thermodynamic limit.} a distribution width $2 \Delta {\gtrsim} J$. Bipartite antiferromagnets are argued to remain stable through large $\Delta$.\cite{Sandvik2002,Rieger2006}
 We note that recent work\cite{Chernyshev2017}  on Hamiltonians relevant to YbMgGaO$_4$ found that  adding a particular spin-orbit-coupled term ($J_{\pm \pm}$) with small magnitude but random sign $\{{\pm} \Delta\}$ melts the orientational part of the magnetic order.  Relatedly, phenomenologically adding disorder with  $2 \Delta {\sim} J$ in a spin wave theory was found\cite{Wen2017} to capture the continuum features of the structure factor. 
The  structure factor features were also discussed in the context of short ranged singlets\cite{Gegenwart2017}. 
Based on our analysis and the conjectured disordered-LSM restriction, if a short ranged valence bond glass is formed, for example by adding strong randomness to a magnetic phase, then  in the thermodynamic limit it would become unstable to spin--1/2 defects,  which would exhibit the strong-disorder physics discussed above.

Random singlet physics has  been proposed to describe the dilute impurity spins (Cu-Zn substitution sites) in the spin liquid candidate Herbertsmithite\cite{Singh2010,Lee2016}. It was also studied theoretically\cite{Biroli2008} on  random graphs with fixed connectivity (related to the Bethe lattice), in a large-$N$ limit, which found that spin excitations were gapless (``pseudo-gapped''). 
It is not obvious how a parametrically broad distribution of `bare' coupling strengths can naturally arise in crystalline lattice magnets without dilution, so at first glance it is not clear how this kind of strong disorder physics can play a role.  Counterintuitively, we will show that a broad-randomness phenomenology can arise even in a tractable {weak}  disorder regime for certain 2D lattice magnets.

\smallsection{Overview}
We will start in the following section (Sec. \ref{VBSdis}) with VBS phases subjected to weak disorder. 
The weak disorder limit $\Delta  \ll \Delta_S, \Delta_{VB}$  provides a clear separation of length scales, and allows predictions that are independent of the precise choice of microscopic  Hamiltonian. We will argue subsequently that many phenomenological conclusions are relevant also in the intermediate disorder limit $\Delta_{VB} \lesssim \Delta \ll \Delta_S $. 

Recall that VBS phases have no magnetic moment and preserve time reversal symmetry, but spontaneously break some  crystal symmetries\footnote{While in the early literature the term ``valence bond solid'' (VBS) was used for singlet phases that  preserve lattice symmetry, such as the AKLT state\cite{Tasaki1988},  it has since become standard to use it only for singlet phases that break some lattice symmetries. We follow this convention, i.e.\  ``valence bond solid'' is synonymous with ``valence bond crystal'' or ``spin-Peierls phase''.} (for a simple example see Fig.~\ref{fig:VBS}). Various VBS phases have been observed numerically\cite{Sandvik2007,White2013,Fisher2014} as well as experimentally\cite{Cheong2000,Cheong2002,Motome2003,Kato2007,Takagi2008} in quantum magnets whose classical magnetic configurations are frustrated.

Since the VBS order parameter transforms under lattice
symmetries, bond randomness
couples to it as a random \textit{field}. Consequently VBS phases are highly sensitive  to  disorder. The fate of the system may  be analyzed in successive steps,  associated with distinct scales of renormalization group flow. 

The first step is the destruction of long range VBS order by disorder. The second step, at lower energies,
involves topological defects in the VBS order parameter: in the simplest cases these are vortices which carry spin--1/2 (in some cases low energy states in domain walls are also important).  These moments are nucleated even though the scale $\Delta$ of disorder is much smaller than the spin gap $\Delta_S$: the local disorder is not strong enough to break a singlet, but the rigidity of the VBS order parameter allows the effects of disorder to accumulate over large scales.
The  spin--1/2 defects  form a random network  
with broadly distributed couplings, leading to formation of long range singlets and spin clusters which is captured,  at least up to a large lengthscale, by a strong disorder RG.  

In Sec. \ref{1d_section} we warm up by reviewing, and clarifying some aspects of, this physics in 1d where it is known that the random singlet phase is one possible fate. In 2d,  the final long length scale  physics is subtle and depends on the details of the lattice, as we discuss for the square lattice in Sec.~\ref{square_lattice_section} and the triangular lattice in Secs.~\ref{tri_lat} and \ref{other_triangular_vbs_section}. At the longest length scales the strong disorder RG  breaks down (in 2d), with the likely fate of the system being very weak spin glass order (on the triangular lattice) or  N\'eel order (on the square lattice). We will substantiate this picture in the following. The theory enables strong randomness to arise from weak disorder through a multi-scale renormalization group flow, which is initiated by magnetic frustration at the lattice scale and remains protected by geometrical frustration down to ultra-low energy scales.

In Sec. \ref{other_triangular_vbs_section} we provide an alternate point of view on these results by formulating a quantum Landau-Ginzburg--like framework  to discuss the effects of disorder on a valence-bond solid. 
This theory is not a standard Landau-Ginzburg theory, which is an   expansion in powers of the order parameter and its gradients. 
  Such a `naive' Landau-Ginzburg expansion should really be thought of as an expansion about a trivial symmetry-preserving state. In quantum magnets with, say, an odd number of spin-$1/2$ per unit cell,   Lieb-Schultz-Mattis (LSM)   theorems  \cite{Mattis1961,Oshikawa2000,Hastings2004}
   prohibit such a trivial symmetry preserving phase.  Thus the standard Landau-Ginzburg expansion is not available for such a quantum magnet. 

How can we construct a useful effective theory for a quantum magnet in the absence of a trivial symmetry preserving phase?  There is a well-known alternative\cite{SenthilFisher2000} that we will briefly describe and utilize. Though there is no trivial symmetry-preserving state for a magnet with spin-1/2 per unit cell, there do exist symmetry-preserving gapped phases. These states necessarily have topological order. The simplest one  that is consistent with LSM restrictions is a $\mathbb{Z}_2$ quantum spin liquid.\footnote{Its universal physics is described by a deconfined $\mathbb{Z}_2$ gauge theory.}  The best we can then do, in the spirit of Landau-Ginzburg theory, is to develop an effective `quantum' Landau-Ginzburg theory for the quantum magnet in terms of the excitations of this $\mathbb{Z}_2$ quantum spin liquid. This takes the form of a $\mathbb{Z}_2$ gauge theory coupled to $\mathbb{Z}_2$ charges.  In the presence of spin $SO(3)$ and time reversal symmetries, the symmetry properties of the  excitations of the $\mathbb{Z}_2$ spin liquid are severely constrained.\cite{Sachdev1991,Vojta1999,Chen2016c,Fang2015,Bonderson2016}
In Sec. \ref{other_triangular_vbs_section} we use the resulting `quantum' Landau-Ginzburg theory  as a framework to discuss disorder effects.

In Sec. \ref{strong_disorder_section} we address the   intermediate-disorder regime (Fig.~\ref{fig:vbginstability}). We use the dimer model alluded to above to discuss pinning of singlets (Sec. \ref{rdmdmr}) and to argue that a defect-free `valence-bond glass' is unstable.
We confirm using the `quantum' Landau-Ginzburg theory that the results hold more generally for a strongly pinned VBS order parameter  (Sec. \ref{pinnedVBS}).  We study the  random dimer model on the triangular lattice numerically, and provide a simple explanation for the necessity of a finite density of monomers in this limit. Our results for the effect of disorder on triangular-lattice classical dimers may also be of independent interest.

Using insights from these analyses, in Sec. \ref{LSM_section} we present some natural  conjectures regarding LSM--like restrictions on the possible ground states of disordered magnets with spin symmetry and statistical translation symmetry.

In Sec. \ref{ymgo_app} we  discuss the phenomenology of experimental observables from the standpoint  described in this paper. 
For concreteness we focus on YbMgGaO$_4$ though the results are more general. 
This material or any other of course may not be in the controlled theoretical limit $\Delta \ll \Delta_S$, and indeed here there is no experimental access to the clean limit that is the conceptually useful starting point in the theory. 
However we note that a quantum paramagnetic regime has been found\cite{Sheng2015,White2015,Chernyshev2017}  in numerical work on the triangular lattice Heisenberg model  for weak second-neighbor exchange $J_2 \gtrsim 0.07 J_1$. 
Assuming  a qualitative theoretical description motivated by our results in the theoretically controlled limits, in terms of  randomly pinned short range singlets together with larger scale excitations arising from defects, we discuss the expected phenomenology for a variety of experimental probes: specific heat $C(T)$, magnetic susceptibility $\chi(T)$, thermal conductivity $\kappa(T)$, and dynamical spin structure factor $S(q,\omega)$, as well as their behavior in an applied magnetic field. We also briefly discuss NMR and $\mu$SR.
We compare all of these theoretical expectations to available experiments.  We also discuss the isostructural compound YbZnGaO$_4$  and various other material candidates for investigating   random valence bond physics.

We conclude with a summary of our results, and a discussion of the questions raised, in Sec. \ref{disc}. 
Several Appendices contain additional details. 


\bigsection{Disordering a valence bond solid}
\label{VBSdis}
We consider spin Hamiltonians featuring  both frustration and  bond  randomness. We begin by considering the simple case where the Hamiltonian has spin $SO(3)$ invariance,
\begin{equation} 
H= \frac{1}{2}\sum_{i, j} J_{i j} \vec{S}_i \cdot \vec{S}_j, \qquad  J_{i j} \equiv \bar{J}_{i j} + \Delta J_{i j}. 
\label{eq:hamiltonian}
\end{equation}
Later we will relax this requirement and consider more general Hamiltonians\footnote{These will take the form  ${H= \frac{1}{2}\sum_{i, j} S_i^\mu J_{i j}^{\mu \nu} S_j^\nu}$ with ${J_{i j}^{\mu \nu} \equiv \bar{J}_{i j}^{\mu \nu} + \Delta J_{i j}^{\mu \nu}}$, {\em i.e} the J should be taken to be $3 \times 3$ matrices in spin space.}
 appropriate for spin-orbit coupled systems like YbMgGaO$_4$.
 Here $S^\mu \equiv \sigma^\mu /2$ is the spin-half moment 
 and  $J_{i j}  $ is the    exchange interaction  for the pair of sites  $(i,j)$ on some lattice, for example the triangular lattice.  In this section we consider the case where the bond randomness scale $\Delta J $ is  much weaker than the mean value $\bar{J} $.

We assume that the ground state is a paramagnetic VBS state when $\Delta J =0$. In contrast to conventional magnetically ordered phases, where bond randomness is irrelevant, the linear coupling to the VBS order parameter means that  VBS order is destroyed even by arbitrarily weak randomness in dimensions $d\leq 2$.
The Imry-Ma argument\cite{Ma1975} shows that  domain walls in the VBS appear on a length scale $\xi \sim (J^2/ {\Delta J^2})^{1/(2-d)}$ when $d < 2$. In $d = 2$ a more sophisticated real space renormalization group treatment of the domain walls\cite{Binder1983, nattermann1988random} yields a length scale $\xi_{2d} \sim \exp [J^2/ {\Delta J^2}]$ at which the long range order is broken up.

Let us now consider the physics that results at lower energies or on length scales larger than $\xi$. Since the order parameter is pinned by the coarse-grained random field on these scales, one might at first sight expect that  there will be no remaining modes at low energies. This would be the case if the order parameter was, say, the Ising spin in  the random field Ising model (modulo rare region effects). But, we argue below,  this `trivial' RG endpoint is forbidden on topological grounds for a VBS order parameter on the square or triangular lattice. The random field necessarily introduces topological defects in the VBS order parameter, in particular certain vortex defects. Crucially, these topological defects carry spin--1/2 degrees of freedom that are not bound into short-range singlets. The physics on scales $\gtrsim \xi_{2d}$ is that of a random network of these defect spins, leading to a strongly disordered regime even when the bare disorder strength $\Delta J$ is small.  The low temperature response is then dominated by this network of broadly distributed defect spins.

 Before describing this physics for triangular lattice magnets, we  consider 1d chains and 2d square lattice magnets, which are interesting in their own right.

\smallsection{Spin-1/2 chain}
\label{1d_section}
A spin--1/2 chain in the spontaneously dimerized phase shows the simplest version of this physics.\cite{Girvin1996, 
yang1996effects,lavarelo2013localization,shu2016properties} We imagine perturbing such a Hamiltonian (e.g.\ the $J_1$-$J_2$ chain with $J_2 > 0.2411 J_1$ \cite{Zeng1995, Sorensen2013}) by adding weak, short-range correlated bond randomness. A given realization of the randomness will break translational symmetry, but we assume that translational symmetry is preserved on average: i.e. that the probability distribution of the disorder is translation invariant.

Weak disorder induces static domain wall  defects in the VBS order parameter with typical separation\cite{Ma1975} ${\xi_{1d} \sim J^2/\Delta J^2}$, so VBS order is lost at this scale. However each domain wall   carries a single unpaired spin--1/2 moment. Virtual processes  induce random  couplings between the domain wall spins, with the strength of the coupling falling off exponentially in the domain wall separation $r$: schematically, $|J_\text{eff} | \sim e^{-r /\eta}$, where $\eta$ is a spin correlation length associated with the dimerized region. This leads to a `renormalized' spin chain with random couplings. 

The exponential sensitivity to separation, combined with the random locations of the defects, means that these couplings are very broadly distributed.  A strong disorder RG approach is therefore appropriate for the next stage of RG flow.\cite{Ma1980,Hu1979,Fisher1994} 

What happens in this flow depends on the sign structure of the effective couplings, as we clarify below. In the simplest case all the effective couplings between adjacent defects are antiferromagnetic. As is well known, such a random AF chain flows to the random singlet fixed point.\cite{Ma1980,Fisher1994} Roughly speaking, this is a state where each spin is paired into a singlet with another spin which may be arbitrarily far away. This yields excitations at arbitrarily low energies that are associated with breaking weak, long--distance singlets.

Note that the above is in stark contrast to the fate of a 2-leg  ladder with antiferromagnetic couplings. Again let us start in a columnar VBS phase, where parallel valence bonds form within each chain of the ladder. The VBS order parameter is again Ising--like, since there are two degenerate VBS tilings. However domain walls between these two tilings now host two unpaired spins. These spins will form a singlet, so that no modes survive to longer lengthscales. This difference between the single chain and the 2-leg ladder is indicative of a  Lieb-Schultz-Mattis--like constraint on disordered magnets that we will return to later.

Now let us discuss the signs of the effective couplings $J_\text{eff}$ in more detail. The essential question is whether the signs of the effective couplings $J_\text{eff}$ are deterministic or random in the weak disorder limit $\Delta J/J \ll 1$. In fact, both scenarios are possible without fine-tuning. Which one occurs  depends on the nature of the spin correlations in the VBS phase of the clean system, as we discuss in Appendix~\ref{sign_structure_appendix}. The amplitude of these spin correlations of course decays exponentially with distance, but the sign structure may be either commensurate or incommensurate with the lattice. The two possibilities are exemplified in the $J_1-J_2$ chain. In the region of the VBS phase   close to the phase boundary to the gapless phase, for $J_2/J_1\gtrsim 0.2411$, the sign of $\langle\vec S(0).\vec S(r)\rangle$ alternates with period 2. On the other hand spin correlations become incommensurate for   $J_2/J_1\gtrsim 0.52$.\cite{Zeng1995}

In the former case $J_\text{eff}$ between  two adjacent defects, which necessarily occupy opposite sublattices, will always be antiferromagnetic. Interestingly, these couplings are  also guaranteed to be antiferromagnetic for a class of Hamiltonians that permit a sign-free Monte-Carlo treatment: a general argument for this is in Appendix~\ref{sign_structure_appendix}. On the other hand, when the spin correlations of the clean system are incommensurate, the large random separations of defects imply that the signs of the couplings $J_\text{eff}$ will be independently random in the limit of weak disorder, i.e. the limit of large $\xi_{1d}$ (Appendix.~\ref{sign_structure_appendix}).

The two possibilities lead to different behaviour at large lengthscales when the VBS is perturbed by weak disorder. As noted above, the antiferromagnetic case leads to the random singlet phase. Conversely, when ferromagnetic couplings are present, strong disorder RG shows that a different fixed point is reached, involving the generation of large effective spins during the RG procedure.\cite{Lee1995,Lee1997} (This  fixed point is not at infinite randomness, so the SDRG treatment of it is not strictly controlled.) Physically, the resulting phase  should perhaps be thought of as having `quasi--long range' spin glass order.  We will encounter related issues  in two dimensions.
 
Some of these theoretical expectations for $J_\text{eff}$ are manifested experimentally in  quasi-one dimensional S=1/2 two-leg-ladder materials\cite{Takagi1997,Mendels2009} such as SrCu$_2$O$_3$ and BiCu$_2$PO$_6$. Hole doping via Cu to Zn substitution, at the level of less than a few percent, shows a signal of spin freezing into a pattern with three dimensional antiferromagnetic N\'eel correlations. The ordering occurs among the moments that are induced by a nonmagnetic impurity, which roughly speaking removes one of the Cu electrons from  a singlet bond. Due to the well-defined sublattice sign structure, the coupling between these dangling moments is unfrustrated, producing antiferromagnetic correlations with spins up on one sublattice and down on the other.

\smallsection{Spin-$1/2$ square lattice}
\label{square_lattice_section}
Now let us consider square lattice spin-$1/2$ magnets in a  columnar VBS phase. This fourfold-degenerate VBS pattern can be associated with four cardinal directions of a planar vector $\vec \varphi$. In addition to domain walls, it admits a discrete $Z_4$ vortex defect which carries an unbound spin--1/2 in its core\cite{Senthil2004}.  This vortex defect is a junction between four `elementary' domain walls, across which $\varphi$ rotates by $\frac{\pi}{2}$. The VBS order will be disrupted at the Imry-Ma lengthscale $\xi_{2d}$ where such domain walls appear.  We make the natural assumption that `elementary' domain walls are less energetically costly than `composite' domain walls where $\vec \varphi$ flips sign.

One may argue that the breakup of the VBS into domains necessarily also introduces VBS vortices with typical separation $\xi_{2d}$. This is because, when $\xi_{2d}$ is large,  the core energy cost of a vortex is negligible in comparison with the typical energy cost of domain rearrangements  on this scale. See   Appendix~\ref{app:nucleation} for more detail (similar arguments have been made in the context of the random field XY model\cite{gingras1996topological}).  

The spin$-1/2$ moments in the vortex cores will then determine the eventual fate of the system. As in 1d, a key feature of this system is that the   distribution of couplings $J_\text{eff}$ for adjacent vortex spins is extremely broad, despite the fact that the bare disorder $\Delta J$ is weak. Since the magnitude $|J_\text{eff}|$ depends exponentially on the separation of the vortices, while the distribution of separations has mean and width both of order $\xi_{2d}$, the distribution of $J_\text{eff}$ is broad even on a logarithmic scale.  

On the square lattice, a key geometrical fact is that core spins of $Z_4$ vortices and anti-vortices are associated with opposite sublattices. Therefore a given defect has a well-defined sublattice assignment, even when there are quantum fluctuations in its precise position. Further, we show in Appendix.~\ref{sign_structure_appendix} that for a natural class of models the {\em sign} of the effective interaction is determined solely by whether the two core spins are on the same or opposite sublattices. (This class includes, but is not limited to, sign-problem-free models for VBS phases such as the much-studied $J$--$Q$ model\cite{Sandvik2007}.) The effective Hamiltonian for the defect spins then takes the form 
\begin{equation}
H_\text{eff} = - \sum_{r,r'} J^\text{eff}_{rr'} \epsilon_r \epsilon_{r'} \vec S_r \cdot \vec S_{r'},
\end{equation}
where $r, r'$ form a random  (but correlated) selection of sites of the square lattice with typical separation $\xi_{2d}$, and $J^\text{eff} > 0$ is exponentially small in the separation between $r$ and $r'$, with the sign $\epsilon_{r} = \pm 1$ for the two sublattices/vorticities. Note that, despite the strongly random sign and magnitude,  this interaction is unfrustrated. 

For energetic reasons, the closest defect  pairs will predominantly be of opposite vorticity (and thus antiferromagnetically coupled), so a significant fraction of the defect spins will be bound into singlets of size $\sim \xi_{2d}$ in the very first step of strong disorder RG. But in this 2d problem, the strong disorder RG is expected eventually to break down, 
due to a flow \textit{away} from infinite randomness.\cite{Lee1982, Igloi2003, Fisher2000,Troyer2012}  
The breakdown of the strong disorder RG, at late stages of the singlet and cluster formation process, does not automatically rule out nontrivial disordered fixed points at finite randomness. \cite{beachtalk}
\footnote{\label{note:noteadded}
Note Added: 
excitingly, it seems this may be possible. Shortly before publication of the present manuscript an extensive numerical study appeared\cite{forthcomingAnders} reporting evidence for a random-singlet phase in a disordered $J-Q$  model on the square lattice.}
\cite{forthcomingAnders}
However since the interaction here is unfrustrated, a  very simple alternative scenario on the square lattice is that at long scales the disordered VBS could transform itself into a dilute very weakly N\'eel ordered state. 
(On frustrated lattices very weak spin-glass order is the natural possibility at the very longest scales, beyond the random-singlet-like regime, as we discuss below.)
 As in 1d spin chains, depending on the initial clean Hamiltonian on the square lattice, we can also in principle obtain effective couplings with fully randomized signs and no sublattice sign structure (Appendix.~\ref{sign_structure_appendix}); we discuss this case in the following section.
 
\begin{figure}[]
\includegraphics[width=\columnwidth]{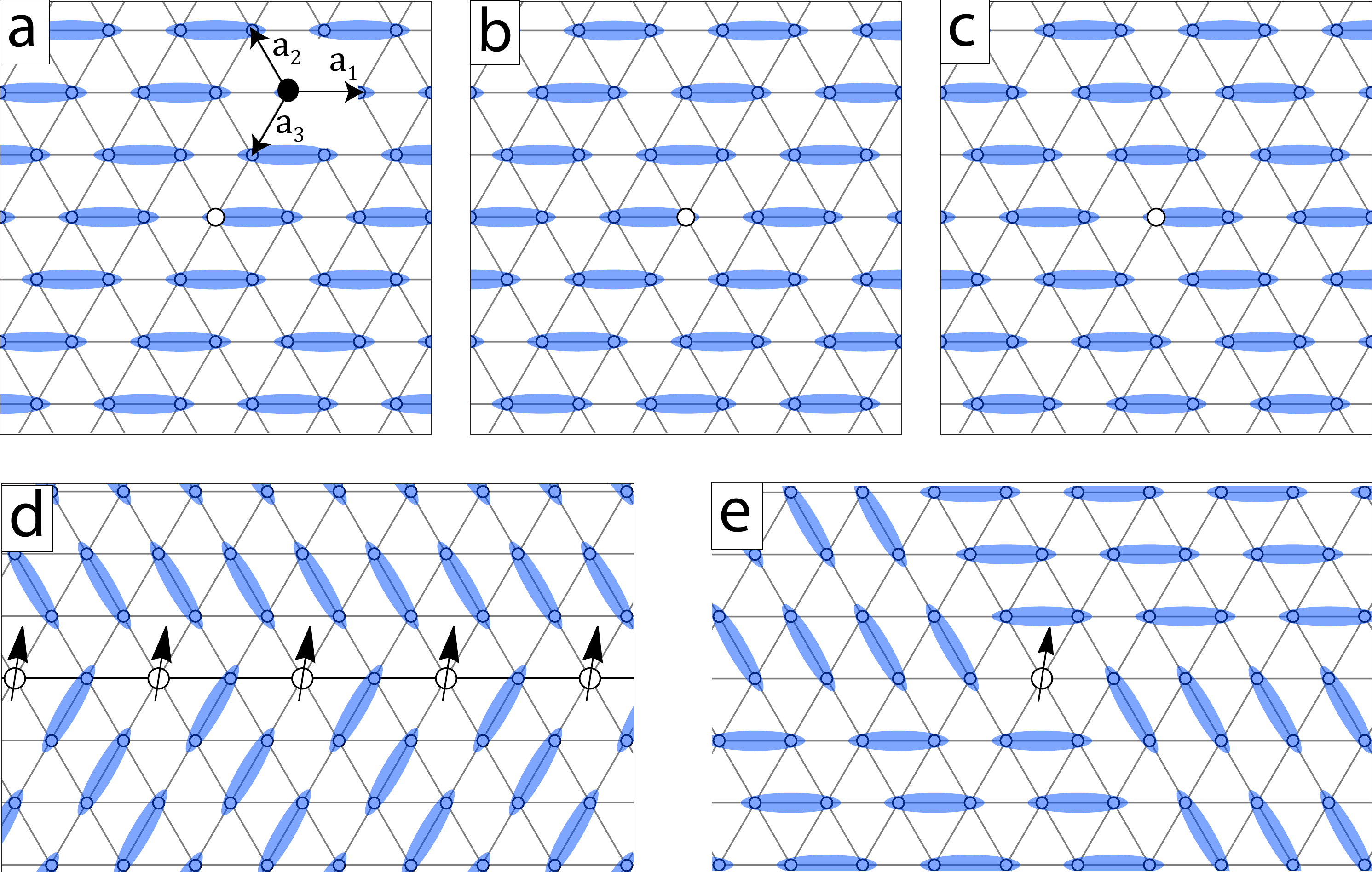}
\caption[]{ \textbf{Triangular lattice columnar VBS order and its topological defects.} 
\textbf{(a,b,c)} 
Frustration from the triangular lattice endows the columnar VBS order with 2-fold chirality as well as 6-fold orientation: the 12 domains are given by $2\pi/6$ rotations of (a) [note common origin in (b)] as well as its opposite-chirality partner (c).
\textbf{(d)} Superdomain walls  (black lines) between domains of different sets necessarily host spin--$1/2$ chains  when their path takes any orientation except for a single defect-free orientation.  In particular this implies that coarse grained superdomain walls host broken-up spin chains.
\textbf{(e)} Within a given superdomain, topological defects similar to  square lattice $Z_4$ vortices can occur.
} \label{fig:VBSdomain}
\end{figure}

\smallsection{Spin-1/2 triangular lattice: columnar VBS}  
\label{tri_lat}
We are now ready to consider the triangular lattice. We start with columnar VBS order, since this is the simplest to visualize. The infra-red fate in this case is fairly intricate, involving several lengthscales. In Sec.~\ref{other_triangular_vbs_section} we comment on another VBS order with a larger unit cell.

For the columnar VBS there are 12 symmetry-related ground states. These can be labelled by a pair of vectors $(a:b)$. The first vector $a$ specifies the direction of the valence bond out of some fixed base point $r_0$ on the lattice, and can take the values $a_i$ for $i=1,2,3$ (the three vectors shown in  Fig.~\ref{fig:VBSdomain}a) and also $\bar a_i \equiv - a_i$ for $i=1,2,3$. The second vector $b$ specifies the axis along which the columns line up, and along which the pattern is invariant by a unit translation. Since the sign of $b$ has no meaning, it is enough to let $b$ be one of the three vectors $a_{1,2,3}$. We will denote the various possibilities by $(i: j)$ or $(\bar i: j)$ for $(a_i: a_j)$ or $(\bar a_i:a_j)$ respectively.  Since $j\neq i$, there are 12 columnar VBS patterns in total. For example the domains in Fig.~\ref{fig:VBSdomain}(a,b,c) are labelled $(1{:}3)$, $(\bar 1{:}3)$, $(1{:}2)$ respectively. 

As we will see, it is natural  to group the 12 VBS patterns into three sets of 4. We denote these sets $[1,2]$, $[2,3]$, $[3,1]$. 
These groupings have a simple geometric meaning: 
each set corresponds to one of three ways of viewing the triangular lattice as a square lattice with an extra  diagonal bond. For example, removing all the bonds parallel to $a_3$ yields one possible square lattice. The corresponding VBS patterns $(a:b)$ are those that can be drawn on this square lattice: they  make use only of the vectors $a_1$ and $a_2$. This gives the set $[1,2]$, which includes the 4 VBS patterns  $(1{:}2), (2{:}1), (\bar{1}{:}2)$ and $(\bar{2}{:}1)$.

We distinguish two kinds of domain walls.  By the above construction, domain walls between  patterns \textit{within} a given set $[i,j]$ are mapped to the  VBS domain walls we already encountered on the square lattice. 
They can similarly be identified as composite and elementary, with four elementary domain walls emanating from a VBS vortex. Such a vortex is shown in Fig.~\ref{fig:VBSdomain}(e). 
Each type of domain allows a straight boundary with two possible orientations. These allowed orientations are the same for all the domains belonging to a given set $[i,j]$, 
and are parallel to the vectors $a_i$ and $a_j$.
This means that an elementary `intra-set' domain wall can alternate between these two orientations without incurring unpaired dangling spins. As a result, after coarse-graining these domain walls can take an arbitrary path without incurring `frustration'.

Domain walls between patterns from \textit{different} sets, which we dub \textit{super}domain walls,  are more complicated. An example can be seen in Fig.~\ref{fig:VBSdomain}(d).
Each superdomain wall generically has a finite density of dangling unpaired spins along its length. More precisely, it  can avoid this fate only when it lies exactly parallel to a single preferred orientation.   A superdomain wall between superdomains $[i,j]$ and $[i, k]$ can be `unfrustrated' only if it lies parallel to the  lattice vector $a_i$ that is shared by the two superdomains.
 For the superdomain wall in Fig.~\ref{fig:VBSdomain}(d), the unfrustrated orientation would be along $a_2$. A non-straight path, as is required generically between two superdomains, must deviate from this single allowed orientation, and hence will  nucleate a finite density of dangling spins along the path. This contrasts with the \textit{intra}-set domain walls, which are able to avoid dangling spins even if they are not straight. 

Now let us turn on bond randomness. The VBS pattern will fragment into domains at a scale $\xi_{2d}$. This is the lengthscale at which the `cheapest' domain walls proliferate. The most natural possibility is that these are the `unfrustrated' elementary domain walls within a given set $[i,j]$. The superdomain walls will then proliferate only at a parametrically longer length scale $\xi_S \gg \xi_{2d}$.

More precisely, the line tension of domain walls decreases  under coarse graining, due to energy optimization from short-scale disorder pinning.\cite{nattermann1988random} (A detailed RG treatment in the present case would have to take account of the anisotropy of the line tension, with different preferred directions for different domain walls.) $\xi_{2d}$ is the lengthscale where the line tension of the cheapest domain walls can no longer compete with the energy gain from disorder when domains are introduced. On longer scales, the line tension of the more expensive superdomain walls continues to renormalize downwards, and superdomains eventually appear on the lengthscale $\xi_S$. Since the Imry-Ma lengthscale depends exponentially on the ratio of the bare line tension to disorder, the larger bare line tension for superdomain walls means that $\xi_S/\xi_{2d}$ will be exponentially large in the limit of weak disorder.
 
This results in a range of scales $\xi_{2d} \ll L \ll \xi_S$ where (after averaging over scales of order $\xi_{2d}$) translation and reflection symmetries are restored, but 60 degree lattice rotation remains broken. On this scale there is a patchwork of domains all belonging to one set $[i,j]$, for example the set $[1,2]$. In this regime there is therefore long range order for a lattice \textit{nematic} order parameter which picks out one of three unoriented vectors $a_i$ (in this example $a_3$, i.e. the vector not included in the set $[1,2]$). This phenomenon, where lattice nematic order is longer ranged than lattice translation order, has been dubbed  `{vestigial order}'.\cite{Kivelson2014}  

The patchwork of domains from a single set $[i,j]$ can be related to domains on the square lattice, as noted above. As on the square lattice, this leads to vortices which have spin-$1/2$ moments at their cores, and a broad distribution of exchange strengths for the effective interactions $J_\text{eff}$ between defects,
\begin{equation}
H_\text{eff} = \sum_{r,r'} J^\text{eff}_{rr'} \vec S_r \cdot \vec S_{r'}.
\end{equation}
What are the signs of these couplings? For the triangular lattice,  fully randomized signs seem the most natural possibility, since this is what happens when the VBS order in the clean system is not too strong: this is argued for heuristically in Appendix~\ref{sign_structure_appendix}. In this regime, the non-bipartiteness of the triangular lattice ensures the signs are scrambled for large $\xi_{2d}$. We will focus on this case in the following.  (Somewhat surprisingly, there is in principle another possibility: for an appropriate choice of clean Hamiltonian, the couplings can have a definite sign structure \textit{within} a given superdomain.\footnote{At a minimum this requires the VBS order to be strong, strongly breaking the symmetry of the triangular lattice locally. As we noted above, a given superdomain type (e.g. $[1,2]$) corresponds to  a way of  embedding a square lattice into the triangular lattice. If the `extra' diagonal bonds were deleted we would be left with the square lattice problem. In this case there is  a  mechanism for  stabilizing an unfrustrated sign structure for $J_\text{eff}$ which we discussed in Sec.~\ref{square_lattice_section}. (This mechanism relied on the fact that vortices and antivortices are assigned to definite sublattices of the bipartite square lattice.) There is no obstacle in principle to this  sign structure surviving when the diagonal bonds are reintroduced; this depends on nonuniversal features of the Hamiltonian (Appendix.~\ref{sign_structure_appendix}). In this situation the ultimate fate of a superdomain may be `N\'eel' ordered, with the N\'eel pattern inherited from the effective square lattice. The definition of the N\'eel order parameter differs between distinct superdomains, so on scales larger than $\xi_S$ the N\'eel--ordered superdomains are coupled together in a frustrated manner, likely leading eventually to a spin glass fixed point.})

For the case of random signs, within a superdomain,  
\begin{equation}
H_\text{eff} = \sum_{r,r'} J^\text{eff}_{rr'} \vec S_r \cdot \vec S_{r'},\qquad
\operatorname{sign} J^\text{eff}_{rr'}= \pm 1.
\end{equation}
We may contemplate a strong-disorder RG treatment for the vortex defects, in which the strongest couplings (either F or AF) are sequentially integrated out. The breadth of the distribution of $J_\text{eff}$ will ensure that this procedure is accurate up to a parametrically large lengthscale. 

During the procedure the presence of F couplings will lead to the formation of large, randomly coupled effective spins\cite{Lee1982,Lee1995,Lee1997, Igloi2003}, 
 and the eventual fate is very likely to be spin glass order. The corresponding lengthscales would require more detailed analysis. 

In the present context (columnar VBS) we must also remember that on the scale $\xi_S$ the superdomain walls appear. These make up a network of 1d spin--1/2 chains, subject locally to weak disorder. These chains will also be gapless at the lowest energies, for example exhibiting the 1d `large spin' phase discussed in Sec.~\ref{1d_section}.  The relative abundance of different types of excitations (associated with vortex spins versus domain walls) at different energy scales is likely to depend on nonuniversal parameters such as the ratio $\xi_S /  \xi_{2d}^2$ and the energetics on the domain walls.

Just as on the square lattice, the basic point is that the Imry-Ma instability of the  VBS  led to a proliferation of spinful defects.  This is true also for  other VBS patterns on the triangular lattice which are not so easily visualized.

\subsection{Other VBS states via proximate $\mathbb{Z}_2$ spin liquid}
\label{other_triangular_vbs_section}
The necessity of `spinful' topological defects, which we saw directly at the lattice level for the columnar VBS, is a more general phenomenon. It is useful to develop a coarse-grained approach that can be applied to  VBS orders with  a more complex structure, e.g. a large unit cell, or strong quantum fluctuations, for which defects are not easily visualized at the microscopic level.

For  ordered phases in classical magnets, the natural coarse-grained language  is Landau-Ginzburg theory which is an expansion of the free energy density in powers of the order parameter and its gradients. As explained in the Introduction, for  quantum magnets with  spin-$1/2$ per unit cell, an analogous Landau-Ginzburg theory is not available. The best we can do is a more sophisticated `quantum' Landau-Ginzburg theory\footnote{This is often also called a `starred' Landau-Ginzburg theory.} which uses a description in terms of the excitations of a gapped topologically ordered $Z_2$ quantum spin liquid.

Consider a gapped $\mathbb{Z}_2$ spin liquid on the (clean) triangular lattice, with a spin--1/2  `spinon' excitation and a spinless  `vison' excitation $v$, both of which are bosonic. Each sees the other as a $\pi$  flux (they are mutual semions). Condensing either of these excitations confines the other one, and destroys the topological order.  The spinon and vison, together with their bound state, which is a fermion, exhaust the three nontrivial anyon types in the spin liquid.

In any such spin liquid the vison field $v$ must transform nontrivially under lattice symmetries, for a reason touched on above. If $v$ was trivial under lattice symmetries, then a symmetric gapped state, without topological order, could be obtained by condensing $v$. This would be inconsistent with the  LSM theorem. When the spin Hamiltonian has spin $SO(3)$ symmetry, it is known, in fact, that the vison --- defined to live on the dual honeycomb lattice ---  sees a background $\pi$ flux through each plaquette.\cite{Sachdev1991,Vojta1999,SenthilFisher2000, Sondhi2001, Chen2016c,Fang2015,Bonderson2016}  Physically one can simply think of the microscopic spin-$1/2$ magnet as having one `background' spinon at each site of the triangular lattice.  The visons see this spinon as $\pi$ flux. This restriction then completely fixes the action of lattice symmetries on the vison field.  Lattice translation, for example, acts projectively. 

An immediate  result of the nontrivial action of lattice symmetry on $v$ is that  its condensation breaks spatial symmetry to give a VBS phase. Therefore a quantum Landau-Ginzburg theory for $v$ allows us to discuss the VBS order and its defects. (We will put this theory on a lattice, since this makes it easier to handle the $\mathbb{Z}_2$ gauge structure.) Starting from the spin liquid, a variety of VBS phases can be described by various condensation patterns for $v$ (See Appendix \ref{appendix:QSLVBStransition} for details).  For concreteness we will restrict to a specific one.

A  mean field  treatment of the vison field on the triangular lattice suggests that  one natural possibility for the condensed state is a  plaquette VBS with a 12 site unit cell.\cite{Sondhi2001,Mila2008} In this continuum treatment, the vison field is  viewed as a 4--component vector $\vec v$, whose four components  arise from four  low-lying modes in the vison's microscopic dispersion.\footnote{Other VBS phases described in the literature\cite{Xu2014} are described in exactly the same way, but with a 6---component vector $\vec v$.}  The field $\vec v$ changes sign under $\mathbb{Z}_2$ gauge transformations, and the physical VBS order parameter is a bilinear in $\vec v$. There is a discrete set of 48 possible directions for ordering of the vector $\vec v$ and 24 for the physical VBS order parameter.

Let us write down a schematic energy functional for the vison. We are interested in a  regime where quantum fluctuations of the (pinned) VBS order are irrelevant, so it suffices to think about minimizing  a classical energy. For convenience we regularize the vison condensate vector $\vec v$ on a (fictitious) coarse-grained lattice, giving a description for the low energy vison field $\vec v$ through a lattice gauge theory:
\be\label{vison_energy}
E = - t \sum_{\<ij\>}  \sigma_{ij} \vec v_i \cdot \vec v_j  -  d \sum_i (\vec h_i \cdot \vec v_i)^2 + \ldots 
\ee
The $t$ term captures the stiffness of the VBS order (and is  invariant under the gauge transformation $\vec v_i \rightarrow \chi_i \vec v_i$ and $\sigma_{ij}\rightarrow \chi_i \chi_j \sigma_{ij}$ with $\chi_i =\pm 1$). The $d$ term is the simplest gauge invariant coupling to weak disorder. The precise form of the interactions is not important, but the ellipses $\ldots$ must include the anisotropy terms which select out a set of \textit{discrete} ordering directions for $\vec v$, corresponding to the discrete set of degenerate VBS states. The $\ldots$ may also include an energy penalty for gauge flux excitations, which are plaquettes of the lattice in which the background flux seen by the bare vison field ($\pi$ flux per spin--1/2 site of the original lattice) is modified by the presence of an additional $\pi$ flux, seen by the low-energy vison condensate field $\vec v$.
 In the gauge theory language these gauge flux excitations are plaquettes of the lattice where the product of the gauge fields on the links is equal to $-1$:
\be
\prod_{{\<ij\>\\\in\text{plaquette}}} \sigma_{ij} = -1.
\ee
These gauge fluxes, seen by the vison, are precisely the spinon excitations of the spin liquid.  Therefore Eq.~\ref{vison_energy} must be supplemented with the crucial information that these spinons carry spin--1/2. The interactions of these spinful degrees of freedom are of course neglected in the  classical Hamiltonian above. 

We now consider topological defects in the VBS order parameter, showing that appropriate defects bind a spinon (see also Appendix \ref{appendix:QSLVBStransition}).
 It is simplest first to consider  an artificial continuum limit where the anisotropies are switched off, and $\vec v$ is allowed to `order' anywhere on the sphere $S^3$. The VBS order parameter then lives in projective space, $\mathrm{RP}^3= S^3/\mathbb{Z}_2$. (The above energy functional is then simply the standard Ising gauge theory representation for  an order parameter in projective space\cite{Toner1995}.) When the disorder $d$ is small in comparison to the stiffness $t$, this order parameter varies slowly and smoothly, except at the locations of point vortex defects. These are allowed as a result of the nontrivial homotopy group ${\pi_1 (\mathrm{RP}^3) = \mathbb{Z}_2}$. 

As we traverse a loop surrounding such a defect, the VBS order parameter smoothly traverses the topologically nontrivial cycle in $\mathrm{RP}^3$.  But the hallmark of such a  topologically nontrivial trajectory is that the vison field $\vec v$ acquires a minus sign on traversing the loop around the defect --- i.e.  that $\vec v$ has a branch cut ending at the defect. Energetics dictates that this branch point terminates at a gauge flux. This is because, for the terms $\sigma_{ij} \vec v_i. \vec v_j$ to remain positive on the links $\<ij\>$ that cross the branch cut, $\sigma_{ij}$ must be $-1$. The termination of this line  of negative $\sigma$s is a plaquette where $\prod\sigma=-1$. Therefore the presence of $\mathrm{RP}^3$ vortex defects is equivalent to the presence of gauge fluxes, i.e. spin-1/2 spinons.

Returning to the case with nonzero anisotropy, these gauge fluxes are inevitable on the Imry-Ma lengthscale.
 Fluxes necessarily accompany appropriate discrete `vortices' in the VBS order parameter, see below.  These vortices will necessarily be nucleated 
since as noted above (Appendix~\ref{app:nucleation}; see also Ref.~\cite{gingras1996topological}) the core energy of a defect cannot compete with the large energy scale associated with domain rearrangements on the Imry-Ma lengthscale.\footnote{In the random field XY model, which corresponds to the case with $\vec v = (v_1, v_2)$ and without any anisotropy, there can be a parametric separation between the lengthscale for pinning of the order parameter and the lengthscale for vortex nucleation,\cite{gingras1996topological}  but we do not expect that when anisotropy is nonzero and the order parameter is  discrete (Appendix~\ref{app:nucleation}).} Roughly speaking, a vortex will be nucleated when the random field, coarse-grained on this scale, itself displays such a vortex configuration.

In the weak disorder limit, the discreteness of the VBS order parameter must be taken into account in classifying defects. (In a given microscopic model the anisotropies that select the discrete ordering directions might happen to be weak,  manifesting themselves only on a large lengthscale. But  in the weak disorder limit this lengthscale is still  much smaller than the size of domains, so discreteness is important.) In the discrete setting a defect is a junction between $k>2$ domains, and types of defect are labelled by the list of  ordering directions $\vec v^{(1)}, \ldots, \vec v^{(k)}$ that surround the defect, in anticlockwise order. The set of possible defects depends on the set of domain walls that are generated in the Imry-Ma process. So long as each type of domain wall has a unique gapped ground state when regarded as an effective 1D system, one may argue heuristically that the flux  trapped at the defect is fully determined by the defect type. In the above model, a nontrivial flux is trapped whenever the sign of the gauge invariant quantity $\prod_{i=1}^k \vec v^{(i)} \cdot \vec v^{(i+1)}$ is negative.   One may check explicitly in this model that such defects indeed exist. These fluxes are the discrete analogues of the $\mathrm{RP}^3$ vortices, and trap localized spinons.

\bigsection{Strong pinning of valence bonds}
\label{strong_disorder_section}
Unpaired spins are energetically expensive. Nevertheless, in the controlled limit of weak disorder we saw that defects in the singlet pattern were inevitable on the Imry-Ma lengthscale. As a result, the ground state could not remain paramagnetic on the longest scales. A ground state made only of short-range, static singlets was impossible.

In this section we argue that this remains true, in two dimensions, in the regime where the disorder is no longer weak, and has a strong pinning effect on valence bonds. More precisely, we argue that a state of static  short-range singlets \textit{without} defects ---   which is what we refer to as a `valence bond glass' --- is \textit{not} a possible ground state for a square or triangular lattice magnet with short-range correlated disorder.  Defects are again nucleated, and the natural possibilities for the physics at the longest lengthscales are again those discussed in the  previous section. 

In Sec.~\ref{LSM_section} we use this result and that of the previous section to motivate Lieb--Schultz--Mattis--like conjectures on allowed ground states for magnets with 1/2--odd integer spin per unit cell.

\subsection{Instability of  valence bond glass to defects}
\label{rdmdmr}
When the spin gap is finite, we expect that the ground state can be described in terms of valence bonds with a finite typical size. In general the ground state is a superposition of  valence bond configurations. 
Here we discuss the extreme limit of valence-bond pinning, in which quantum fluctuations are completely suppressed, and the ground state is  a \textit{single} frozen configuration of nearest-neighbour singlets. In this limit there is no meaningful local VBS order parameter (but see also Sec.~\ref{pinnedVBS}). 

In the present regime the system can be mapped into a  \textit{classical} dimer model with random bond energies, at zero temperature.  Hard-core dimers living on bonds of the lattice represent the singlets.
Each lattice bond is assigned a random energy, which we draw from a distribution 
(e.g. uniform and bounded) 
with width of order $\Delta$.   In the putative `paramagnetic' state, which would correspond to a valence bond glass, the dimer covering is complete (every site belongs to a dimer) and the  energy is the sum of energies of dimer-occupied bonds.   We also allow  unpaired \textit{monomer} sites, at a large energy cost $K$.

In a given finite sample, 
the  ground state when ${K=\infty}$ is a unique complete dimer covering selected by the disorder. This energy minimization problem is nontrivial due to the dimer constraint. We ask whether this state is stable when monomers are allowed with a large but finite cost $K$. 

The universal properties of the $K=\infty$ state depend on whether the lattice is bipartite. For bipartite lattices such as the square lattice the question of stability has been addressed,\cite{Fisher1999,Middleton2000} and is closely connected to the stability of the random field XY model to vortices\cite{Ostlund1982} and the stability of the Bragg glass\cite{LeDoussal1995} describing an elastic medium subject to pinning. Numerical studies in this context\cite{Fisher1999,Middleton2000} found that, while introducing a fixed monomer  in a finite system costs  a \textit{positive} average energy $\overline{\delta E} >0$ which grows without bound with system size, the standard deviation of $\delta E$ also grows. The net result is that the typical energy cost of an \textit{optimally} placed defect in a system of size $L$ is \textit{negative} at large size and of order $\delta E_\text{opt} \sim -(\log L)^{3/2}$. At large $L$ this overwhelms the core cost $K$, so that defects are nucleated and the `Bragg glass' is destroyed.

For nonbipartite lattices the mapping to an elastic medium does not apply, and stability of the pinned `dimer glass' does not appear to have been studied. We first give a numerical treatment and then a very simple theoretical explanation.

\begin{figure}[]
\includegraphics[width=\columnwidth]{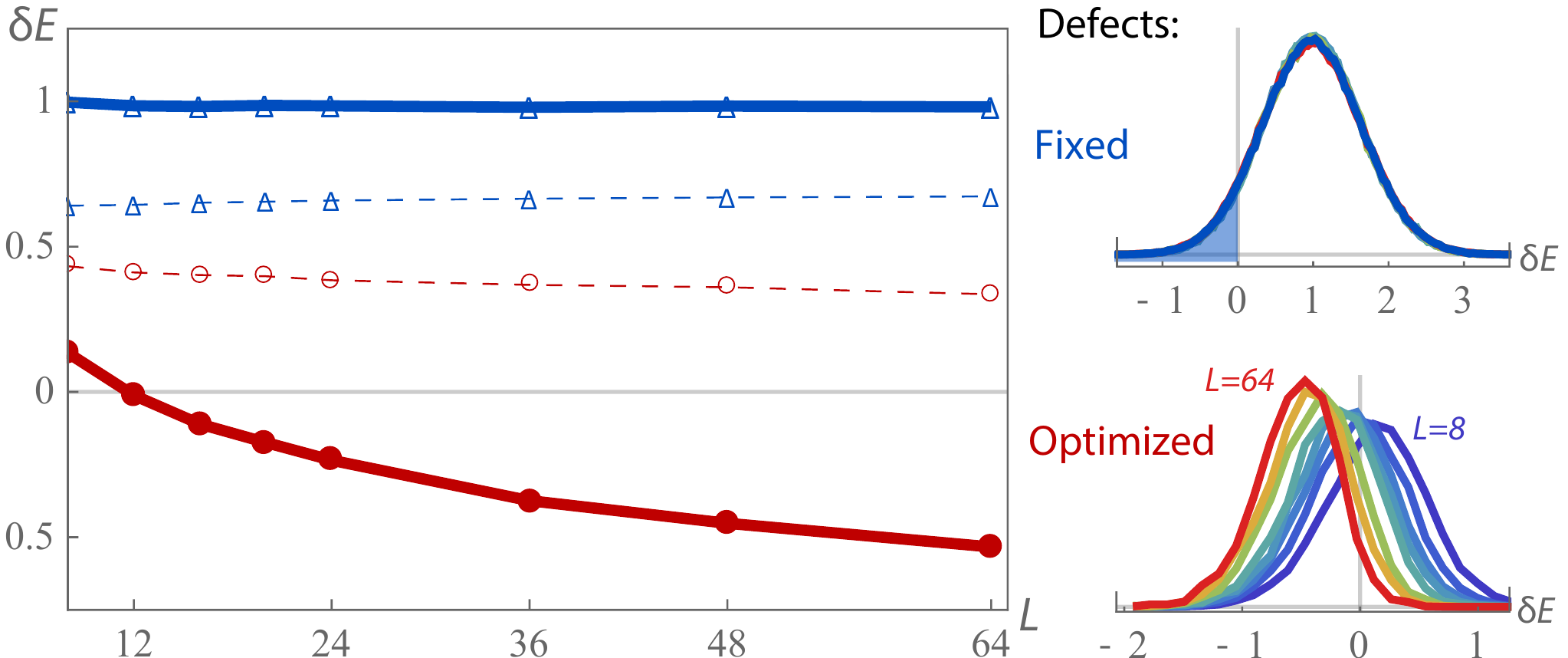}
\caption[]{ \textbf{Defect nucleation in the random dimer model.}
We obtained optimal dimer coverings numerically for an ${L\times L}$ triangular lattice, with random bond energies drawn from the interval $[-1,1]$. Monomer defects were introduced in two ways: (a) at \textit{fixed} sites (blue triangles, top); and (b) at sites chosen, from $L$ possibilities, to \textit{optimize} the defect energy cost $\delta E$ (red circles, bottom). 
The histograms of the $\delta E$ distributions for various system sizes are shown on the right. Left: their  means (solid curves) and standard deviations (dashed). (a) Fixed defects typically cost energy $\delta E > 0$: the   distribution of $\delta E$ converges to a fixed $L=\infty$ shape, with a nonzero tail at $\delta E < 0$ (shaded). (b) Optimized defects occupy these rare regions, giving an energy gain $- \delta E > 0$ which appears to increase without bound as $L\rightarrow \infty$.
} \label{fig:nucleation}
\end{figure}

To determine the stability of the dimerized state we must determine the probability distribution of the energy cost $\delta E$  for introducing a defect ($\delta E$ is measured relative to the defect--free ground state, and may be negative). We have simulated the model on the triangular lattice and studied the behavior of monomers. Results are shown in Fig.~\ref{fig:nucleation} (see Appendix \ref{app:nucleationnumerics} for details and comparison with the square lattice). 

  \text{Fixed} defects on the triangular lattice are found to have an energy  distribution which converges to a fixed form as $L\rightarrow \infty$, with mean and width of order one, unlike the square lattice case where the mean and width diverge with $L$. Varying $K$ simply shifts the mean of this distribution. If the tail of the distribution extends to arbitrarily negative values, this will  be sufficient to guarantee nucleation of defects, as we discuss below.

For a  direct check, we compute the energy of a `partially optimized' defect pair, where the optimization is over $O(L)$ configurations. We find that whereas the mean energy of unoptimized defects is finite as ${L\rightarrow \infty}$, the optimized energy becomes increasingly negative  at large $L$, in a manner consistent with the form 
  ${\delta E_\text{opt} \sim -(\log L)^{1/2}}$.

Thus despite the differences between the two classes of geometries, the ultimate conclusion is the same. By allowing previously-impossible dimer configurations that better conform to the disordered potential landscape, optimally placed  monomers reduce the system's energy at sufficiently large sizes (no matter how large the core energy) yielding  a finite density of spin--1/2 defects  in the thermodynamic limit.

These results can be simply interpreted. We conjecture that the universal properties of the defect free $K=\infty$ state are related by duality to those of the 2D Ising spin glass, which exists at zero temperature though not above.\cite{bray1984lower}  One way to motivate this conjecture is to note that if we relax the constraint that one dimer touches each site, so as to allow any odd number of dimers, there is an exact duality mapping to a standard nearest-neigbour Ising spin glass on the honeycomb lattice.\footnote{Let the dimer occupation number on a link $\ell$ of the triangular lattice be represented with a variable $\mu_\ell= \pm 1$ which is $-1$ if the link is occupied. Take the energy to be $E=\sum_\ell \epsilon_\ell \mu_\ell$, with the random bond energies $\epsilon_\ell$  independent and symmetrically distributed around zero. Let $\mu^0_\ell$ be an arbitrary fixed reference configuration. We may define an Ising spin configuration on the sites of the dual honeycomb lattice (which are located at the centres of triangular plaquettes) satisfying $S_i S_j = \mu^0_\ell \mu_\ell$, where $i$ and $j$ are the dual sites either side of link $\ell$. If the number of dimers touching each triangular lattice site is constrained only to be odd, then $S_i=\pm 1$ is unconstrained. Defining similarly ${J_{ij} = \mu^0_\ell \epsilon_\ell}$, the energy is $E = \sum_{\<ij\>} J_{ij} S_iS_j$. For the above distribution of bond energies, the $J_{ij}$ are independent and symmetrically distributed around zero, giving a standard nearest-neigbour spin glass.} 
More heuristically, note that, as in the Ising spin glass, the elementary excitations of the ground state are supported on  loops. In the Ising spin glass these (unoriented) loops are the boundaries of domains which are flipped relative to the ground state. Here, a loop excitation is supported on a chain of bonds which are alternately occupied and unoccupied, and consists in reversing the occupation of each bond on the loop compared to its value in the ground state.

In a system of size $L$, the lowest energy excitation is expected to be a loop of linear extent $\sim L$ which visits $\sim L^{d_f}$ bonds (with $d_f>1$), and which has energy $L^{\theta}$ for \textit{negative} $\theta$. (We have computed the exponent $d_f=1.28(1)$ numerically to confirm it is consistent with the Ising spin glass value\cite{bray1987chaotic,middleton2001energetics,hartmann2002large}.) The  decrease of the energy $L^{-|\theta|}$ with $L$ is equivalent to the instability of this glass to finite temperature.\cite{Huse1986,fisher1988equilibrium} The negative value of $\theta$ also explains  the $L$--independence of  the distribution of energy of defects, as explained in Appendix \ref{app:dimerdefects}.  Introducing a defect introduces an \textit{open} excitation string, whose energy can be parsed into contributions at different distances from the defect; the negative value of $\theta$ means that contributions from large distances are subleading.

The convergence of the distribution at large $L$ means that introducing a defect at a typical site costs an energy $\sim K$ with fluctuations only of order one. At first sight one might think that this implied the state was stable for large $K$. However this is not correct, because we must take into account rare locations where the energy gain $|\delta E|$ from the defect is much larger than average. If there  exists a finite density $\rho$ of rare locations with $\delta E \sim - K$, then no matter how small this density,  the glass will be destroyed on scales $\gg 1/\rho$. 

A simple rare region argument (Appendix \ref{app:dimerdefects}) indeed shows that such locations always exist, with a density which is {at least} of order ${\rho_\text{min}\sim \exp[- c (K/\Delta)^2]}$ at large $K$, with $c$ a numerical constant. This scaling arises because a rare region of area $\sim (K/\Delta)^2$ is required to overcome the core energy of the defect. (This simple picture gives $\delta E_\text{opt} \sim - (\log L)^{1/2}$ at large $L$.)  As a result of the nonzero value of $\rho$, the glass is ultimately unstable. In our toy model the lengthscale for this instability can be made  large by tuning $K$, but it is finite for any $K<\infty$.

In conclusion, the natural attempt to use disorder to  produce a paramagnetic state, a valence bond glass without topological order or long range VBS order, does not work.

\subsection{VBS subjected to strong pinning}
\label{pinnedVBS}
In the previous section disorder was applied at the level of individual singlets. In this section we point out that the same conclusions hold in a slightly different limit, where local VBS order is formed  on short scales and then  pinned on slightly longer scales. 

As  discussed above, various VBS order parameters on the triangular lattice can be viewed as living on a manifold of the form $\mathrm{RP}^{n-1}=S^{n-1}/\mathbb{Z}_2$. Let us consider the pinning of such an order parameter using the Ising gauge theory formalism described in Sec.~\ref{other_triangular_vbs_section} \cite{Toner1995}:  
\be
E = -t  \sum_{\<ij\>} \sigma_{ij} \vec v_i \cdot \vec v_j -d \sum_i (\vec h_i \cdot \vec v_i)^2 + \ldots
\ee
As above, the $n$-component unit vector $\vec v\in S^{n-1}$ is a redundant parametrization of $\mathrm{RP}^{n-1}$, and the classical energy functional must be supplemented with the information that a gauge flux (a plaquette where ${\prod_{\<ij\>} \sigma = -1}$) is accompanied by a spinon. We assign an energy cost $K$ to such plaquettes. In contrast to Sec.~\ref{other_triangular_vbs_section} we now take the disorder strength $d$ to be large.\footnote{Anisotropy favouring a discrete set of points on $\mathrm{RP}^{n-1}$ should also be included but does not affect the following.}

Consider the limit of {strong} pinning on the sites, so that $\vec v_i = \tau_i \vec h_i$ is enforced, with $\tau_i = \pm 1$. This gives 
\be
E = -t  \sum_{\<ij\>} \sigma_{ij} (\vec h_i \cdot \vec h_j) \tau_i \tau_j + \ldots
\ee
When spinons are forbidden by hand (at $K=\infty$) we can choose the gauge $\sigma_{ij}  = 1$. We then have an Ising model with effective couplings $J_{ij} = \vec h_i \cdot \vec h_j$. These couplings are randomly frustrated due to plaquettes where the gauge-invariant quantity $\prod_{\<ij\>}(\vec h_i \cdot\vec h_j)$ has negative sign.  Similar models, without the gauge field, have been studied in the context of Heisenberg models with random anisotropy \cite{jayaprakash1980random,vicari2007zerotemperature}. Ising spin glass behaviour is expected at zero temperature \cite{vicari2007zerotemperature}. This agrees with the above picture for the dimer model.

The stability to introduction of spinons is then the question of whether a large core energy cost $K$ for gauge fluxes in $\sigma$ can be outweighed  by their efficacy (when optimally placed) in relieving frustration in $J_{ij}$. As above, a simple rare region argument indicates that spinons \textit{will} be nucleated, with a density at least $\exp[- c (K/\Delta)^2]$.

In principle we could wonder about the possibility of nontrivial phases at weaker pinning that are not captured by the above treatment. But it seems unlikely that other stable phases exist in 2d.

We have found, in three separate regimes, that pinning of singlets by disorder inevitably nucleates spinful defects in 2d. In the previous sections we considered a VBS---ordered state subject to arbitrarily weak disorder. In this section we have considered a limit with strong pinning of singlets on individual bonds, and also a limit in which there is  a well-defined VBS order parameter on short lengthscales that is strongly pinned on lengthscales of the same order. In Sec.~\ref{LSM_section} we will synthesize these observations.

\bigsection{Lieb-Schultz-Mattis--like conjectures for disordered magnets}
\label{LSM_section}
Above we started with a clean system in either a VBS phase, breaking translation symmetry,  or a spin liquid phase with topological order. We then added enough disorder to remove symmetry-breaking VBS order via Imry-Ma, or to destroy topological order via vison condensation. Naively one might have expected the resulting phases to be spin--gapped and featureless (with no symmetry breaking or topological order). But this fate was averted by the appearance of the spin--1/2 defects.

In this section we suggest that this is a consequence of more general constraints on ground states of random magnets which can prevent them from having a `trivial'  disordered ground state. That is, some Lieb-Schultz-Mattis--like constraints survive in the random setting, so long as the Hamiltonian preserves translation symmetry  `on average'.
 Recall that for non-random (i.e.\ translationally invariant) magnets with an odd number of spin--1/2 per unit cell, the Lieb-Schultz-Mattis (LSM) theorem\cite{Mattis1961}, extended by Oshikawa\cite{Affleck1997,Oshikawa2000} and Hastings\cite{Hastings2004}, implies that a unique gapped ground state is not possible in a system with periodic boundary conditions in the thermodynamic limit. In effect, the ground state must have a broken symmetry or topological order with its associated ground state degeneracy on the torus; or, otherwise, it must have gapless excitations.

Here we make some initial conjectures for magnets with quenched randomness in one, two and three dimensions, with spin-1/2 per unit cell.
After defining the setting in Sec.~\ref{sec:setting}, in Sec.~\ref{spectrum}  we discuss the spectrum and correlation functions.
  We  conjecture that in 2d a disordered paramagnet, without topological order or translational symmetry breaking,  must have gapless spin excitations. 
We also show that in 1d, if statistical translational symmetry is not broken,  the spin correlation function must decay sufficiently slowly in space ($1/r^\alpha$ with $\alpha \leq 2$). We then suggest that 3d admits states that are rather different to those in lower dimensions.
Finally, in Sec.~\ref{entanglement1} 
we  propose that the entanglement structure of the ground state provides a useful alternative perspective.

The importance of `statistical'  lattice symmetries in protecting topological insulator surface states from localization was appreciated in Refs.~\cite{FuKaneTopology2012,FulgavanHeckStatistical}.
Also, Ref.~\cite{Hastings2010} considered a different approach to Lieb-Schultz-Mattis like constraints in disordered systems.\footnote{
Even for clean systems with strict translational invariance, the Lieb-Schultz-Mattis-Hastings-Oshikawa theorems may in certain cases give a gapless spectrum but do not give constraints for its spin/charge quantum numbers; issues related to such quantum number restrictions were explored in Ref.~\cite{Oshikawa2003}.}

\smallsection{Setting}
\label{sec:setting}
Consider Hamiltonians, or rather statistical ensembles of Hamiltonians, that preserve exact spin symmetry and  statistical translation symmetry. For the purposes of this discussion the spin symmetry will be taken to be $SO(3)$.\footnote{We expect these statements also to apply for $O(2)$ symmetry i.e. $S_z$ rotations plus a discrete spin flip that reverses $S_z$. Similar restrictions should also apply for discrete symmetries, e.g. $\mathbb{Z}_2 \times \mathbb{Z}_2$, or time reversal.} `Statistical' translation symmetry means that the \textit{probability distribution} of the disorder, which is assumed to be short-range-correlated, is translationally invariant.  This statistical translation symmetry allows us to define a unit cell. Crucially, we assume that there is half-odd-integer spin per unit cell.

It is important that the spin symmetry  is exact rather than merely statistical: for example the Hamiltonian $H= - \sum_i \vec h_i \cdot\vec S_i$, with the random fields $\vec h_i$ uniformly distributed on the sphere $|\vec h_i|=h$, has a manifestly trivial ground state.

We consider systems on an $L\times L\times \ldots \times L$ torus of even size $L$.  First we conjecture that the averaged energy gap $\overline{\Delta E}$ must vanish with $L$ 
 at least as a power law in this situation.
Note that this conjecture is only possible with both statistical translation symmetry and short-range correlated disorder: without these conditions it is easy to construct counterexamples. For example, we may obtain a gapped Hamiltonian by explicitly dimerizing the couplings in the pattern of a defect-free dimer configuration; but this does not correspond to short-range correlated disorder.\footnote{It is also useful to check that  the case of a VBS state in 3d, with weak disorder,  is consistent with our conjecture. In 3d the VBS order is stable to weak disorder. In the clean system the reason for the small gap is the existence of  multiple symmetry-equivalent VBS states. In the disordered system these states are split by an amount scaling as the square root of the volume. Instead the gap is closed by rare region effects: a small density of patches where the VBS order parameter can be flipped with low energy.}

We now consider more detailed constraints. To avoid trivial counterexamples to some of the statements, let us restrict to ensembles for which the ground state on the torus is unique with probability one (and therefore a singlet).  From now on, let us also restrict to ground states that do not spontaneously break either the spin symmetry or the statistical translational symmetry. 
We discuss  how to define this in the disordered setting below using appropriate local order parameters. 
Note that in  a {generic} 1d or 2d  disordered system translational symmetry breaking is prevented by Imry-Ma.

\smallsection{Spin gap and correlation functions}
\label{spectrum}

First consider the spectrum in 1d and 2d. 
Strengthening the above conjecture regarding the vanishing of the gap $\overline{\Delta E}$, we conjecture that in 1d the average gap to \textit{spinful} excitations, $\overline{\Delta E_S}$, also vanishes at least as fast as a power law at large $L$. This is natural given the argument in Sec.~\ref{sec:1dLSM} for power law spin correlations.

In 2d we must allow for the possibility of topological order: a spin liquid, with a spin gap, is of course a possible ground state.
However we conjecture that a spin gap is possible only in the presence of topological order. In the absence of topological order we conjecture that  $\overline{\Delta E_S}$ again vanishes at least as fast as a power law at large $L$. Therefore,  either  $\overline{\Delta E_S}$  is bounded by $L^{-a}$ for some $a>0$, or the gap to all excitations, $\overline{\Delta E}$, is exponentially small in $L$ as a result of topological ground state degeneracy on the torus.

In a disordered system, especially when the disorder distribution is not bounded, the gap may of course close for relatively trivial reasons as a result of Griffiths (rare region) effects. But we may contrast the above with the case of a weakly disordered magnet with an even number of spin-1/2 per unit cell, for example a ladder with dimers on the rungs, or more nontrivially, the featureless gapped states that can arise on e.g.\ the  honeycomb lattice.\cite{Vishwanath2013,Ran2016} For weak, bounded disorder, these states remain gapped. 

Nevertheless it is worth also looking for additional diagnostics in the disordered setting. One possibility is  to examine correlation functions. See also Sec.~\ref{entanglement1} where we discuss entanglement.

In fact, for the 1d chain it is possible to show that if statistical translation symmetry is unbroken (this is guaranteed by Imry-Ma for a generic 1d disordered system) then the disorder-averaged $\<S^zS^z\>$ correlation function  decays no faster than $1/(\text{distance})^2$.
A version of this follows from a general theorem about classical probability distributions for spins or charges on the line\cite{aizenman2001bounded}, and for ground states of the translationally-invariant (clean) Heisenberg chain this statement is proved in Ref.~\cite{aizenman1994geometric,affleck1986proof}. Related ideas have also been discussed in Ref.~\cite{Hastings2010}.
In Section~\ref{sec:1dLSM} we give a self-contained argument for any singlet  states --- which  applies even if the states are not ground states of local Hamiltonians --- which conforms to the definition of statistical translation breaking given below. 

Since the Imry-Ma mechanism prohibits translation breaking in a generic random 1d model, this implies that the random singlet phase, for which $|\<S^z_i S^z_j\>|\sim |i-j|^{-2}$, exhibits the fastest possible power law for decay of spin correlations.

Given the above, a natural question is whether in 2d, if the system is not a spin liquid, there must again be a spinful local operator $O_S(x)$ whose two-point function decays slower than exponentially with distance.

In 3d, we again conjecture that, in the absence of topological order, the spin gap $\overline{\Delta E_S}$ must vanish as ${L\rightarrow \infty}$. (Recall that we are restricting to states that do  not break statistical translation symmetry, so for example the VBS state is excluded.) However  we speculate that a much weaker scaling is possible than in lower dimensions, with  $\overline{\Delta E_S}$ tending to zero only logarithmically with $L$. The mechanism for this is as follows. 
We have seen that a valence bond glass is unstable in 2d. 
In 3d magnets, however, it is possible to consider a wider range of paramagnetic states built from singlets. We will discuss these in  Ref.~\cite{forthcoming3dglass}.
Unlike in 2d, stable glassy states can be constructed. These   allow for rare regions in the form of defect loops, with long loops being exponentially rare. These loops carry  spinful excitations whose gap is power-law small in the loop length. This necessarily closes the spin gap. But if the ground state of each loop is a singlet,  the gap closes weakly.  The longest loops in a sample of size $L$ are logarithmically large in $L$, giving a logarithmically small  spin gap.

Finally let us consider what it means to break spin or translational symmetry in the disordered system. We will not attempt to be mathematically precise (for example we  neglect the possibility that the large $L$ limits below are ill-defined). Spin symmetry is unbroken if for every `spinful'  local\footnote{$O_S(x)$ is supported on a finite number of sites around $x$.} operator $O_S(x)$, transforming in a nontrivial representation of $SO(3)$, the following two-point function tends to zero for large $|x-y|$, no matter how the limit of large $|x-y|$ is taken:
\be
\lim_{L\rightarrow \infty} \overline{\left|\< O_S(x) O_S^\dag(y) \>\right|^2}.
\ee
Here the operator $O_S(x)$ depends on $x$ only by simple translations. The angle brackets are the zero-temperature quantum average and the overline is the disorder average. The above rules out for example spin-glass order as well as uniform orders.

Translational symmetry is broken if there exists a local operator $O(x)$ (again, depending on $x$ only via simple translations) for which the correlation function
\be
C(x-y) = \lim_{L\rightarrow \infty} \overline{\< O(x) O(y)^\dag\>}
\ee
has non-decaying oscillations, with a nonzero wavevector, at large distance.\footnote{In principle these oscillations may be present only along certain directions. For example 
 consider a clean 2d system that splits into decoupled 1d chains that are in the VBS phase. The ground state is unique since the ground state of each chain is  the superposition of its two VBS configurations. Spatial symmetry is broken but this can only be detected by $C(r)$ if $r$ is parallel to the chains.}

\smallsection{Restriction on spin correlations in 1d chains}
\label{sec:1dLSM}

In this section we construct an argument restricting the spin correlations in one-dimensional spin chains that have an odd number of spin--1/2 sites per unit cell of statistical translation symmetries.

Take a 1D spin--1/2 chain of arbitrary even length $L$ with periodic BCs. We consider an arbitrary probability distribution on states $\ket{\psi}$ that are global singlets under $U(1)$ spin rotations around the $Z$ axis,
\be
\sum_i Z_i \ket{\psi} = 0
\ee
where $Z_i$ is a Pauli matrix $(\sigma^z)$ for the spin on site $i$.
The distribution could be the distribution of ground states of a local random Hamiltonian (in which case the distribution is translationally invariant although the states are not) but it does not need to be. For example we could also consider a distribution which is supported on a single state for each $L$. In this case the averages below can be dropped. 

We  show that if the averaged $Z$ two-point function,
\be
\overline {C_Z(i,j)} = \overline{\bra{\psi} Z_i Z_j \ket{\psi}},
\ee
decays sufficiently rapidly with distance, then we can define a strictly local operator $V_\eta(x)$  (supported on a finite number of sites) whose averaged two-point function has  oscillations at nonzero wavevector (namely $\pi$)  which do not decay at large distance, so that statistical translation symmetry is broken according to our definition.

To allow for periodic BCs, let us write $(k)_L= k + m L$, with the integer $m$ chosen so that $|k+mL|$ is minimized, and similarly $|k|_L = |(k)_L|$. Then for the above to hold it is sufficient if the correlator is bounded by
\be\label{Cdecay}
\left| \overline {C_Z(i,j)} \right| <c  |i-j|_L^{-\alpha} 
\ee
for $L$-independent $c>0$ and $\alpha >2$. The random singlet phase has $|\overline{C_Z(i,j)}|\sim 1/|i-j|^2$,\cite{Fisher1994} so exhibits the largest $\alpha$ possible for a system that does not break translational symmetry.

The present argument is straightforward using rotations around the $Z$ axis.
A similar statement also follows from a general theorem in Ref.~\cite{aizenman2001bounded}.\footnote{Since we will only consider commuting operators, there is  a close relation to probability distributions for classical spins or classical particles on a line.\cite{aizenman2001bounded} We thank J. Chalker for this observation.} 
Related ideas are discussed in  Ref.~\cite{aizenman1994geometric,affleck1986proof}  for clean systems and in  Ref.~\cite{Hastings2010}, which shows  that for a disordered spin Hamiltonian with a `mobility gap', a certain flux insertion operator (which does not however appear to be a local operator) has oscillations at momentum $\pi$. A local twist operator related to the one considered here was used in Ref.~\cite{Tasaki2017} for a different construction in a clean system.

Label sites $i=1,\ldots L$. Letting $x\in \mathbb{Z}+1/2$ label a bond, and fixing an integer $\eta$, define the unitary operator
\ba
V_\eta(x) &= \prod_{i} \exp\lf\frac{i}{2} \sum_j \phi_j(x) Z_j \ri,\\
\phi_j(x) &= \pi \operatorname{sign}(x-j)_L \max \left\{ \frac{\eta-|j-x|_L }{\eta} , 0 \right\}.
\end{align}
$V_\eta(x)$ is supported on a finite number ($2 \eta$) of sites (so it is a local operator) and it rotates spins by an angle which jumps from approximately $\pi$ to $-\pi$ as the bond $x$ is crossed, and which decays gradually to zero away from bond $x$. We will show that whenever (\ref{Cdecay}) holds with $\alpha>2$, it is possible to choose $\eta$ large enough so that the averaged two-point function of $V$ shows non-decaying oscillations at wavevector $\pi$. In fact for any desired $\epsilon>0$, it is possible to choose $\eta$ large enough so that, for any sufficiently large $L$,
\be
\overline{C_V(x,y)} = \overline{\bra{\psi}  V_\eta (x) V_\eta^\dag (y) \ket{\psi} }
\ee
is within $\epsilon$ of $(-1)^{y-x}$. 

Consider 
\ba
(-1)^{y-x} C_V(x,y) = (-1)^{y-x} \bra{\psi} V_\eta(x) V_\eta^\dag(y) \ket{\psi}. 
\end{align}
The phase of the rotation in $VV^\dag$ now jumps at both $x$ and $y$. We can eliminate both of these jumps by rotating the spins in between $x$ and $y$ by an additional $2\pi$. But such a $2\pi$ rotation is equivalent to the $(-1)^{y-x}$ which is included explicitly above. That is,
\be
(-1)^{y-x} C_V(x,y) = \bra{\psi} \exp\lf \f{i}{2} \sum_j \theta_j Z_j \ri \ket{\psi},
\ee
where  $\theta_j$ increases  from $0$ to $2\pi$ over a region of length $2\eta$ around $x$, and then decreases back to 0 over a similar region around $y$.  Let us define
\be
A(\lambda) = \bra{\psi}  \exp\lf \f{i \lambda}{2} \sum_j \theta_j  Z_j \ri \ket{\psi},
\ee
so that $A(0) = 1$ and $A(1) =(-1)^{y-x} C_V(x,y)$. Differentiating,
\ba
\partial_\lambda A(\lambda) & = \f{i}{2}  
\bra{\psi}  \exp\lf \f{i \lambda}{2} \sum_j \theta_j Z_j \ri  \sum_k \theta_k Z_k \ket{\psi}.
\end{align}
The RHS is proportional to the inner product of $\bra{\psi}  \exp\lf \f{i \lambda}{2} \sum_j \theta_j Z_j \ri$ and $\sum_k \theta_k Z_k \ket{\psi}$. The former has norm one and the latter has norm $N$,
\ba
N^2 &= \sum_{jk} \theta_j \theta_k \bra{\psi} Z_j Z_k \ket{\psi}\\
 & = - \f{1}{2} \sum_{jk} (\theta_j - \theta_k)^2 \bra{\psi} Z_j Z_k \ket{\psi},
\end{align}
where we used $\sum_j Z_j \ket{\psi} = 0$. Therefore 
\ba
|\partial_\lambda A(\lambda) | & \leq \f{N}{2}, 
&
|A(1) -A(0)| & \leq \f{N}{2}.
\end{align}
Averaging,
\ba
\overline{ |C_V(x,y) - (-1)^{y-x}|} \leq 
\frac{1}{2} \overline{N} \leq \frac{1}{2} \sqrt{\overline{N^2}}.
\end{align}
The RHS involves
\ba
\overline{N^2} & = - \f{1}{2} \sum_{jk} (\theta_j-\theta_k)^2 \overline{C_Z(j,k)} \\
& \leq
 \f{1}{2} \sum_{jk} (\theta_j-\theta_k)^2 \left| \overline{C_Z(j,k)}  \right|.
\end{align}
It is now easy to check that if $C_Z$ obeys (\ref{Cdecay}) with $\alpha >2$, the above expression is of  order $\eta^{2-a}$ for large $\eta$.  (We examine separately the cases where one, both, or neither of $j$ and $k$ lie within a given region of size $2\eta$ where $\theta$ is varying.) 
Therefore $\overline{ |C_V(x,y) - (-1)^{y-x}|}$, and a fortiori ${|\overline{C_V(x,y)} - (-1)^{y-x}|}$, can be made as small as desired by choosing $\eta$ large enough.

\smallsection{Entanglement structure of ground states}
\label{entanglement1}

In this section we give a heuristic picture for the kinds of states that are allowed.  (From now on when we refer to `states' we  implicitly assume the absence of long-range order.) We would like to classify ground states according to how `nonlocal' they are. In a disordered system there are more possibilities than in a clean one. 

First, it may or may not be possible to construct the state using a `finite' depth unitary circuit (in a disordered system this concept must be extended to allow for rare regions\footnote{The depth of the circuit would not  be strictly finite. Instead there would be a requirement on how the accuracy with which the ground state could be approximated converged with the circuit depth. In a disordered system it is important that such a definition should allow for rare (Griffiths) regions: \textit{locally} the required circuit depth should be allowed to be arbitrarily large, so long as the probability for this is sufficiently small.}). For example, a ground state which is a product of nearest-neighbour singlets can be constructed from a product of `up' spins by a quantum circuit of depth one.\footnote{Here we are not imposing conditions on the transformation properties of the initial product state, or the unitaries, under spin --- we are only constraining the final state. Other protocols (e.g. starting from a product of spin singlets) could be considered.} Let us call  states that can be constructed using finite-depth circuits `constructible'.  An example of a state which is \textit{not} constructible is the ground state of a quantum spin liquid: this requires a quantum circuit of depth proportional to the system size.\cite{BravyiHastingsVerstraete,haah2016invariant} Another example is a random singlet ground state in 1d, which requires a circuit of  $O(L)$ depth to produce the largest singlets.

Second, if the state \textit{is} `constructible', there is still a distinction to be made according to whether or not the local structure of the circuit depends on the disorder far away. If it does not, we call the state `constructible with local information'. For a disorder ensemble whose ground state is constructible with local information, there exists a protocol for defining the unitary circuit, given the Hamiltonian, in which the local structure of the unitary circuit depends exponentially weakly on values of the random couplings far away. 

We only attempt to make this distinction at a heuristic level. Consider first the example of  a 2-leg ladder with nonzero, weakly random AF couplings on the rungs, and all other couplings zero. This has two spin-1/2s per unit cell and we do not have any LSM restriction. The ground state is the product of singlets on the rungs and is  evidently constructible with local information: the unitary circuit that constructs this state has no dependence on disorder at all.  By contrast, consider a magnet with spin-1/2 per unit cell in high spatial dimension. Assume that in 3d and above, as discussed in Sec.~\ref{spectrum},  there exist stable `glassy' states built only out of short-range singlets (neglecting rare regions). Since these states are close to a product of local singlets,  they are also constructible. However, unlike the example of the ladder, they are not constructible with local information. A slight change to the disorder at one location can affect the singlet pattern at a distant location (with a probability that is small, but not exponentially small).


Let us now consider one, two and three spatial dimensions in turn.

In 1d, we propose that there are no constructible states for magnets with  spin-1/2 per unit cell --- even if we do not require them to be ground states of local Hamiltonians. As we have discussed (Section~\ref{sec:1dLSM})  it is impossible to write down a singlet wavefunction in which spin correlations are rapidly decaying and in which translational symmetry is unbroken. Therefore there are no constructible ground states for disordered Hamiltonians with statistical translation symmetry. Recall that we are restricting here to states without LRO.

In 1d we also  make a conjecture in terms of the amount of entanglement entropy between two halves of a system of size $L$.   We conjecture that the disorder-averaged entanglement entropy necessarily diverges at least logarithmically with $L$ in any ground state without LRO. The random singlet phase\cite{Moore2009}, and the gapless ground state of the clean antiferromagnetic chain, are examples consistent with this conjecture.

In 2d we conjecture that there are again no constructible ground states for magnets with spin-1/2 per unit cell. In Sec.~\ref{rdmdmr} we discussed a putative `valence bond glass' in 2d. This discussion shows that constructible \textit{wavefunctions} can be written down  (unlike in 1d), but we argued that they could not be \textit{ground states} of local Hamiltonians, as they are unstable to the nucleation of spinful defects.

In 3d and above we suggested the possibility of constructible `valence bond glass' ground states for spin-1/2 magnets. However, these states are not constructible with local information. This distinguishes them from more `trivial'  ground states that are constructible with local information. We propose that the latter can only exist for magnets with integer spin in the unit cell.


\bigsection{Application to experiments}
\label{ymgo_app}
Let us now apply these theoretical ideas to an experimental setting, focusing on the particular case of  YbMgGaO$_4$. A schematic partial summary is offered in Table~\ref{tab:measurables}.

Before comparing with experimental data, we note two differences from the theoretical scenario described above. First, in YbMgGaO$_4$ bond randomness is likely not weak, since the differing charges on Mg$^{2+}$ and Ga$^{3+}$ modify the oxygen charge-transfer energy as well as the Yb-O-Yb bond angle\cite{Zhang2017}; in iridium oxides with equivalent edge-sharing oxygen octahedra such variations are known\cite{Khaliullin2009,Vishwanath2014} to impact the SOC-based magnetic exchanges. Of course since the material is intrinsically disordered, the energy scales $\Delta_S$ and $\Delta_{VB}$, which are defined only in the theoretical clean limit with vanishing disorder strength $\Delta \rightarrow 0$, cannot be accessed directly.

Second, we note that strong SOC breaks spin SU(2) rotation symmetries; 
\footnote{Spin-orbit coupling can produce multiple spatially-varying anisotropic exchanges. In YbMgGaO$_4$, crystal symmetries restrict nearest-neighbor interactions $J_{i j}^{\mu \nu} $ to four independent parameters; these are sometimes denoted by $J^z, J^{xy},J^{z \pm}, J^{\pm \pm}$ in the literature\cite{Zhang2015}, but may be fruitfully rewritten in terms of the Kitaev-type exchanges that occur in iridates with analogous edge-sharing oxygen octahedra, namely $J^z, J^{xy}$, Kitaev $K$, and pseudo-dipole $I $ interactions; see Appendix \ref{appendix:exchanges} and also Ref~.\onlinecite{Chernyshev2017}. } 
however VBS phases remain well defined as quantum paramagnets that preserve time-reversal symmetry but break some lattice symmetries. It is also conceptually helpful to consider two-spin singlet states, which remain eigenstates even with SOC as long as  $J_{i j}^{\mu \nu} $ is symmetric, as may be enforced by inversion symmetry across a bond midpoint, since the singlet state is odd under inversion (see Appendix \ref{appendix:SOC} for details).
More generally, time reversal symmetry and Kramer's degeneracy is enough to protect the S=1/2 Kramer's doublet at the core of an isolated defect, even without any spin rotation symmetry.

\begin{table}[]
\centering
\begin{tabular}{l|l}
\hline\hline
Observable \ \ \ \ & Prediction (Sec.~\ref{ymgo_app})  \\ \hline\hline
$C[T]$ & $T^\alpha , \ 0<\alpha<1$  \\ \hline
$\chi[T]$ & $T^{\alpha-1} + c_0 T^{-1} $\\ \hline
$\kappa[T]$ & $T^{\rho} ,  \ 1.8 \lesssim \rho \lesssim 2 $ \\ \hline
$C[H,T]$ & Eq.~\ref{eq:CTHscaling}, \ Fig.~\ref{fig:Cfield}  \\ \hline
 $S[H;q,\omega]$ & Eq.~\ref{eq:Sfield}, \ Fig.~\ref{fig:Sfield}  \\ \hline\hline
\end{tabular}
\caption{Shorthand summary of theoretical results for a few experimental observables within the framework of a $d>1$ lattice-emergent random-singlet regime following the ideas discussed above for a random frustrated Heisenberg model Eq.~\ref{eq:hamiltonian}. These predictions, namely heat capacity, spin susceptibility, thermal conductivity, scaling of heat capacity in magnetic field, and in-field dynamical structure factor, are expected to be seen in YbMgGaO$_4$ and related systems at low temperatures.}
\label{tab:measurables}
\end{table}

To discuss the consequences of the pinned singlets and their instabilities, let us consider in turn two distinct regimes of temperature and energy.

\begin{figure}[b]
\includegraphics[width=0.75\columnwidth]{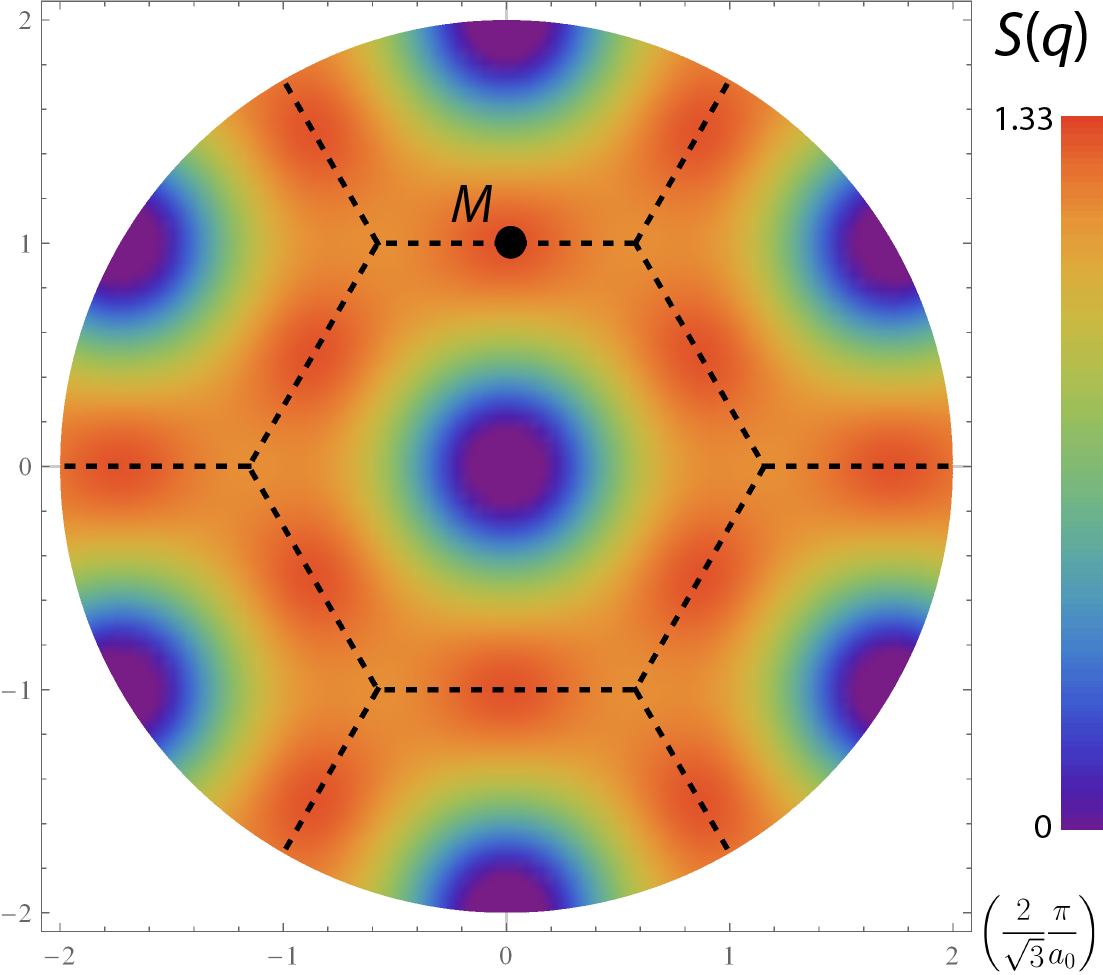}
\caption[]{ \textbf{Spin structure factor for pinned singlets.} Equal time spin correlations $\langle S^+ S^-\rangle$  for spins frozen into randomly-oriented short-ranged valence bonds  show a deep minimum at the Brillouin zone center; a $4{:}1$ ratio of first/second-neighbor bonds also give a broad maximum at the edge midpoint $M$. Both features are seen in neutron scattering\cite{Zhao2016,Mourigal2016,Gegenwart2017,Rueegg2017}.} \label{fig:corr}
\end{figure}

\smallsection{Structure factor at low energies}
At low or intermediate temperatures and energies of order $J$, inelastic neutron scattering data\cite{Zhao2016,Mourigal2016,Gegenwart2017,Rueegg2017} suggests that the  random singlets scenario is on the right track. 
Two distinctive features appear at the lowest frequencies $\omega \gtrsim 0.1$ meV in low temperature: (a) increasing intensity from the BZ center to the BZ edge, and (b) a broad maximum at the BZ edge midpoint, position $M$. 
The features persist across the energies $0.1<E\ (\text{meV})< 1$; for comparison the magnetic exchange scale has been estimated\cite{srep,Zhang2015,muSR} to be roughly 0.2 meV. 
Randomly oriented short-ranged singlets  capture both features, as we show  in Fig. \ref{fig:corr}. 
Note that feature (b) clearly implies some second-neighbor correlations; the wavevector $M$ is oriented towards second neighbors on the triangular lattice.  Since the distribution of short ranged  $S=0$ resonances can vary even within a particular VBS phase through different resonances, we take the ratio of first- to second- neighbor singlets as a free fitting parameter.
In this short-ranged-singlets phenomenological model, the equal time spin correlations $\langle S^+ S^-\rangle$ (for dynamics see Eq.~\ref{eq:Sfield} below) are easily computed and are given by 
\begin{equation}
S(q) = 1 - \frac{1}{3}\sum_{i=1}^3 \left( f_1 \cos(q \cdot a_i) + f_2 \cos(q \cdot b_i) \right)
\end{equation}
for a fraction $f_1$ ($f_2$) of spins participating in first (second) neighbor singlets;  $a_i$ ($b_i$) are first (second) neighbor vectors. A good fit is found for a ratio $f_1{:}f_2=4{:}1$, as used (with $f_1{+}f_2{=}1$) for Fig. \ref{fig:corr}.
The experimental observations at $T \lesssim J$ are well reproduced by random short-ranged-singlets. 
%
%

\begin{figure*}[]
\includegraphics[width=\textwidth]{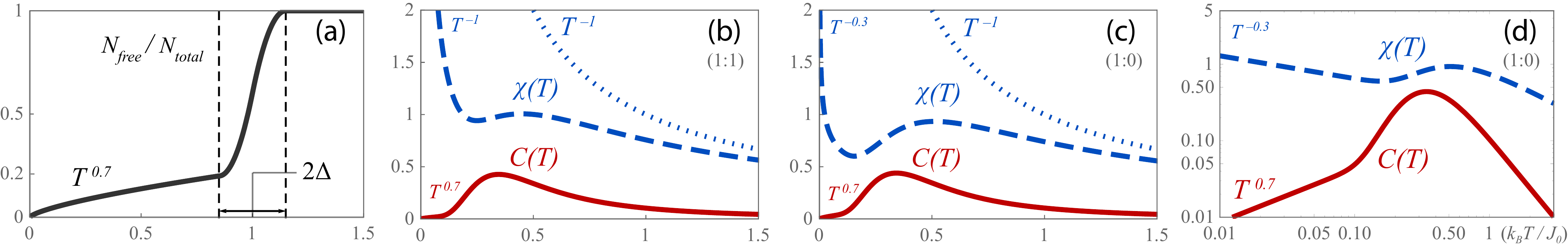}
\caption[]{ \textbf{Susceptibility $\chi(T)$ and specific heat $C(T)$ in a simple two-regime model.} \textbf{(a)} The model is defined by a temperature-dependent function $N_\text{free}/N_\text{total}$, the fraction of singlets/triplets with energy less than $k_B T$. Here, 80\% of sites form singlets at the narrow energy window $J_0 \pm \Delta$  ($\Delta {=} 0.15 J_0$), while remaining 20\% sites form singlets and triplets with ratio 1:1 (panel (b)) or 1:0  i.e.\ only singlets (panels (c, d)), with energy distribution $T^{\alpha}$ with $\alpha=0.7$. \textbf{(b, c, d)} Linear and log-log plots of magnetic susceptibility $\chi(T)$ (dashed blue) [c.f.\ free spin $\chi(T)=1/T$ (dotted blue)] and specific heat $C(T)$ (red) in the spin pair approximation. 
Units are $J_0/ k_B$ for temperature (x-axes), $\mu_B^2/k_B J_0$ for $\chi(T)$, and 
$k_B$ for $C(T)$.  
While $\chi(T)$ is sensitive to emergent FM bonds (panel (b)),  $C(T) \sim T^{\alpha}$ is robust.
}  \label{fig:plots}
\end{figure*}

\smallsection{Heat capacity at ultralow energies}
At ultralow energies and temperatures below $J$, the magnetic specific heat $C(T) \sim T^{0.7}$ provides direct evidence for a power-law density of states of strong randomness. To compare specific heat as well as magnetic susceptibility across a wider temperature range, we introduce a two-regime model which, despite its simplicity, can still capture the experimental data semi-quantitatively as seen in Fig.~\ref{fig:plots}.  The model is defined by considering a density of states consisting of two distinct regimes, whose integrated density of states is plotted in Fig.~\ref{fig:plots}(a). 
For this model we let a fraction $f$ of the sites form singlet pairs at energies tightly distributed (in a triangular distribution) within the narrow range $(J_0-\Delta, J_0 + \Delta)$, while at lower energies the remaining $1{-}f$ sites form either singlet pairs, or alternatively  coexisting singlets and larger-spin triplets, in either case with energies drawn from the power-law distribution $\rho(E) \sim E^{\alpha - 1}$ with $\alpha=0.7$.  (The exponent $\alpha$ is non-universal in the theory; for example, a phenomenological fit for YbZnGaO$_4$ would instead entail $\alpha=0.59$.)  The effective average magnetic energy scale\footnote{Note that $\chi(T)$ and $C(T)$ can in general require different effective $J_0$ within the spin pair approximation, depending on the eigenstate content of the spectrum.} $J_0$  may be estimated\cite{srep,Zhang2015,muSR} as $J_0 \sim 1.5$K. The results of the model are largely insensitive to the value of $\Delta$; here we took $\Delta = 0.15 J_0$ to match the $g$-factor broadening extracted from crystalline electric field excitations\cite{Zhang2017}. 
By construction, the model captures the observed magnitude and scaling of specific heat,
 $C =  1.41\ T^{0.7}\ \text{J/mol K} =  0.17\ T^{0.7} \ k_B$  
 with temperature measured in Kelvin. Within a zeroth-order approximation of frozen isolated singlets (see Appendix \ref{app:spinpairapprox} for details), the computed specific heat captures the experiments at the parameter value  $f{=}4/5$; a better approximation including (model-dependent) singlet fluctuations would likely increase $C(T)$ and require a larger $f$ for fitting the experimental $C(T)$ data.
When the renormalized spin coupling is antiferromagnetic such that only singlet pairs form at low temperature, the susceptibility $\chi(T)$ is tied to $C(T)/T$ and shows a soft divergence   $\chi \sim T^{\alpha-1}$ at ultralow temperatures  below a pronounced peak  (Fig.~\ref{fig:plots} (c)).
Admixing a fraction of large-spin clusters,  likely to develop from any ferromagnetic bonds in the low-energy distribution, will contribute an additional $1/T$ Curie term, flattening out the peak to produce an apparent saturation before the rise (Fig.~\ref{fig:plots} (b)).
Experimentally an apparent saturation is seen\cite{srep,muSR} around $T=0.3$ K. 
   The lowest temperature measurements of susceptibility are thus consistent with the random singlet mechanism; the random singlet phase would predict a slow rise of susceptibility at very low temperatures.\footnote{Ideally the $q=0$ susceptibility at ultralow temperatures should be measured directly;  measurements of local suscepibility, through the $\mu$SR Knight shift, may be further complicated by interactions of the muons with the charged Mg/Ga disorder.}  
   At even lower temperatures, if large-spin clusters develop and freeze into a spin glass --- as is the likely $T{=}0$ RG fixed point fate of the random spin network --- then the resulting glassy moment formation would serve as a cutoff for an otherwise-divergent magnetic susceptibility, and  simultaneously would be visible for standard probes of spin glasses.

\smallsection{Thermal conductivity}   
Next let us consider the ultralow temperature thermal conductivity. Since the gapless spin excitations are all localized, heat is carried primarily by phonons. 
At ultralow temperatures the distribution of random singlets (as well as any frozen spin-glass moments, if those arise at lower temperatures) both produce a collection of quantum two-level systems, associated with local rearrangements of the singlets and their levels or spin configurations, which scatter acoustic phonons via resonant absorption at a rate $\Gamma$ that is linear in the phonon frequency\cite{Varma1972,Thompson2002}. 
The leading contribution to low temperature thermal conductivity $\kappa = C v^2 / 3 \Gamma$ (with $v$ the acoustic velocity) arises from this phonon scattering mechanism, leading to the anomalous scaling $\kappa \sim T^2$.
   This  ultralow-$T$ thermal conductivity, in particular the vanishing $\kappa/T \rightarrow 0$ as $T \rightarrow 0$, presents a contrast with the finite  $\kappa/T$ at zero temperature that would be expected from a spinon-Fermi-surface with itinerant spinons. Experimentally\cite{Li2016} $\kappa/T$ was found to vanish with  $T \rightarrow 0$. 

The  $\kappa \sim T^2$ power law naively expected here is a familiar theoretical result for structural as well as spin glasses, where log corrections and other effects are known\cite{Thompson2002}\footnote{We thank Chandra Varma for raising this point.} to result in experimentally observed effective power laws $\kappa \sim T^\alpha$ with a slightly reduced exponent 
$1.8 \lesssim \alpha \lesssim 2$. 
Thermal conductivity measurements\cite{Li2016} in YbMgGaO$_4$ found a power law behavior $\kappa \sim T^{\alpha}$ with a  $\alpha=1.85$, and measurements\cite{Wen2017} in YbZnGaO$_4$ found $\kappa \sim T^{\alpha'}$ with a  $\alpha'=1.97$, both in excellent agreement with this full theoretical description.

\begin{figure}[]
\includegraphics[width=0.96\columnwidth]{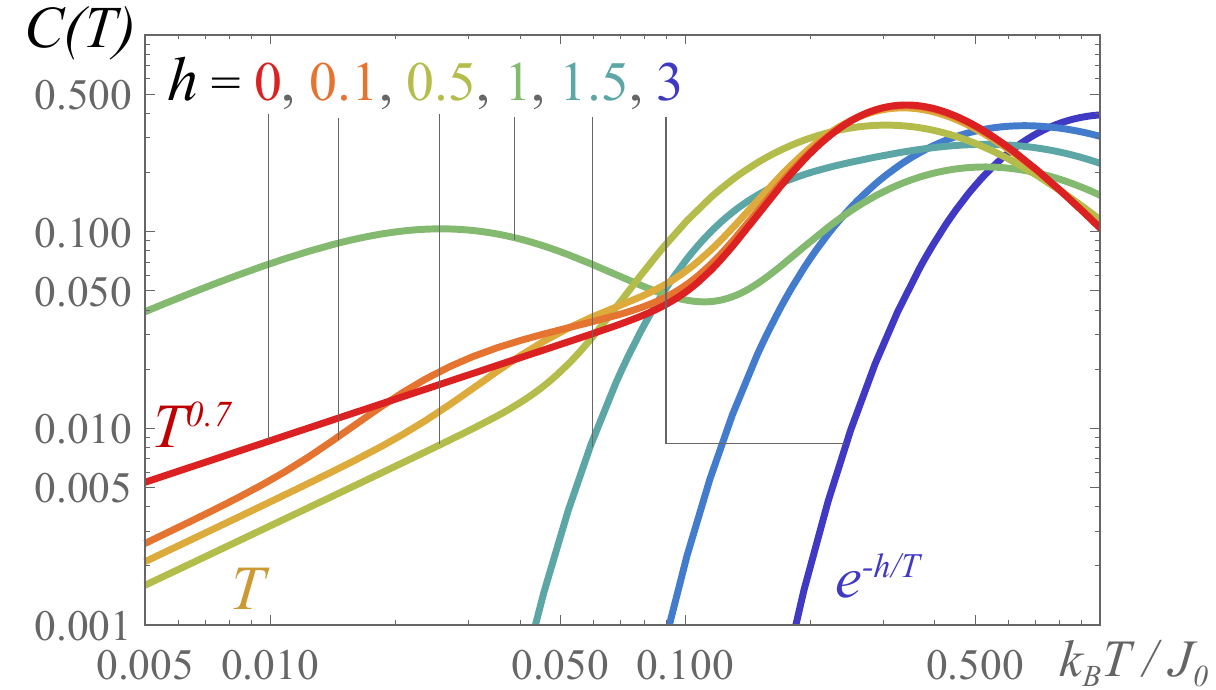}
\caption[]{ \textbf{Specific heat in a magnetic field.} 
Specific heat $C(T)$ is computed within the two-regime model with low energy singlets, as in Fig.~\ref{fig:plots}(c,d), here with an application of a magnetic field $H$ with dimensionless energy $h  \equiv g \mu_B H/J_0$ taking the values $h= 0, 0.1, 0.2, 0.5, 1, 1.5, 2, 3$   (red to violet colors). The anomalous curve at $h=1$ is due to a resonance with the narrow distribution chosen for Fig.~\ref{fig:plots}(a). Applying a small field changes the low temperature scaling of specific heat from $C(T) \sim T^{0.7}$ to a linear form $\sim T/H^{0.3}$, then turning to an exponential form at larger fields. 
(For modifications of the scaling by spin-orbit coupling see also Ref.~\cite{Scaling2018}).
} \label{fig:Cfield}
\end{figure}

\begin{figure}[]
\includegraphics[width=\columnwidth]{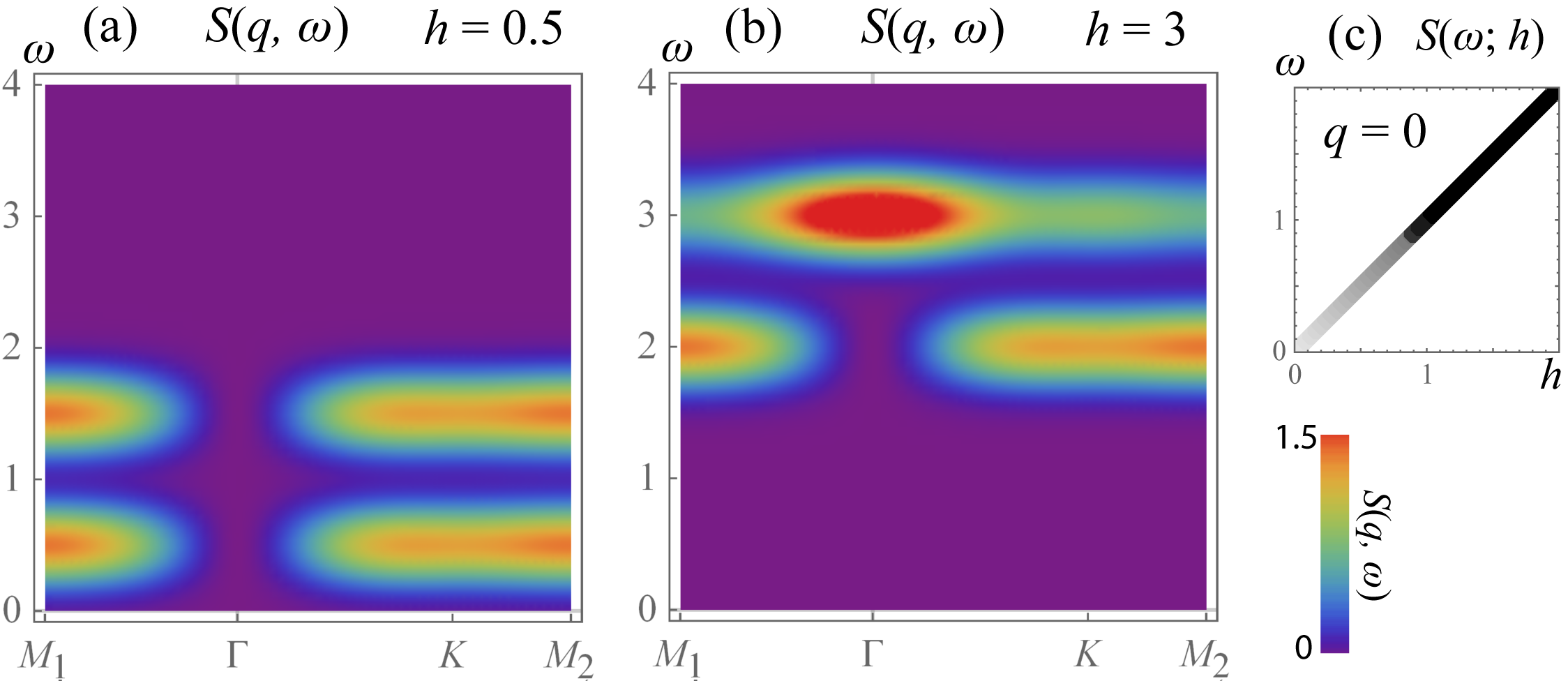}
\caption[]{ \textbf{Spin dynamics in a magnetic field.} 
The dynamical structure factor for isolated singlets, randomly distributed as in Fig.~\ref{fig:corr}, in an applied magnetic field (Eq.~\ref{eq:Sfield}). \textbf{(a,b):}  Transverse spin fluctuations S(q,w) at fields $h=0.5$ [(a)] and $h=3$ [(b)], with $h  \equiv g \mu_B H/J_0$. Dirac $\delta$-functions are plotted as $\sigma{=}0.2$  Gaussians.
\textbf{(a):} 
For fields $h<1$ smaller than the singlet-triplet splitting, low intensity features are visible at the $\Gamma$ point ($q=0$): fixing frequency at $\omega = \omega_\pm \equiv 1 \pm h$ (here $ \omega_\pm  = 0.5,1.5 $) and taking $q \rightarrow 0$ shows the intensity \textit{decreasing}. The spread of intensity in $q$ and $\omega$ away from these points can manifest as ``constriction points'' of minimum intensity.
\textbf{(b):} 
For fields $h>1$ that start to polarize a given singlet, a $q=0$ peak is observed at a frequency $\omega_0 \equiv h$, which scales linearly with field. 
Coupling the isolated singlets will connect the $\omega=h,h-1$ features into a magnon dispersion.
\textbf{(c):} 
Given some broad distribution of energy splittings, e.g.\ as in Fig.~\ref{fig:plots}, the structure factor at $q=0$  shows a single peak, at $\omega = h$,  with an intensity that increases with $h$ and saturates for $h \gtrapprox 1$.  This $q=0$ feature, as well as the $q\rightarrow 0$ decrease in intensity approaching the ``constriction points'', are observed in experiments\cite{Mourigal2017,Zhao2017}.
} \label{fig:Sfield}
\end{figure}

\smallsection{Signatures in a magnetic field}   

Applying an external magnetic field $H$ will produce signatures in both low-temperature thermal transport as well as heat capacity.
We will discuss the effects of an  external magnetic field $H$ on three physical observables:  low-temperature thermal transport;  heat capacity; and the dynamical structure factor.

First we consider how the form $\kappa \sim T^2$ for  thermal conductivity is modified by a magnetic field. 
When the Zeeman energy grows larger than the singlet gap for a given set of singlets or frozen moments, their ground state becomes spin polarized and no longer participates in scattering acoustic phonons. This reduction in scattering mechanism will be directly observable as an enhancement of thermal conductivity in a magnetic field. For large enough fields of order a few times the average exchange $J$, the spins will all be essentially polarized and $\kappa$ will saturate at a value substantially larger than its zero-field form. This enhancement and saturation is indeed seen\cite{Li2016} in YbMgGaO$_4$ at fields of a few tesla. 

Second we consider effects of magnetic field on the heat capacity and magnetic susceptibility. 
The behavior of specific heat in a magnetic field, computed within the two-regime model discussed above, for the case of low energy singlet formation, is plotted in Fig.~\ref{fig:Cfield}. Within the low-temperature power law tail, applying a small magnetic field $H$ changes the scaling of $C(T)$ from $T^{0.7}$ to linear in $T$, following the scaling function
\begin{align}
\label{eq:CTHscaling}
&C[H,T] \sim T^\alpha   \quad \text{for} \ \  T \gg  g \mu_B H \ \text{ or } H=0
\\
&C[H,T] \sim \frac{T}{H^{1-\alpha}}   \quad \text{for} \ \  T \ll  g \mu_B H \nonumber
\end{align}
which holds for a power law distribution of pure spin-singlets. (A modification of this scaling function under particular forms of spin-orbit coupling will be discussed elsewhere\cite{Scaling2018}). 

Scaling in a magnetic field is also seen in other observables. Susceptibility $\chi[H,T]$ is cut off to scale as a constant proportional to $H^{-0.3}$ at low $T$. The scaling limit requires $H$ and $T$ to both be sufficiently smaller than the lattice energy scales, so as to have a chance of being in the emergent power law regime. Fields with Zeeman energies larger than the lattice exchange energy scales $J_0$ polarize the spins, leading to a gapped spin polarized state with asymptotically exponential form ($C(T) \approx \left( (h-1)^2 / 2 T^2 \right) \exp[(1-h)/T]$ with $T$ in units of $J_0$, with dimensionless Zeeman energy $h \equiv g \mu_B H/J_0$). 
Adding triplets or larger spin clusters into the low energy spectrum would modify the susceptibility sharply and specific heat somewhat more mildly. 
In YbMgGaO$_4$, a scaling regime may be difficult to reach, experimental studies of $C(T)$ in fields above $1$ tesla have been reported\cite{srep} and are consistent with this sharp decrease of $C(T)$ in a field. 

Third we briefly consider the dynamical spin structure factor in a field. A spin singlet with some effective splitting $J_0$ will transition  from the singlet ground state into a polarized state when the Zeeman energy overcomes the splitting $J_0$,  at $H=J_0$ (here $H$ is the Zeeman energy, i.e.\ the magnetic moment is set to unity).  The spin dynamics $\langle S^+ S^- \rangle$ transverse to the applied field, for a valence bond involving two sites separated by the vector $R$, are  straightforwardly computed by considering the dynamics in the singlet or spin-polarized ground state of the two-spin Hilbert space. This yields the structure factor contribution of a singlet bond with energy and separation $\{J_0,\vec R\}$. The expression for the structure factor, which holds for finite field $H>0$ as well as zero field $H=0$, is then
\begin{align} 
S(q,\omega) &=  \theta[J_0-H] \frac{1-\cos(\vec q \cdot  \vec R)}{2}\sum_\pm \delta(\omega - J_0 \pm H)
\nonumber\\ & + \theta[H-J_0]\bigg(  
\frac{1-\cos(\vec q \cdot \vec   R)}{2}\delta(\omega +J_0 - H) 
\nonumber\\
& \ \ \ \ \ \ \ \ \ \ \ \ \ \ \ \ \ \ \ \ \ + 
 \frac{1+\cos(\vec q \cdot \vec R)}{2}\delta(\omega - H)  \bigg)
 \label{eq:Sfield}
\end{align}
with $\theta[x]$ the Heaviside step function, assuming full spin rotation symmetry. In the presence of spin-orbit coupling, the $H=J$ transition is smeared and full polarization is achieved only asymptotically at large fields.
The signatures in the spin dynamics are shown in Fig.~\ref{fig:Sfield}.
The distribution of singlets used for Fig.~\ref{fig:Sfield}, namely $4{:}1$ first:second neighbor singlet bonds, is the same as that extracted from the zero-field structure factor as shown in Fig.~\ref{fig:corr}. 
At wavevectors near the Brillouin zone center, $q=0$, the dynamical response shows a constriction at low fields into a feature of minimum intensity (Fig.~\ref{fig:Sfield} (a)). Sitting at $\omega$ fixed  to the frequency of this ``constriction point'', and taking $q$ from the BZ edge to $q=0$, shows the intensity decreasing. Interestingly within a spinon Fermi surface\cite{Chen2017,Zhao2017} interpretation the expectation for this spectral-crossing-like ``constriction point'' is for an opposite feature, namely a sharp increase in intensity as $q\rightarrow 0$. Experimentally, the intensity approaching the ``constriction point'' or crossing-like feature is observed to decrease\cite{Zhao2017}. 
 At large enough fields (Fig.~\ref{fig:Sfield} (b)) the dynamics show a delta-function peak at a frequency proportional to the magnetic field, $S(q{=}0,\omega) = \delta(\omega - h)$ for $h > J$. The distribution of effective $J$ implies that this field-gapped delta function response, $S(q{=}0,\omega) = \delta(\omega - h)$, will be observed at any magnetic field, but with an intensity that grows with field and then saturates above $h \approx J_0$ (Fig.~\ref{fig:Sfield} (c)), again in apparent agreement with recent experiments\cite{Zhao2017,Mourigal2017}.

At very low frequencies, the expression $S[J_0,\vec R](q,\omega)$ of Eq.~\ref{eq:Sfield} must be integrated against the density of states $\rho[E]$ to yield the power-law form of the low frequency structure factor, 
\begin{align} 
S(q,\omega) = \int_E \rho[E] S[E,\vec{R}(E)](q,\omega)
 \label{eq:Sfieldint}
\end{align}
where $\vec{R}(E)$ is the singlet size as a function of its energy splitting, which can depend on the lattice and other details; in the 1d spin chain random singlet fixed point\cite{Fisher1994}, $\log E \sim \sqrt{R}$. 
   
\smallsection{Other experimental signatures}   
   Let us also consider signatures of this theoretical scenario in other measurements.   Inelastic neutron scattering at ultralow frequencies would distinguish the present picture from a spinon-Fermi-surface, expected to show some $2k_F$ structure, as well as from any magnetically ordered domains, which would have magnetic Bragg peaks visible in scattering and ordered moments visible in e.g.\ $\mu$SR. Short ranged VBS order may be observed, through high-order  crystalline symmetry breaking effects which can be seen in sensitive structural probes; for columnar VBS, the lattice modulation would be visible at wavevector $M$, though we note that the (non-infinitesimal) microscopic randomness present in YbMgGaO$_4$ means that short-ranged VBS order is not required by the theory. 
NMR spin-lattice relaxation is expected to exhibit a stretched exponential associated with a  distribution of $1/T_1$ decay coefficients.  Finally, though the ultimate fate of the RG flow for the nucleated defect spins is unknown, on the triangular lattice as discussed above one reasonable expectation is that large ferromagnetic spin clusters form, and then freeze into a spin glass;  signatures of spin freezing, with several unusual properties, were recently observed\cite{Wen2017} in both YbMgGaO$_4$ and YbZnGaO$_4$.

   \smallsection{Relevance to other materials}   
The simplest starting point for the theoretical treatment presented above is a clean Hamiltonian in a VBS phase. Such a setting is well approximated experimentally by the organic material
EtMe3P[Pd(dmit)2]2, a triangular lattice $S=1/2$ magnet with an antiferromagnetic Heisenberg coupling of order $J$=250 K, which has been shown\cite{Kato2007} to exhibit valence bond solid order below an ordering temperature $T$=25 K.
Given such a clean system in a VBS phase, one can hope to introduce quenched disorder in a controlled manner. Indeed irradiation by x-rays has been shown\cite{Kanoda2015} to produce non-magnetic disorder that can nevertheless substantially modify the magnetic properties of organic compounds.  For EtMe3P[Pd(dmit)2]2 or similar systems, controlling the irradiation time would allow experimentalists to interpolate between the well-understood theoretical limit of weak disorder and the relatively strong disorder seen in YbMgGaO$_4$ and  YbZnGaO$_4$.

The physics discussed here is expected to be quite general. For very weak disorder, the very small fraction of spins that participate in the emergent power laws may make detection difficult even as the materials tend closer to the theoretically controlled limits; and in other compounds, the presence of site disorder in addition to bond randomness, such as dilution or random fields, gives an obvious source of magnetic impurities that does not require the analysis provided here. But in many materials time-reversal-symmetry is well preserved microscopically even by disorder, and there are numerous possible microscopic sources of spatial variations in the energy, which corresponds to bond randomness, that then permit an analysis based on the theory provided here.
 
Indeed since this work was first released a variety of other magnetic insulators have been found which appear to be well described by the present theory. The essence is a heat capacity that exhibits power law scaling with a fractional exponent, with a magnitude consistent with a contribution arising from a portion of the local moments in the material.
At the time of writing these compounds already include the layered spin--1/2 Mott insulators H$_3$LiIr$_2$O$_6$,\cite{Takagi2018} LiZn$_2$Mo$_3$O$_8$,\cite{McQueen2012,McQueen2014,Broholm2014,Scaling2018} synthetic herbertsmithite\cite{Lee2010,Lee2016,Scaling2018} and 1T-TaS$_2$.\cite{Kanigel2017} All of these compounds show power law heat capacity, and moreover a data collapse of heat capacity in a magnetic field, into a scaling function of the single variable $T/H$, analogous to Eq.~\ref{eq:CTHscaling}. Indeed 1T-TaS$_2$ appears\cite{Kanigel2017,forthcomingDagan} to exhibit exactly the $C\sim T/H^{1-\alpha}$ scaling function of Eq.~\ref{eq:CTHscaling}. The other three materials exhibit a similar scaling function but with an additional factor of $T/H$ at low temperature; the complete analysis of this modified scaling, based on a particular role of spin-orbit coupling, will be discussed elsewhere\cite{Scaling2018}.


\bigsection{Discussion}
\label{disc}

Quenched randomness allows for new kinds of quantum ground state with interesting entanglement structures, and destroys states that would otherwise be stable.  One of the main results of this paper is a study of the path by which weak disorder transforms a paramagnetic valence-bond solid into a state with spinful excitations, starting with the nucleation of spin--1/2 vortex defects.  Spin--1/2  defects were also inevitable in a stronger disorder regime where the starting point was a `glassy' covering of short-range valence bonds rather than an ordered one. Together, both of these results motivated our disordered-LSM conjectures.

The starting point for this analysis requires us to make the distinction between a state of frozen short ranged singlets, which we denote a valence bond glass,  and valence bond states with gapless spin excitations such as random singlet phases with singlets at arbitrarily long length scales and low energies. 
We exploited a weak disorder limit in which the renormalization group flow could be partly tracked. In this limit, the relevant energy scales are 
exponentially small, likely precluding any observation of this regime in experiments. Here we chose to focus on experimental applications to YbMgGaO$_4$, which shows complex phenomenology that is captured by the theory for stronger disorder. Indeed disorder arises in YbMgGaO$_4$, as well as the isostructural and phenomenologically analogous compound YbZnGaO$_4$,  through the nonmagnetic layers which in both cases appear to be fully amorphous. We expect the  theory to also describe other strongly frustrated compounds with much milder bond randomness; in those cases, the small emergent energy scales may require further care to observe.

We have focused here on systems in one and two spatial dimensions. 
In 3d a VBS is  stable to weak disorder, but similar issues could  arise in the presence of sufficiently strong disorder. Distinct types of paramagnetic states are also possible in 3d: we will discuss these separately \cite{forthcoming3dglass}. In the context of 3d systems with random valence bond patterns it is interesting to consider the 3d magnet Ba$_2$YMoO$_6$, where degenerate orbital and spin--1/2 moments residing on a face-centered cubic lattice are seen to freeze into a disordered pattern of orbital dimers and spin singlets.\cite{Bos2010}  Defect spin--1/2 moments are observed at finite density within the dimerization pattern, and these defects appear to freeze into a dilute spin glass at ultralow temperatures $T_g \sim 0.01 \theta_{\text{CW}}$ with $\theta_{\text{CW}}=$160 K the Curie-Weiss temperature.\cite{Bos2013}
A recent theory\cite{Jackeli2017} suggests that this dimer-singlet phase can arise from an unusually robust extensive degeneracy due to the orbital dimers; these properties of the orbital moments, together with a lack of observed bond randomness but apparent presence of some magnetic site dilution disorder\cite{Bos2010,Cranswick2010}, suggest that the physics in Ba$_2$YMoO$_6$ may involve additional ingredients beyond those considered here.

Returning to 2d magnets, it is worth noting that even for a layered material with vanishing magnetic coupling between two layers, some three-dimensionality will be required in describing the VBS order in the real material, due to phonons. The 3d phonons couple to the lattice modulations of the VBS order, thereby giving an effective coupling between VBS order parameters across different layers. This gives a three dimensional VBS order in the clean system, which then melts only under finite, but very weak, disorder.

Some of the phenomena discussed here could be studied numerically. Sign-free Quantum Monte Carlo can be used to numerically study the fate of the weakly disordered VBS on the unfrustrated square lattice\cite{Sandvik2007,shu2016properties}. 
  However the very small energy scale of defect interactions ---  in the theoretically controlled ultraweak disorder limit of Sec.~\ref{VBSdis}, it is exponentially small in $\xi_{2d}$ --- could be an obstacle to numerical studies of the low energy regime.
\footnote{However see footnote \ref{note:noteadded} and Ref.~\onlinecite{forthcomingAnders}.}
  (In principle, an indirect `multi-scale' approach might be possible, with separate simulations to establish the geometry of the defect array and to treat their interactions.) Further studies of the phase diagrams for experimentally relevant Hamiltonians, even in the clean limit, would also be useful as a starting point for understanding the disordered materials.  We note that when bond randomness suppresses a given magnetic ordering without immediately producing a spin glass, 
the resulting low energy excitations may be well described by the RG flow from pinned singlets discussed here.

A particularly interesting aspect of our considerations is the restriction on ground states of disordered quantum magnets with spin-$1/2$ per unit cell and with statistical lattice symmetries. 
The restriction naturally explains the emergence of low energy excitations triggered by disordering the VBS symmetry breaking order.
We hope that the physical arguments presented in this paper lead to mathematically rigorous scrutiny of such Lieb-Schultz-Mattis like restrictions in disordered systems.
The specific conjectures we formulated should be a good target for such future studies.

\textbf{Acknowledgements.} We thank Leon Balents, John Chalker, Radu Coldea, Fabian Essler, David Huse, Mehran Kardar, Chris Laumann,  Patrick Lee, Max Metlitski, Martin Mourigal, Siddharth Parameswaran, Anders Sandvik, Chandra Varma, and Minoru Yamashita for helpful discussions. I.K. was supported by the Pappalardo Fellowship at MIT. A.N. was supported by EPSRC Grant No. EP/NO28678 and by the Gordon and Betty Moore Foundation under the EPiQS initiative (grant No. GBMF4303). T. S. was supported
by NSF grant DMR-1608505, and partially through a Simons
Investigator Award from the Simons Foundation.

%

\appendix
\section*{Appendices}

\appendixsection{Sign structure of defect spin couplings}
\label{sign_structure_appendix}

We consider the effective interactions $J_\text{eff}$ between the spin-1/2s in the cores of defects, in either 1d or 2d. We restrict to the limit of weak disorder $\Delta$, where the typical defect separation $\xi$ diverges, and where $J_\text{eff}$ is exponentially small in $\xi$. In particular we are interested in whether the signs of these effective couplings are necessarily randomized due to disorder in this $\Delta\rightarrow 0$, $\xi\rightarrow \infty$  limit or whether they can retain a definite sign structure. 

For simplicity, let us assume that the vortex spins are well-localized at a single site.  In the large $\xi$ limit, interactions involving more than two vortex spins are strongly suppressed relative to the two-spin interactions, so it suffices to focus on a pair of vortices in an infinite system,  with Pauli operators $\vec \sigma_1$ and $\vec \sigma_2$. A convenient way to define their effective Hamiltonian is
\be\label{Heff}
e^{-\beta H_\text{eff}[\vec\sigma_1, \vec\sigma_2]} 
=
\Tr_\text{other spins} e^{-\beta H}.
\ee
The resulting $H_\text{eff}$ will be approximately $\beta$-independent for sufficiently large $\beta$  (apart from a trivial constant term) and by $SO(3)$ symmetry will be of the form ${H_\text{eff} = J_\text{eff}\, \vec \sigma_1.\vec\sigma_2}$.

The effective coupling $J_\text{eff}$ is determined by the exponentially decaying spin correlations of the `other spins' in the background configuration of pinned VBS domains.
This can be seen in linear response, where we treat the coupling between each defect spin and its neighbours as a small parameter (removing this approximation would only dress the operators appearing in the correlator below, and would not change the basic behaviour).
Let us write the couplings in $H$ that involve $\vec \sigma_1$ as $\epsilon_1 \vec \sigma_1. \vec S_1$ and similarly for $\vec \sigma_2$. Here $\vec S_1$ is a weighted sum of the magnetizations of the sites surrounding the defects (the coefficients depending on the microscopic Hamiltonian) and the coupling strength $\epsilon_1$ is treated as a small parameter. We have
\be
H =\sum_{i=1,2} \epsilon_i \vec \sigma_i. \vec S_i + H_\text{other spins}.
\ee
From (\ref{Heff}), expanding in $\epsilon$ gives\footnote{`Sufficiently large' $\beta$ in Eq.~\ref{Heff} means that the time integral of the 2-point function can be treated as running from $-\infty$ to $\infty$. In a clean paramagnetic background the necessary order of magnitude  $\beta_\text{min}$ scales with the square root of the separation between the defects, as can be seen from the exponential decay of the 2-point function as a function of space-time distance, or from a directed path expansion. This power is probably unaffected by disorder.}
\ba\label{linear_response}
H_\text{eff} & \simeq J_\text{eff}  \vec \sigma_1 \cdot \vec \sigma_2,
 &
 J_\text{eff} & = - \f{\epsilon_1\epsilon_2}{3}\int_{-\infty}^\infty \dd \tau  
 \<  \vec S_1(0) \cdot \vec S_2(\tau) \>.
\end{align}
Here it must be remembered that the expectation value is taken not in the clean system, but in a  configuration in which VBS domains of scale $\xi$ are pinned in a specific pattern that depends on the  disorder realization.

In Sec.~\ref{sign_free_subsection} we consider a class of `sign-free' models on bipartite lattices, where $J_\text{eff}$ can be shown directly to have a non-random sublattice sign structure, in any disorder realization. While this class includes various important models, sign-freeness is a fine-tuned property which is not present in generic models. Therefore we must ask whether the sign structure of $J_\text{eff}$ can survive (in the limit of interest where disorder is present but small, and $\xi$ is parametrically large) when the sign-free property is broken. In  Sec.~\ref{generic_subsection} (which is independent of Sec.~\ref{sign_free_subsection}) we argue that   the sign structure found for the sign-free models \textit{can} survive in generic models in the relevant limit $\Delta\rightarrow 0$, $\xi\rightarrow \infty$.

Our discussion of generic models uses the fact that correlation functions of gapped degrees of freedom can be understood, very generally, in terms of directed path ensembles.\cite{mehran_book}   We also point out that generic models exist where the sign of $J_\text{eff}$ is completely randomized at weak disorder.

\subsubsection{Sign-free models on bipartite lattices}\label{sign_free_subsection}

The sign-free class we consider includes, among many other models, Heisenberg models on bipartite lattices with (perhaps random) nearest-neighbour AF couplings.\footnote{More generally, AF couplings between opposite sublattice sites and F couplings between same sublattice sites.} The class also includes the  (randomized) `JQ model',\cite{Sandvik2007} which includes a 4-spin resonance term which can be used to drive the clean system into a VBS phase.  For these models $J_\text{eff}$ has a definite sublattice sign structure: $\operatorname{sign} J_\text{eff} = - \chi$, where we define $\chi$ to be $+1$ if the two defects are on the same sublattice and $-1$ if they are on opposite sublattices. This sign structure is  well known in the context of strong-disorder RG for the  1d  AF Heisenberg chain and generalizations.\cite{Lee1981,shu2016properties} For various models it also follows directly from Eq.~\ref{linear_response} and standard `loop' representations (see e.g. Ref.~\cite{evertz2003loop})  of correlation functions. However, let us give a simple general argument which holds for arbitrary disorder strength and dimensionality and does not require the loop representation. 

The defining property of this class of sign-free models is that all off-diagonal $H$ elements are negative when the Hamiltonian is written using the single-site basis $\ket{\alpha}$, $\alpha = 1,2$, defined by
\begin{align}
& \text{A sublattice:} & \ket{1} & = \ket{\uparrow}, & \ket{2} & = \ket{\downarrow}, \\ 
& \text{B sublattice:} & \ket{1} & = \ket{\downarrow}, & \ket{2} & = - \ket{\uparrow}
\end{align}
(so that a singlet between opposite sublattice sites is $\sum_\alpha \ket{\alpha}_A\ket{\alpha}_B$). This means that Trotterizing $\Tr e^{-\beta H}$ in this basis maps the partition function to a classical statistical mechanics problem, with positive weights, for degrees of freedom $\alpha = 1,2$ in spacetime. By Eq.~\ref{Heff}, the partition function of interest to us, which yields matrix elements of $H_\text{eff}$, is $\lf \Tr_\text{other spins} e^{-\beta H} \ri_{\alpha'\beta', \alpha\beta}$. This has periodic BCs in time at all sites except for the defect sites, where there are temporal `boundaries' at $t=0, \beta$ where the `colour index' is fixed by the matrix element considered.

Consider two defects, one on the A and one on the B sublattice, and label their states by $\alpha$ and $\beta$ respectively.  There is only one nontrivial term in $H_\text{eff}$, which can be written in terms of the singlet projector $\mathcal{P}$, so we can always write (for some constants $A$, $B$)
\ba
\lf e^{-\beta H_\text{eff}} \ri_{\alpha'\beta', \alpha\beta} & =
\lf \Tr_\text{other spins} e^{-\beta H} \ri_{\alpha'\beta', \alpha\beta} \\ &= 
A \delta_{\alpha'\alpha}\delta_{\beta'\beta} + B \delta_{\alpha'\beta'} \delta_{\alpha\beta}.
\end{align}
The first term is the identity and the second term is proportional to $\mathcal{P}$. 
If we instead consider defects on the same sublattice, with states labelled by $\alpha_1, \alpha_2$, then the singlet projector may be written $\mathcal{P}=(1-\mathcal{S})/2$, where $\mathcal{S}$ is the swap operator, which in components is the final term below:
\ba
\lf e^{-\beta H_\text{eff}} \ri_{\alpha_1'\alpha_2', \alpha_1\alpha_2} &=
\lf \Tr_\text{other spins} e^{-\beta H} \ri_{\alpha_1'\alpha_2', \alpha_1\alpha_2} \\ &= 
A \delta_{\alpha_1'\alpha_1}\delta_{\alpha_2'\alpha_2} + B \delta_{\alpha_2'\alpha_1} \delta_{\alpha_1'\alpha_2}.
\end{align}
If we take  $\beta\gg \beta_\text{min}$ and $\beta J_\text{eff} \ll 1$, then $B/A$ is proportional to $\beta J_\text{eff}$ in the first case and $-\beta J_\text{eff}$ in the second case. So to show that the coupling is fixed by the sublattice (AF for opposite sublattices and F for same sublatice) it is sufficient to show that  the constants $A$ and $B$ are always positive for the sign-free models considered.

Consider the opposite-sublattice case (the same-sublattice case is similar). We have
\ba
\lf \Tr_\text{other spins} e^{-\beta H} \ri_{12, 12} &= A, \\
\lf \Tr_\text{other spins} e^{-\beta H} \ri_{11, 22}& = B.
\end{align}
The LHS of each formula maps to a `classical' partition function with fixed boundary conditions at $t=0, \beta$ for the defect sites, and is manifestly positive. This establishes the claim about the sign structure of $J_\text{eff}$ for this class of sign free models.

\subsubsection{Generic models}\label{generic_subsection}

It is clear from Eq.~\ref{linear_response} that models exist where  ${\operatorname{sign} J_\text{eff}}$ is completely randomized in the weak disorder limit. As mentioned in the text, this will certainly occur in 1d whenever the exponentially decaying spin correlations in the \textit{clean} VBS have incommensurate sign structure, as a result of the large random separations between defects.   For a 2d model where the  spin correlations in the clean VBS are incommensurate,  randomized $\operatorname{sign} J_\text{eff}$ is also the natural possibility (although it does not strictly follow because of a caveat discussed at the end of this section).  

It is less obvious whether the non-random sign structure for certain bipartite models described in Sec.~\ref{sign_free_subsection} can survive in models that are not fine-tuned (in the limit of interest,  where $\Delta\rightarrow 0$ and $\xi\rightarrow 0$ in the manner prescribed by Imry-Ma). First recall that
the exponentially decaying correlations of gapped degrees of freedom (here the spin) can be understood  in terms of directed path expansions.\cite{mehran_book} A simple classical example is the Ising model in the disordered phase, where the high-temperature expansion relates the two-point function to an effective partition function for a path which connects the two points.
 This formalism is essential for understanding the effect of disorder on correlation functions of disordered degrees of freedom. Disorder is an RG-relevant perturbation for sums over paths of this type.

Similar path expansions may be written for the quantum magnet. Heuristically, these paths are tunneling trajectories for gapped spin-1 excitations. 
The details will not concern us, but it is useful to have in mind a  toy model. Let us take the valence bond covering to be determined  by strong explicit dimerization in the Hamiltonian. This is \textit{not} of course the regime of interest to us, but it illustrates the basic features and makes the path expansion simple.
Fix a dimer covering  that is complete except for the two defects, and take the  couplings on dimerized bonds  to be of order  $J_\text{strong}$ and couplings on undimerized bonds to be of order $J_\text{weak}$ with $J_\text{weak}/J_\text{strong}\ll 1$. 

In the  limit of small $J_\text{weak}$, the perturbative expression for $J_\text{eff}$ is dominated by directed paths which traverse the minimal number $k$ of weak bonds, and is of order $J_\text{weak}(J_\text{weak} / J_\text{strong})^{k-1}$. More precisely, in this limit
\be
J_\text{eff} = -\sum_\text{paths} (-1)^{\# \text{length}} \frac{ \prod_\text{weak bonds $\ell$} J_\ell }{\prod_\text{strong bonds $\ell'$} 2J_{\ell'} },
\ee
where we sum over paths of minimal $k$. In the numerator the product is over all weak bonds traversed by a given path. In the denominator the product is over all the strong bonds that the path visits: the path can visit a strong bond either by traversing it, or by taking two successive steps on weak bonds that are adjacent to the strong bond. `Length' is the total number of bonds traversed by the path.  The simplest case is 1d, where the above sum reduces to a single term. 

On a bipartite lattice in any number of dimensions, every term in the above sum has the same sign (if the couplings are all positive). This is because on a bipartite lattice all paths between two points have the same length modulo two.  In this case $J_\text{eff}$ is given by  a sum of positive terms, once we extract the sign factor $\operatorname{sign} J_\text{eff} = -\chi$ which depends only on the sublattices of the two defects. 
After extracting this sign factor, the expression for $J_\text{eff}$ above defines an effective classical partition function for a path or `string' whose energy depends on which links it visits. 

A similar picture, in terms of a partition function for a directed path, will be valid at a coarse-grained level even away from the  artificial limit above, so long as spin correlations are exponentially decaying.\cite{mehran_book} For a sign-free model,  the Boltzmann weights for the effective classical partition function will always be positive.
 The  exponential decay of $J_\text{eff}$ with separation  is equivalent to the nonzero free energy per unit length, or `line tension', of this path.  In a clean 2d system, this line tension ensures that the sum is dominated by  trajectories that are straight on scales of order $\xi_{2d}$.\footnote{If the VBS breaks rotational symmetry the line tension can depend on orientation.} In a given disordered VBS background, the geometry of the  path may not be straight on this scale, but a section of the path away from VBS domain walls will be approximately straight.\footnote{Rotational symmetry can be broken  in different ways in different VBS domains, so that the orientation-dependence of the line tension differs in different domains. The path is of length $\sim \xi_{2d}$, so it can visit more than one (but order one) VBS domains. In a given VBS background its energy may be minimized by a non-straight trajectory crossing through more than one domain. The path could also prefer to be `glued' to a VBS domain wall, as discussed below.} By standard results on paths in random media, such a section will deviate from being straight on a parametrically smaller scale $\sim \xi_{2d}^{2/3}$ (neglecting logs\footnote{Logarithms in $\xi_{2d}$ arise because the limit of interest is where disorder $\Delta\rightarrow 0$ and $\xi_{2d} \rightarrow\infty$ with $\xi_{2d} \sim e^{c/\Delta^2}$, i.e. ${\Delta \sim (\ln \xi_{2d})^{-1/2}}$.})  but this will not concern us.\cite{HuseHenleyFisher1985respond,kardar1985commensurate,kpz}

In a model that is not sign-free some of the paths that are summed over in the effective classical partition function determining  $J_\text{eff}$ will acquire negative `Boltzmann weights'. The question is whether $J_\text{eff}$ retains a fixed sign when the breaking of sign-freeness is weak.
There are two  distinct ways to `weakly' break the sign-free condition on the off-diagonal matrix elements of $H$:

 (1) First, we can introduce \textit{weak} sign-rule violating couplings \textit{everywhere}. For example, in our toy model we may introduce AF second-neighbour couplings with a small magnitude $J_2\ll J_\text{weak}$. Such weak violations of the sign rule will be present generically even in the clean system so we must consider them. 
 
 (2) Second, we can introduce \textit{rare} locations where the sign rule is \textit{strongly} violated. For concreteness, we can imagine flipping the sign of $J_\text{weak}$ in our toy model for a small fraction $p$ of bonds. This is a spatially random effect associated with bonds where the disorder $|\Delta J|$ is locally strong. More generally, in the following we may take $p$ to be the probability of having a sign-rule violating bond whose strength is above some order one threshold.

We take the weak disorder limit by rescaling the probability density for $\Delta J$ on each bond, so that it has standard deviation $\Delta$. That is we take ${P(\Delta J) = \Delta^{-1} f(\Delta J / \Delta)}$, where the distribution $f$ has standard deviation 1. If the distribution $f$ has bounded support, then effect (2) cannot occur in the weak disorder limit of interest to us: the probability $p$ above is strictly zero for small $\Delta$. However, if the distribution $f$ does not have bounded support, then $p(\Delta)$ tends to zero in some way as $\Delta$ tends to zero. This gives a small but nonzero density of negative bonds in the directed path partition function. In this case we must check whether this small density of negative bonds can affect the very long paths which are relevant in the weak disorder limit.

In 1d, for any distribution with a finite variance, $p(\Delta)$ scales to zero faster than $\Delta^2$. This means that  rare strong bonds are not seen on the Imry-Ma scale $\xi_{1d}\sim \text{const.}/\Delta^2$ and effect (2) does not play a role.

In 2d, the much larger Imry-Ma lengthscale means that, depending on the disorder distribution, there may be many negative bonds  in a patch of size ${\xi_{2d}\sim e^{-\text{const}./\Delta^2}}$. If the tail of $f(u)$ decays as $e^{-\text{const}. u^2}$ with a sufficiently large constant, this is narrow enough to banish negative bonds on this scale. This is certainly sufficient to ensure that effect (2) does not play a role. This is not in fact necessary --- a weaker condition on $f$ would certainly also be sufficient as we discuss below. However, it may be natural to impose this stronger condition on the distribution $f$ anyway.  If a patch of size $\xi_{2d}$ contains a significant number of rare bonds with large $|\Delta J|$, there will likely be \textit{other} disorder effects that make the signs of $J_\text{eff}$ moot.
For example, a strong ferromagnetic bond can nucleate a spin-1, and when these spin-1s have larger density than the vortex defects the large-scale physics is no longer dominated by the latter.  

Nevertheless, let us briefly consider the possible effect of  (2). The effect of rare negative bonds (with density $p\ll 1$) on directed path partition functions has been  studied extensively, with motivations from various contexts.\cite{nguyen1985tunnel,medina1989interference,WangExact,medina1992quantum,roux1994interference,NguyenCrossover,husesigntransition,laumannsigntransition} In 1d the sign of the partition function is of course randomized whenever the length $\xi$ of the path is much greater than the crossover scale  ${\xi_* = 1/p}$, i.e. whenever the path encounters a large number of negative bonds. 
The 2d case is considerably more subtle. The sign is again randomized at large scales whenever $p>0$.  The crossover scale $\xi_*$ depends on what other disorder is present, in addition to the negative bonds. The scaling is $\xi_*\sim  1/p$ if, in addition to the negative bonds, there is disorder of $O(1)$ strength in the positive bonds.\cite{husesigntransition,laumannsigntransition} The case where the negative bonds are the only source of disorder has recently been addressed in Ref.~\cite{laumannsigntransition}, and an exponentially larger lengthscale, $p^{-\text{const.} p^{-1/3}}$, was found.
Here we are interested in yet another case, where there is disorder of parametrically small strength $\Delta$ on the positive bonds, in addition to the  parametrically small density $p(\Delta)$ of negative bonds. By the  rare-region logic of Ref.~\cite{laumannsigntransition}, it is clear that the crossover scale $\xi_*$ [which we must compare with the path length $\xi_{2d}(\Delta)$] will be exponentially large at small $\Delta$, with the functional form depending on the breadth of the $\Delta J$ distribution. This will \textit{at least} ensure that effect (2) is banished for distributions with tails weaker than than $f(u)\sim e^{-\text{const.} |u|^x}$, where $x$ is some constant smaller than 2. So effect (2) is certainly avoided even for distributions somewhat broader than a Gaussian. A more careful analysis would be required to establish whether this effect can ever play a role for broader distributions (with finite variance).\footnote{A naive estimate of $\xi_*$ suggests that effect (2) may be absent for any distribution $f$ with finite variance, but this estimate may be too crude. The initial disorder $\Delta$  is small but grows under the RG. It will cause a crossover from random walk behaviour to disorder-dominated behaviour at a lengthscale $\Delta^{-4}$ measured longitudinally, or equivalently a lengthscale $\Delta^{-2}$ measured transverse to the path.\cite{kim1991finite} Naively one might think that in the present case this transverse lengthscale $\Delta^{-2}$ should play the role of $\ell_*$ in  the rare-region argument of Ref.~\cite{laumannsigntransition}. This would give $\xi_*\sim p(\Delta)^{-\text{const.} \Delta^{-2}}$, which is sufficiently large to ensure that $\xi_*\gg \xi_{2d}$, so that effect (2) cannot play a role at all. However this estimate may be invalid because for a sufficiently broad power law distribution  it gives a $\xi_*$ that exceeds the previous result for the case $\Delta =0$, $p\neq 0$.}

It remains to check that mechanism (1) does not affect the sign of $J_\text{eff}$ if the sign-rule-violating couplings are sufficiently weak. This is in fact straightforward to see in the path picture. Microscopically, a wrong-sign $J_2\vec S_{i}.\vec S_{i+2}$ coupling in 1d, say, will allow the path to take steps of length 2 that incur a negative Boltzmann weight. However if the  weight associated with these wrong-sign steps is sufficiently small, then after some slight coarse-graining the effective Boltzmann weight will be  positive for all coarse-grained steps. (For example, the negative weight associated with the step of length 2 allowed by $J_2$ will be outweighed by the positive-weight trajectory between the same points given by two steps of length 1.) This is true also in higher dimensions, and is  not affected by weak disorder, which  modulates the local coarse-grained Boltzmann weights by an amount of order $\Delta$ but does not change their sign if they are initially always positive and of order 1.

Above we noted that if the clean system had incommensurate spin correlations then random $\operatorname{sign} J_\text{eff}$ is the natural expectation. This is clear in 1d but we note a caveat in 2d. Recall that in a given disorder realisation $J_\text{eff}$ is mediated by a coarse-grained `optimal' path $P$ between the defects which has a definite geometry on scales of order $\xi$. Note that in 2d the defects are also connected by VBS domain walls. The line tension of the optimal path $P$ depends on the local VBS environment. Therefore it could be the case that the effective free energy  of the path $P$ is minimized when it is `glued' to a VBS domain wall. That is, spin correlations could be mediated principally by one of the domain walls connecting the defects. Further, in this situation it is conceivable (although it sounds rather contrived) that the correlations mediated by such a domain wall could have a commensurate sign structure even when the correlations in the bulk of a VBS domain are incommensurate.

Finally we comment on the triangular lattice case. Recall that for columnar VBS order, within a superdomain, there is a correspondence with the columnar VBS and its defects on  the square lattice. We noted in the text that in principle $J_\text{eff}$ for these defects could have the bipartite sign structure that is natural on this embedded square lattice, but that this cannot happen when VBS order is weak. For weak VBS order (and weak quenched disorder) the sign structure of the spin correlations on short scales should respect the symmetries of the triangular lattice. These symmetries are incompatible with the sublattice sign structure associated with the embedded square lattice. If this sign structure is broken on short scales it is unlikely that it will be restored on longer scales.

\appendixsection{Vortex nucleation in the weak disorder limit}
\label{app:nucleation}
Here we argue that in the weak disorder limit VBS vortices are necessarily nucleated on the Imry-Ma lengthscale $\xi_{2d}(\Delta)\sim \exp (J^2/ {\Delta^2})$, where  $\Delta$ is the strength of disorder.  Introducing a  pair of VBS vortices into a vortex-free state allows a rearrangement of the domain pattern on the lengthscale $\xi_{2d}(\Delta)$. If the disorder configuration is favourable this rearrangement can reduce the energy by an amount of order $\Delta \times  \xi_{2d}(\Delta)$. Since $\xi_{2d}(\Delta)$ grows exponentially as $\Delta \rightarrow 0$, this energy gain is large when $\Delta$ is small, and overwhelms the $O(1)$ core energy cost associated with the vortex core.  

The exponential growth of $\xi_{2d}$ for small  $\Delta$ also means that the typical energy scale for the  defect-spin couplings $J_\text{eff}$ is doubly-exponentially weak, since $J_\text{eff}$ itself is exponentially small in defect separation.  For the application to experiments one must consider stronger disorder, and the couplings between unpaired spins will not be exponentially small in any sense.

\appendixsection{VBS order as O($n$) vison condensate from $Z_2$ spin liquid}
\label{appendix:QSLVBStransition}

In this appendix section we give more details for the discussion of the generalized VBS phases on the triangular lattice through the transition from the proximate $Z_2$ quantum spin liquid. This is a ``starred''-Landau-Ginzburg theory\cite{Sachdev1994}, involving condensation of  the  $Z_2$ vison. 
The VBS phases we discuss include not only the columnar-VBS state of Fig.~\ref{fig:VBS} but also the known\cite{Sondhi2001,Mila2008,Xu2014} plaquette-VBS phases, whose resonances involve various 4-spin, 12-spin, or 16-spin clusters.  It is more difficult to visualize a point defect in these complicated phases; nevertheless, we shall now show that these VBS phases all host point topological defects which are $\mathbb{Z}_2$ vortices (i.e.\ the vortex is its own anti-vortex), and that these $\mathbb{Z}_2$  vortices again necessarily host unpaired spin--1/2s. 

 The spin liquid variables enable  a natural description of the VBS topological defects, in a unified language; the present section will be concerned with various aspects of this description. 
At the vison condensation transition, the various VBS phases are unified into a continuous manifold of parent VBS states.  As described below (see also Refs.\onlinecite{Sondhi2001,Xu2014}), for the two simplest vison condensation transitions 
this manifold is a unit sphere in $n$-dimensions ($n=4,6$ respectively) but with antipodal points identified, i.e.\ a headless $n$-dimensional unit vector. 
The point defects are then easily understood by analogy  to 120-degree magnetic order and nematic orders, whose order parameters are similarly headless: these defects are the $Z_2$ vortices.\cite{Miyashita1985}. Here a VBS vortex necessarily requires a $S=1/2$ spinon to be nucleated in the vortex core; the  vortex $S=1/2$ modes may be considered as a relic of the fractionalized $Z_2$ spinon. 

Condensing the spinon leads to a magnetically-ordered phase, such as the 120$^\circ$ ordered phase of the triangular lattice Heisenberg antiferromagnet. The vison does not carry magnetic quantum numbers, so its condensation does not lead to a magnetic (or any time-reversal-broken) phase. Instead, it leads to a lattice-symmetry-broken phase, such as a singlet or plaquette valence bond crystal. 

To analyze the manifold of VBS order parameters that results from vison condensation, let us consider the vison condensation in more detail. The vison sees the spin-half spinon as a $\pi$ flux. Each site of the triangular lattice carries a single spin-$1/2$ degree of freedom: the number of spinons (in total for both spin flavors) at each lattice site is constrained to be exactly one in the ground state. So the vison experiences a $(-1)$ phase when hopping around any triangular lattice site. The vison hopping can thus be modeled as hopping on a fully frustrated honeycomb lattice, where the honeycomb is the dual lattice formed by triangular plaquettes, and the frustration implies an effective uniform magnetic field experienced by the vison, with flux $\pi$ per hexagon plaquette.  The fully frustrated honeycomb hopping problem is known\cite{Sondhi2001,Xu2014} to produce different possible solutions, allowed by the spin liquid PSG, depending on the particular pattern of vison hopping. The simplest vison hopping produces four degenerate minima [real modes from wavevectors $\pm K/2$, $\pm K'/2$, where  $K$ is a BZ corner], while a more complicated vison hopping (which need not  necessarily be more complicated in terms of original spin variables) produces six degenerate minima [real modes from wavevectors $\pm M_1/2$, $\pm M_2/2$, $\pm M_3/2$, where $M$ is the midpoint of a BZ edge]. The family of VBS orders in the latter scenario includes the columnar order\cite{Xu2014} with wavevector $M$. 
 This four- or six-component vector $\vec{v}$ of the vison condensate describes the resulting VBS order parameter, with  an overall sign redundancy. 
Though any given VBS order parameter is discrete it is thus natural to describe it as a discrete subset of this continuous manifold, where the manifold is a unit sphere in $n$ dimensions with each pair of opposite points identified into a single point,  written as $S^{n-1}/Z_2 \equiv \mathrm{RP}^{n-1}$.  The resulting manifold $\mathrm{RP}^{n-1}$ has point-like defects, characterized by the first homotopy group $\pi_1(\mathrm{RP}^{n-1})=Z_2$, produced by paths from a point to its diametrically opposite point on the sphere. (Without modding by $Z_2$ the sphere homotopy is trivial for $n>2$.) Thus there is a vortex point defect which  is its own antivortex. 

We next argue that the vortex core in the VBS orders carries a protected spin-half degree of freedom. To see the spin-half at the vortex core, consider condensing the vison into a VBS configuration with a vortex. The vortex represents a winding of the vison condensation vector $\vec{v}$ such that $\vec{v}$ winds from some initial value $\vec{v_0}$ at angle $\theta=0$ near the vortex core, to the opposite value $-\vec{v_0}$ at $\theta=2\pi$. (Recall that $\vec{v_0}$ and $-\vec{v_0}$ represent the same VBS order.) The vison field sees the vortex core as a $\pi$ flux. But there is only one such object in the $Z_2$ spin liquid: the spinon, which necessarily also carries a spin-half degree of freedom.

\begin{figure}[]
\includegraphics[width=0.9\columnwidth]{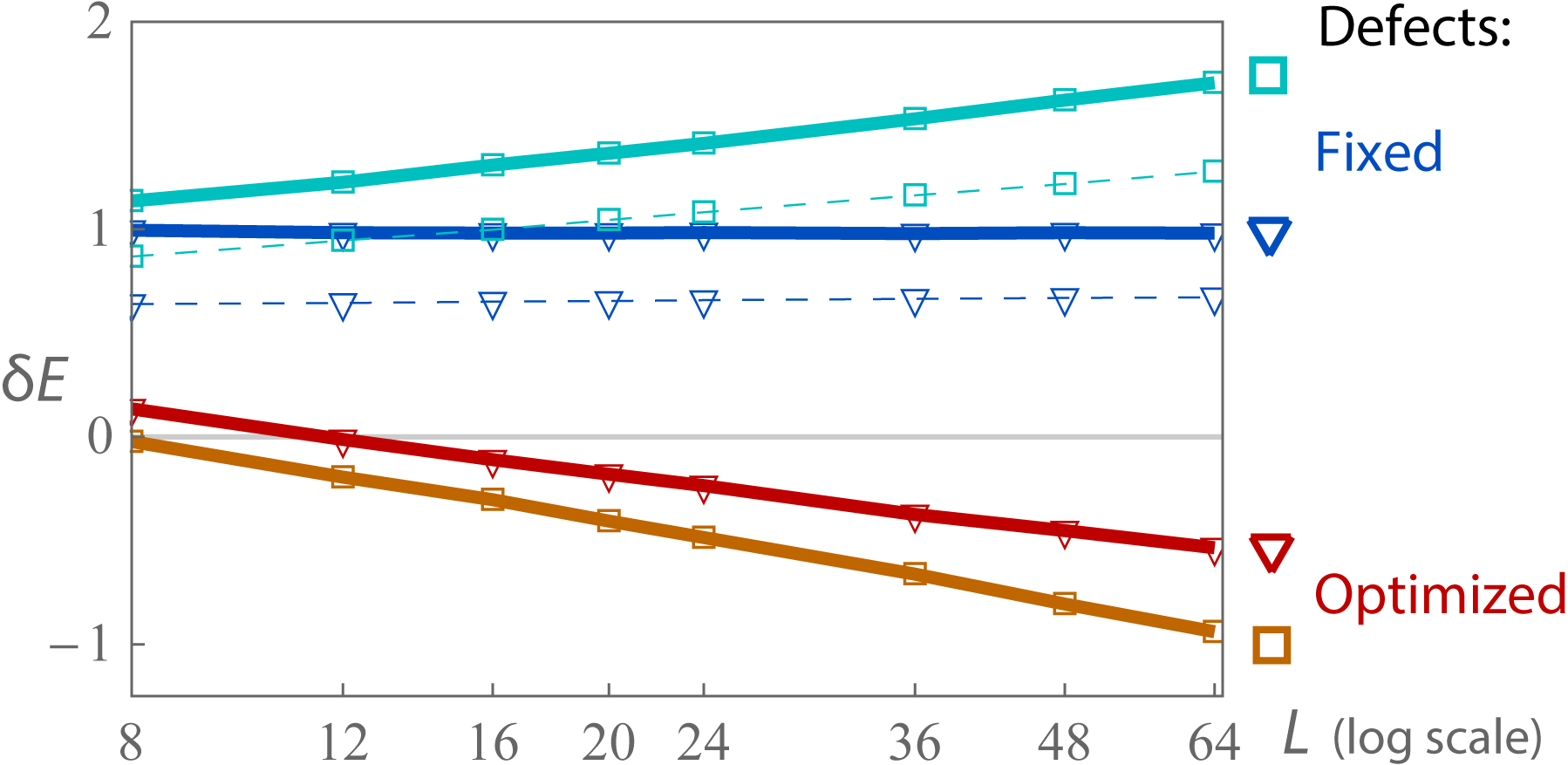}
\caption[]{ \textbf{Energy cost for fixed and optimized monomers on the square and triangular lattices.} 
Here we compare the energy cost for both fixed-site and partially-optimized monomers between the square lattice (square symbols) and triangular lattice (triangle symbols), in a log-linear plot. For fixed defects,  the growth of the standard deviation (dashed lines) ensures that optimized defects gain diverging energy, both for bipartite and non-bipartite lattices.  
} \label{fig:fixedmonomers}
\end{figure}

\begin{figure}[]
\includegraphics[width=\columnwidth]{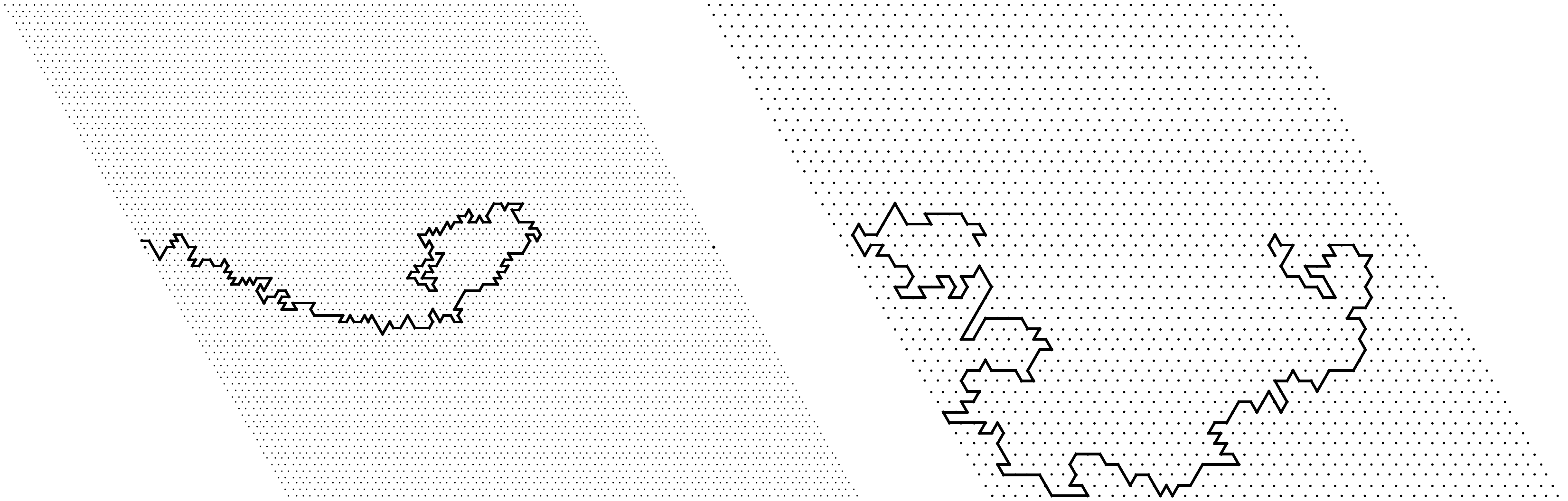}
\caption[]{ \textbf{String of shifted dimers connecting monomers in the random dimer model.} 
The optimal dimer coverings in the random-energy dimer model, computed for the full lattice and for a lattice with a pair of monomer defects, is the same everywhere except along a string connecting the two monomer sites: along the string all dimers are shifted by one. Here we show sample plots of these strings, for fixed defects [$L=80$, left] and optimized defects [$L=64$, right]. The string fractal dimension is computed to be $d_f=1.28(1)$ for the triangular lattice, and  $d_f=1.25(1)$ for the square lattice.
} \label{fig:dimerchainapp}
\end{figure}

\appendixsection{Numerical simulation of random-energy classical dimer model}
\label{app:nucleationnumerics}

Here we discuss details of the numerics  on the random energy dimer model. 
 We used  $L\times L$ lattices with periodic boundary conditions in the $x$ direction (e.g. identifying left and right edges of each lattice in Fig.~\ref{fig:dimerchainapp}), giving a cylinder. Note that this is a planar graph (identical results were found on a torus).  We assigned random energies to the edges of the graph, taken from a uniform distribution on the integers $\{ -\Delta, - \Delta + 1, \ldots, \Delta\}$, with $\Delta = 5000$. Plots of $\delta E$ in the main text are given with $\delta E$ measured in units of $\Delta$. When looking at defects, we removed pairs of sites by removing both a random site from the equator of the lattice ($x=x_0, y=0$) as well as the site $x=x_0+L/2, y=1$ approximately opposite 
  (the shift to $y=1$ ensured defects were on opposite sublattices for the square lattice, as required for finding a dimer covering on the remaining graph; the same shift was implemented on the triangular lattice to enable direct comparisons of the results).
We computed the complete dimer covering with minimum dimer energy using the famous \textit{blossom} algorithm of Edmond\cite{Edmonds1965,Galil1986} as implemented in the NetworkX python package (implementation by Joris van Rantwijk, ``max\_weight\_matching'' with ``maxcardinality=True''; weights are interpreted as negative energies). At least $5\times 10^3$ independent disorder realizations were computed for each parameter set for the case of optimized defects, with many more ($10^5$) disorder realizations for each parameter set for the case of fixed defects.
For the calculation of partially optimized defects,  we again restricted to opposite defect pairs  on the equator and optimized over $x_0$.  This reduced the computational effort of the optimization and also reduced finite size effects from open boundary conditions. The  change in system energy $\delta E$ we found for these partially-optimized defects is  an upper bound on $\delta E$ for fully-optimized defects, so it  suffices to show that partially optimized defects give a negative $\delta E$ at large $L$.

Let us briefly give further technical details for the histograms shown in Fig.~2 of the main text.  These are histograms of the numerically computed distribution of system energy change $\delta E$ under introduction of a pair of monomers spaced $L/2$ apart, that are either fixed [top] or allowed to optimize their position along the equator of the cylindrical system [bottom]. Colors shown correspond to the various system sizes $L= 8, ... , 64$ with rainbow colors from blue to red. 
Histogram bin size is $0.05 \Delta$ and  $0.15 \Delta$ for the fixed and optimized monomers respectively, where $2\Delta$ is the width of the uniform distribution of graph edge weights $(-\Delta,\Delta)$; it is implied that $\delta E$ is in units of $\Delta$. Observe that for fixed monomers, the histograms are all numerically close to Gaussians, with mean and variance that converge quickly as $L\rightarrow \infty$, and with a small tail for negative $\delta E$. For optimized monomers, the histograms (no longer Gaussian-like) narrow and shift towards negative $\delta E$ as $L$ increases towards the thermodynamic limit. The behavior of the peak with increasing $L$ is consistent with the scaling $\sim - (\log L)^{1/2}$ expected for minima obtained by sampling a polynomial-in-$L$ number of times from a fixed-defect distribution whose tail is ${\sim \exp \lf {-\text{const}. \, \delta E^2}\ri}$.

For the sake of comparison with known results on the bipartite case, Fig.~\ref{fig:fixedmonomers} shows the energy cost for fixed and optimized defects on the square and triangular lattices. On bipartite lattices, a fixed monomer costs energy $\log L$ since it is a long-ranged vortex, with $\log L$ interactions in the elastic medium which maps to this problem. This elastic-vorticity effect indeed does not occur for the nonbipartite case. 

 It is also interesting to study more subtle properties of the system with monomers. Introducing defect monomers modifies the optimal dimer configuration along a string which connects the two monomers; as shown in Fig.~\ref{fig:dimerchainapp}, this string is a fractal object. The energy gain due to this domain wall string excitation enables the monomers, on either bipartite or nonbipartite lattices, to be pulled down from arbitrarily high energies and nucleate on the lattice.

\appendixsection{Defects in the classical dimer model on a non-bipartite lattice}
\label{app:dimerdefects}

Here we discuss monomer defects in the classical dimer model, focusing on the case of non-bipartite lattices.
 
A defect can be regarded as one endpoint of an excitation which is an open line instead of a closed loop. The typical energy cost or gain due to disorder can be parsed into contributions from sections  of this line in a series of concentric annuli of radii $\ell_k \sim 2^k$ with $k=1,2,\ldots$. The contribution from the $k$th lengthscale will be of random sign and of order $\ell^\theta = 2^{-k|\theta|}$. The sum of these contributions is convergent, giving a random variable with mean and variance of order one.

 We can also construct a rare region which shows that the density for defects in the classical triangular lattice dimer model, at zero temperature, is at least $\exp[- c (K/\Delta)^2]$ for large $K$, where $K$ is the core energy cost of a monomer defect.  Consider a disc-like region $D$ of size $\ell$, where we have fixed the signs of all the bond energies to favour a particular dimer configuration in $D$ with a defect at the center. The probability of such a region is exponentially small in the area, $e^{-\text{const.}\ell^2}$. To \textit{remove} the defect at the center, we must disturb the favoured configuration inside $D$ along a line of linear extent $\sim \ell$ connecting the center of $D$ to its exterior. This costs an energy of order $\ell \times \Delta$ with $\Delta$ the disorder strength. Therefore, for large $\ell$, such regions correspond to places where the energy `cost' for introducing the defect, relative to the ground state of the defect-free model, is $\delta E \sim K - \text{const.} \ell \Delta$. By taking $\ell\sim K/\Delta$ we obtain a region where it is favourable to introduce a defect. This shows such regions have a density $\rho$ at least as large as $\exp[- c (K/\Delta)^2]$ with $c$ a numerical constant. Note that the rare regions necessary for this argument do not require the disorder distribution on an individual link to be unbounded.

\appendixsection{Defining singlets with spin-orbit coupling}
\label{appendix:SOC}
Here we consider the effects of spin-orbit coupling on microscopic definitions of two-spin singlet states; recall that VBS phases are generally defined as lattice-broken quantum paramagnets, which preserve time reversal symmetry but transform non-trivially under some lattice translations and rotation symmetries. 

The strong spin-orbit coupling  of Yb$^{3+}$ completely breaks the continuous SO(3) spin rotation symmetry, which naively suggests that the spherically-symmetric two-spin singlet state is no longer a well-defined eigenstate. 
However, as long as inversion symmetry is present, this is not the case. The singlet, which is odd under inversion, does not mix with the inversion-even triplet manifold. (Extensions to the case of spin orbit coupling without inversion centers will be discussed elsewhere\cite{Scaling2018}.)
To see this explicitly, consider a given bond $(i,j)$ with its $3\times 3$ matrix of spin interaction coefficients $J_{i j}^{\mu \nu}$. The matrix $J^{\mu \nu}$ is real and, if it is symmetric (e.g.\ if the bond has an inversion center), it can be diagonalized to yield real-valued eigenvectors, known as its principal axes.   (The symmetry condition on $J^{\mu \nu}$ is equivalent to ignoring any Dzyaloshinskii-Moriya interactions, which may arise in particular disorder configurations but are forbidden if inversion symmetry is preserved on average.)  The spin interaction is then written as a sum of exchanges through the principal axes $x,y,z$, i.e.\ $J^x S^x S^x + J^y S^y S^y + J^z S^z S^z$ with different coefficients $J^x, J^y, J^z$. The singlet state $( \left| \uparrow \downarrow  \right\rangle - \left|  \downarrow \uparrow \right\rangle)$ is an eigenstate of each of the terms --- this is clear for $S^z S^z$, and easy to see from the singlet's spherical symmetry also for $S^x S^x$ and $S^y S^y$. Thus if the $J^x, J^y, J^z$ coeffients  are on average sufficiently antiferromagnetic then the ground state of the two spins $(i,j)$ will be the spin singlet state, despite the strong SOC.

\begin{figure}[]
\includegraphics[width=0.85\columnwidth]{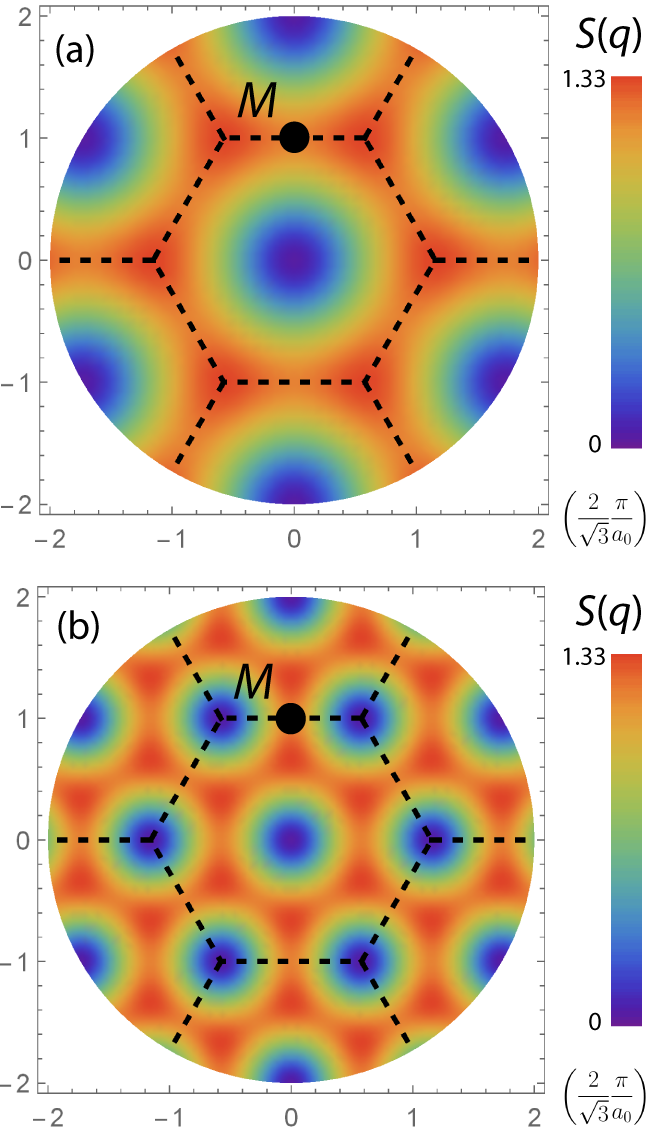}
\caption[]{ \textbf{Spin structure factor for frozen random singlets.} As in Fig. \ref{fig:corr} but here plotted for the cases when the singlet bonds are either purely nearest-neighbor [panel (a)] or purely second-neighbor [panel (b)]. Neither case reproduces the $M$ peak position of Fig. \ref{fig:corr}.} \label{fig:corrNN}
\end{figure}

\appendixsection{Spin-pair approximation}
\label{app:spinpairapprox}
Let $n(E)$ be the number of singlet pairs with energy less than or equal to $E$, i.e.\ the function plotted in Fig.~\ref{fig:plots}. The spin-pair approximation consists of treating each singlet pair as completely decoupled from its environment. The resulting formulas for susceptibility and specific heat are as follows,
\begin{align}\chi(T) = \frac{\mu_B^2}{k_B T} \int_0^\infty  \frac{dn}{dE}\frac{4}{3+e^{E/T}} d E 
\\
C(T) =  \int_0^\infty  \frac{dn}{dE}\frac{3e^{E/T} E^2}{2(3+e^{E/T})^2 T^2} d E \end{align}
When the coupling between two spins has $E<0$, corresponding to a ferromagnetic sign coupling and a triplet ground state, the integrands are given with energy $-E$.

\appendixsection{Microscopic considerations for YbMgGaO$_4$ }
\label{appendix:exchanges}
As mentioned above, the experimentally observed neutron scattering peaks along the BZ boundary (also seen in the corresponding theoretical computation shown in Fig.\ref{fig:corr}) require some small but nonzero amount of second-neighbor correlations. In particular, it is easy to see that pure first-neighbor correlations do not reproduce the correct peak locations even qualitatively (see Fig.\ref{fig:corrNN}). Second neighbor correlations are weak but present. 

They could be supported by a possible second neighbor exchange, which though would be unusual for $f$ moments, could potentially arise via Yb-O-O-Yb exchange pathways via direct overlap of in-plane-oriented p-orbitals on adjacent oxygen ions. 
Even at $J_2=0$ strong second neighbor correlations can arise. Let us consider the nearest neighbor interactions in detail.

The spin Hamiltonian for YbMgGaO$_4$ in the clean limit involves multiple exchange terms whose form is determined by the crystal symmetries. The question of the exact form of the Hamiltonian has not been fully settled.\cite{Chen2016,Chen2016a,Mourigal2016,Rueegg2017,Mourigal2017} We note (see also Ref.\onlinecite{Chernyshev2017}) that the so-called $J^{\pm \pm}$ and $J^{z \pm}$ terms may be re-written through a pseudo-dipole exchange $I$ and a triangular lattice Kitaev term $K$,
\begin{equation}
H_{i j}^{SOC} = I (\vec{S}_i \cdot \vec{r}_{i j}) (\vec{S}_j \cdot \vec{r}_{i j} )
 + K (\vec{S}_i \cdot \vec{\gamma}_{i j}) (\vec{S}_j \cdot \vec{\gamma}_{i j} )
\end{equation}
where $\vec{r}_{i j}$ is the unit vector connecting sites $i$ and $j$, and the Kitaev label $\vec{\gamma}_{i j} \propto \vec{r}_{\text{Yb-O}_1} \times \vec{r}_{\text{Yb-O}_2} $ is defined as the unit vector perpendicular to both Yb$_i$-O$_{1,2}$-Yb$_j$ exchange paths, i.e.\ perpendicular to the plane spanned by Yb sites $i,j$ and the two intervening oxygen ions. (On the triangular lattice, $\vec{\gamma}_{i j}$ is fully determined by the bond orientation  $\vec{r}_{i j}$, and $\{\vec{\gamma}_{i j}\}$ form an orthonormal coordinate system whose $(1,1,1)$ axis is normal to the lattice plane.)

The terms can be related to the original formulation\cite{Zhang2015} as follows:
\begin{align}
 \mathrm{Notation}:\ \  & \bigg(J_z, J_{\pm}, J_{\pm \pm}, J_{z \pm}\bigg)
 \\
 \mathrm{Heisenberg}\ J = & \bigg(1, \frac{1}{2}, 0, 0\bigg)
 \\
 \mathrm{Pseudodipole}\ I = & \bigg(0, \frac{1}{4}, \frac{1}{4}, 0 \bigg)
 \\
 \mathrm{Kitaev}\ K = & \bigg(\frac{1}{3}, \frac{1}{6}, -\frac{1}{6}, \frac{\sqrt{2}}{3}\bigg)
\end{align}
For example, on a bond for which the pseudodipole exchange is along $\hat{x}$, the Kitaev exchange is then along $\sqrt{2/3}\hat{y} +\sqrt{1/3}\hat{z}$.

These interactions are together strongly frustrating, and it is quite conceivable that once some preferred nearest neighbor singlet states are formed, other nearby sites prefer to order into configurations that involve second neighbor singlet states; or, that resonances among multiple sites are formed, which entail both first and second neighbor correlations in some ratio. Regardless of the microscopic origin of the observed weak second-neighbor correlations, the point in the main text regarding interpreting the experimental neutron scattering is that it can be captured by a simple model of short ranged randomly-oriented singlets with a particular small ratio of second to first neighbor correlations. 

Regarding promising regions in parameter space for stabilizing non-magnetic phases, we note the following fact about the Hamiltonian above: the parameter point $K=-2$, $J_z=J_{xy}=+1$ is known (via a transformation into an exactly solvable ferromagnet) to produce a stripy antiferromagnetic order which, at this parameter point, has zero quantum fluctuations\cite{Khaliullin2010,Vishwanath2014a}. Though the stripy phase should therefore extend over a large region of parameter space\cite{Chernyshev2017}, tuning away from the fluctuations-free point may be a useful route for destabilizing magnetic order. In this context it is also useful to observe that when upon adding Kitaev interactions to the Heisenberg antiferromagnet 120-degree order, the resulting classical ground state\cite{Daghofer2016} involves a finite but low density of nucleated $Z_2$ vortices (the topological defects of the 120 degree order), with complicated long-wavelength spin configurations reminiscent of skyrmion crystals. The true quantum fate of such configurations upon further perturbations, though they may be difficult to study using unbiased quantum algorithms,  may result in a quantum paramagnet phase.

\bibliography{YbRefs}

\begin{thebibliography}{145}%
\makeatletter
\providecommand \@ifxundefined [1]{%
 \@ifx{#1\undefined}
}%
\providecommand \@ifnum [1]{%
 \ifnum #1\expandafter \@firstoftwo
 \else \expandafter \@secondoftwo
 \fi
}%
\providecommand \@ifx [1]{%
 \ifx #1\expandafter \@firstoftwo
 \else \expandafter \@secondoftwo
 \fi
}%
\providecommand \natexlab [1]{#1}%
\providecommand \enquote  [1]{``#1''}%
\providecommand \bibnamefont  [1]{#1}%
\providecommand \bibfnamefont [1]{#1}%
\providecommand \citenamefont [1]{#1}%
\providecommand \href@noop [0]{\@secondoftwo}%
\providecommand \href [0]{\begingroup \@sanitize@url \@href}%
\providecommand \@href[1]{\@@startlink{#1}\@@href}%
\providecommand \@@href[1]{\endgroup#1\@@endlink}%
\providecommand \@sanitize@url [0]{\catcode `\\12\catcode `\$12\catcode
  `\&12\catcode `\#12\catcode `\^12\catcode `\_12\catcode `\%12\relax}%
\providecommand \@@startlink[1]{}%
\providecommand \@@endlink[0]{}%
\providecommand \url  [0]{\begingroup\@sanitize@url \@url }%
\providecommand \@url [1]{\endgroup\@href {#1}{\urlprefix }}%
\providecommand \urlprefix  [0]{URL }%
\providecommand \Eprint [0]{\href }%
\providecommand \doibase [0]{http://dx.doi.org/}%
\providecommand \selectlanguage [0]{\@gobble}%
\providecommand \bibinfo  [0]{\@secondoftwo}%
\providecommand \bibfield  [0]{\@secondoftwo}%
\providecommand \translation [1]{[#1]}%
\providecommand \BibitemOpen [0]{}%
\providecommand \bibitemStop [0]{}%
\providecommand \bibitemNoStop [0]{.\EOS\space}%
\providecommand \EOS [0]{\spacefactor3000\relax}%
\providecommand \BibitemShut  [1]{\csname bibitem#1\endcsname}%
\let\auto@bib@innerbib\@empty
\bibitem [{\citenamefont {Li}\ \emph {et~al.}(2015{\natexlab{a}})\citenamefont
  {Li}, \citenamefont {Liao}, \citenamefont {Zhang}, \citenamefont {Li},
  \citenamefont {Jin}, \citenamefont {Ling}, \citenamefont {Zhang},
  \citenamefont {Zou}, \citenamefont {Pi}, \citenamefont {Yang}, \citenamefont
  {Wang}, \citenamefont {Wu},\ and\ \citenamefont {Zhang}}]{srep}%
  \BibitemOpen
  \bibfield  {author} {\bibinfo {author} {\bibfnamefont {Yuesheng}\
  \bibnamefont {Li}}, \bibinfo {author} {\bibfnamefont {Haijun}\ \bibnamefont
  {Liao}}, \bibinfo {author} {\bibfnamefont {Zhen}\ \bibnamefont {Zhang}},
  \bibinfo {author} {\bibfnamefont {Shiyan}\ \bibnamefont {Li}}, \bibinfo
  {author} {\bibfnamefont {Feng}\ \bibnamefont {Jin}}, \bibinfo {author}
  {\bibfnamefont {Langsheng}\ \bibnamefont {Ling}}, \bibinfo {author}
  {\bibfnamefont {Lei}\ \bibnamefont {Zhang}}, \bibinfo {author} {\bibfnamefont
  {Youming}\ \bibnamefont {Zou}}, \bibinfo {author} {\bibfnamefont
  {Li}~\bibnamefont {Pi}}, \bibinfo {author} {\bibfnamefont {Zhaorong}\
  \bibnamefont {Yang}}, \bibinfo {author} {\bibfnamefont {Junfeng}\
  \bibnamefont {Wang}}, \bibinfo {author} {\bibfnamefont {Zhonghua}\
  \bibnamefont {Wu}}, \ and\ \bibinfo {author} {\bibfnamefont {Qingming}\
  \bibnamefont {Zhang}},\ }\bibfield  {title} {\enquote {\bibinfo {title}
  {Gapless quantum spin liquid ground state in the two-dimensional spin-1/2
  triangular antiferromagnet ybmggao4},}\ }\href
  {http://dx.doi.org/10.1038/srep16419} {\bibfield  {journal} {\bibinfo
  {journal} {Scientific Reports}\ }\textbf {\bibinfo {volume} {5}},\ \bibinfo
  {pages} {16419} (\bibinfo {year} {2015}{\natexlab{a}})}\BibitemShut {NoStop}%
\bibitem [{\citenamefont {Li}\ \emph {et~al.}(2015{\natexlab{b}})\citenamefont
  {Li}, \citenamefont {Chen}, \citenamefont {Tong}, \citenamefont {Pi},
  \citenamefont {Liu}, \citenamefont {Yang}, \citenamefont {Wang},\ and\
  \citenamefont {Zhang}}]{Zhang2015}%
  \BibitemOpen
  \bibfield  {author} {\bibinfo {author} {\bibfnamefont {Yuesheng}\
  \bibnamefont {Li}}, \bibinfo {author} {\bibfnamefont {Gang}\ \bibnamefont
  {Chen}}, \bibinfo {author} {\bibfnamefont {Wei}\ \bibnamefont {Tong}},
  \bibinfo {author} {\bibfnamefont {Li}~\bibnamefont {Pi}}, \bibinfo {author}
  {\bibfnamefont {Juanjuan}\ \bibnamefont {Liu}}, \bibinfo {author}
  {\bibfnamefont {Zhaorong}\ \bibnamefont {Yang}}, \bibinfo {author}
  {\bibfnamefont {Xiaoqun}\ \bibnamefont {Wang}}, \ and\ \bibinfo {author}
  {\bibfnamefont {Qingming}\ \bibnamefont {Zhang}},\ }\bibfield  {title}
  {\enquote {\bibinfo {title} {Rare-earth triangular lattice spin liquid: A
  single-crystal study of ${\mathrm{ybmggao}}_{4}$},}\ }\href {\doibase
  10.1103/PhysRevLett.115.167203} {\bibfield  {journal} {\bibinfo  {journal}
  {Phys. Rev. Lett.}\ }\textbf {\bibinfo {volume} {115}},\ \bibinfo {pages}
  {167203} (\bibinfo {year} {2015}{\natexlab{b}})}\BibitemShut {NoStop}%
\bibitem [{\citenamefont {Li}\ \emph {et~al.}(2016{\natexlab{a}})\citenamefont
  {Li}, \citenamefont {Adroja}, \citenamefont {Biswas}, \citenamefont {Baker},
  \citenamefont {Zhang}, \citenamefont {Liu}, \citenamefont {Tsirlin},
  \citenamefont {Gegenwart},\ and\ \citenamefont {Zhang}}]{muSR}%
  \BibitemOpen
  \bibfield  {author} {\bibinfo {author} {\bibfnamefont {Yuesheng}\
  \bibnamefont {Li}}, \bibinfo {author} {\bibfnamefont {Devashibhai}\
  \bibnamefont {Adroja}}, \bibinfo {author} {\bibfnamefont {Pabitra~K.}\
  \bibnamefont {Biswas}}, \bibinfo {author} {\bibfnamefont {Peter~J.}\
  \bibnamefont {Baker}}, \bibinfo {author} {\bibfnamefont {Qian}\ \bibnamefont
  {Zhang}}, \bibinfo {author} {\bibfnamefont {Juanjuan}\ \bibnamefont {Liu}},
  \bibinfo {author} {\bibfnamefont {Alexander~A.}\ \bibnamefont {Tsirlin}},
  \bibinfo {author} {\bibfnamefont {Philipp}\ \bibnamefont {Gegenwart}}, \ and\
  \bibinfo {author} {\bibfnamefont {Qingming}\ \bibnamefont {Zhang}},\
  }\bibfield  {title} {\enquote {\bibinfo {title} {Muon spin relaxation
  evidence for the u(1) quantum spin-liquid ground state in the triangular
  antiferromagnet ${\mathrm{ybmggao}}_{4}$},}\ }\href {\doibase
  10.1103/PhysRevLett.117.097201} {\bibfield  {journal} {\bibinfo  {journal}
  {Phys. Rev. Lett.}\ }\textbf {\bibinfo {volume} {117}},\ \bibinfo {pages}
  {097201} (\bibinfo {year} {2016}{\natexlab{a}})}\BibitemShut {NoStop}%
\bibitem [{\citenamefont {Xu}\ \emph {et~al.}(2016)\citenamefont {Xu},
  \citenamefont {Zhang}, \citenamefont {Li}, \citenamefont {Yu}, \citenamefont
  {Hong}, \citenamefont {Zhang},\ and\ \citenamefont {Li}}]{Li2016}%
  \BibitemOpen
  \bibfield  {author} {\bibinfo {author} {\bibfnamefont {Y.}~\bibnamefont
  {Xu}}, \bibinfo {author} {\bibfnamefont {J.}~\bibnamefont {Zhang}}, \bibinfo
  {author} {\bibfnamefont {Y.~S.}\ \bibnamefont {Li}}, \bibinfo {author}
  {\bibfnamefont {Y.~J.}\ \bibnamefont {Yu}}, \bibinfo {author} {\bibfnamefont
  {X.~C.}\ \bibnamefont {Hong}}, \bibinfo {author} {\bibfnamefont {Q.~M.}\
  \bibnamefont {Zhang}}, \ and\ \bibinfo {author} {\bibfnamefont {S.~Y.}\
  \bibnamefont {Li}},\ }\bibfield  {title} {\enquote {\bibinfo {title} {Absence
  of magnetic thermal conductivity in the quantum spin-liquid candidate
  ${\mathrm{ybmggao}}_{4}$},}\ }\href {\doibase 10.1103/PhysRevLett.117.267202}
  {\bibfield  {journal} {\bibinfo  {journal} {Phys. Rev. Lett.}\ }\textbf
  {\bibinfo {volume} {117}},\ \bibinfo {pages} {267202} (\bibinfo {year}
  {2016})}\BibitemShut {NoStop}%
\bibitem [{\citenamefont {Paddison}\ \emph {et~al.}(2017)\citenamefont
  {Paddison}, \citenamefont {Daum}, \citenamefont {Dun}, \citenamefont
  {Ehlers}, \citenamefont {Liu}, \citenamefont {Stone}, \citenamefont {Zhou},\
  and\ \citenamefont {Mourigal}}]{Mourigal2016}%
  \BibitemOpen
  \bibfield  {author} {\bibinfo {author} {\bibfnamefont {Joseph A.~M.}\
  \bibnamefont {Paddison}}, \bibinfo {author} {\bibfnamefont {Marcus}\
  \bibnamefont {Daum}}, \bibinfo {author} {\bibfnamefont {Zhiling}\
  \bibnamefont {Dun}}, \bibinfo {author} {\bibfnamefont {Georg}\ \bibnamefont
  {Ehlers}}, \bibinfo {author} {\bibfnamefont {Yaohua}\ \bibnamefont {Liu}},
  \bibinfo {author} {\bibfnamefont {Matthew~B.}\ \bibnamefont {Stone}},
  \bibinfo {author} {\bibfnamefont {Haidong}\ \bibnamefont {Zhou}}, \ and\
  \bibinfo {author} {\bibfnamefont {Martin}\ \bibnamefont {Mourigal}},\
  }\bibfield  {title} {\enquote {\bibinfo {title} {Continuous excitations of
  the triangular-lattice quantum spin liquid ybmggao4},}\ }\href
  {http://dx.doi.org/10.1038/nphys3971} {\bibfield  {journal} {\bibinfo
  {journal} {Nat Phys}\ }\textbf {\bibinfo {volume} {13}},\ \bibinfo {pages}
  {117--122} (\bibinfo {year} {2017})}\BibitemShut {NoStop}%
\bibitem [{\citenamefont {Shen}\ \emph {et~al.}(2016)\citenamefont {Shen},
  \citenamefont {Li}, \citenamefont {Wo}, \citenamefont {Li}, \citenamefont
  {Shen}, \citenamefont {Pan}, \citenamefont {Wang}, \citenamefont {Walker},
  \citenamefont {Steffens}, \citenamefont {Boehm}, \citenamefont {Hao},
  \citenamefont {Quintero-Castro}, \citenamefont {Harriger}, \citenamefont
  {Frontzek}, \citenamefont {Hao}, \citenamefont {Meng}, \citenamefont {Zhang},
  \citenamefont {Chen},\ and\ \citenamefont {Zhao}}]{Zhao2016}%
  \BibitemOpen
  \bibfield  {author} {\bibinfo {author} {\bibfnamefont {Yao}\ \bibnamefont
  {Shen}}, \bibinfo {author} {\bibfnamefont {Yao-Dong}\ \bibnamefont {Li}},
  \bibinfo {author} {\bibfnamefont {Hongliang}\ \bibnamefont {Wo}}, \bibinfo
  {author} {\bibfnamefont {Yuesheng}\ \bibnamefont {Li}}, \bibinfo {author}
  {\bibfnamefont {Shoudong}\ \bibnamefont {Shen}}, \bibinfo {author}
  {\bibfnamefont {Bingying}\ \bibnamefont {Pan}}, \bibinfo {author}
  {\bibfnamefont {Qisi}\ \bibnamefont {Wang}}, \bibinfo {author} {\bibfnamefont
  {H.~C.}\ \bibnamefont {Walker}}, \bibinfo {author} {\bibfnamefont
  {P.}~\bibnamefont {Steffens}}, \bibinfo {author} {\bibfnamefont
  {M.}~\bibnamefont {Boehm}}, \bibinfo {author} {\bibfnamefont {Yiqing}\
  \bibnamefont {Hao}}, \bibinfo {author} {\bibfnamefont {D.~L.}\ \bibnamefont
  {Quintero-Castro}}, \bibinfo {author} {\bibfnamefont {L.~W.}\ \bibnamefont
  {Harriger}}, \bibinfo {author} {\bibfnamefont {M.~D.}\ \bibnamefont
  {Frontzek}}, \bibinfo {author} {\bibfnamefont {Lijie}\ \bibnamefont {Hao}},
  \bibinfo {author} {\bibfnamefont {Siqin}\ \bibnamefont {Meng}}, \bibinfo
  {author} {\bibfnamefont {Qingming}\ \bibnamefont {Zhang}}, \bibinfo {author}
  {\bibfnamefont {Gang}\ \bibnamefont {Chen}}, \ and\ \bibinfo {author}
  {\bibfnamefont {Jun}\ \bibnamefont {Zhao}},\ }\bibfield  {title} {\enquote
  {\bibinfo {title} {Evidence for a spinon fermi surface in a
  triangular-lattice quantum-spin-liquid candidate},}\ }\href
  {http://dx.doi.org/10.1038/nature20614} {\bibfield  {journal} {\bibinfo
  {journal} {Nature}\ }\textbf {\bibinfo {volume} {540}},\ \bibinfo {pages}
  {559--562} (\bibinfo {year} {2016})}\BibitemShut {NoStop}%
\bibitem [{\citenamefont {Li}\ \emph {et~al.}(2017{\natexlab{a}})\citenamefont
  {Li}, \citenamefont {Adroja}, \citenamefont {Voneshen}, \citenamefont
  {Bewley}, \citenamefont {Zhang}, \citenamefont {Tsirlin},\ and\ \citenamefont
  {Gegenwart}}]{Gegenwart2017}%
  \BibitemOpen
  \bibfield  {author} {\bibinfo {author} {\bibfnamefont {Yuesheng}\
  \bibnamefont {Li}}, \bibinfo {author} {\bibfnamefont {Devashibhai}\
  \bibnamefont {Adroja}}, \bibinfo {author} {\bibfnamefont {David}\
  \bibnamefont {Voneshen}}, \bibinfo {author} {\bibfnamefont {Robert~I.}\
  \bibnamefont {Bewley}}, \bibinfo {author} {\bibfnamefont {Qingming}\
  \bibnamefont {Zhang}}, \bibinfo {author} {\bibfnamefont {Alexander~A.}\
  \bibnamefont {Tsirlin}}, \ and\ \bibinfo {author} {\bibfnamefont {Philipp}\
  \bibnamefont {Gegenwart}},\ }\bibfield  {title} {\enquote {\bibinfo {title}
  {Nearest-neighbour resonating valence bonds in ybmggao4},}\ }\href
  {http://dx.doi.org/10.1038/ncomms15814} {\bibfield  {journal} {\bibinfo
  {journal} {Nat. Comm}\ }\textbf {\bibinfo {volume} {8}},\ \bibinfo {pages}
  {15814} (\bibinfo {year} {2017}{\natexlab{a}})}\BibitemShut {NoStop}%
\bibitem [{\citenamefont {{T{\'o}th}}\ \emph {et~al.}(2017)\citenamefont
  {{T{\'o}th}}, \citenamefont {{Rolfs}}, \citenamefont {{Wildes}},\ and\
  \citenamefont {{R{\"u}egg}}}]{Rueegg2017}%
  \BibitemOpen
  \bibfield  {author} {\bibinfo {author} {\bibfnamefont {S.}~\bibnamefont
  {{T{\'o}th}}}, \bibinfo {author} {\bibfnamefont {K.}~\bibnamefont {{Rolfs}}},
  \bibinfo {author} {\bibfnamefont {A.~R.}\ \bibnamefont {{Wildes}}}, \ and\
  \bibinfo {author} {\bibfnamefont {C.}~\bibnamefont {{R{\"u}egg}}},\
  }\bibfield  {title} {\enquote {\bibinfo {title} {{Strong exchange anisotropy
  in YbMgGaO$\_4$ from polarized neutron diffraction}},}\ }\href@noop {}
  {\bibfield  {journal} {\bibinfo  {journal} {ArXiv e-prints}\ } (\bibinfo
  {year} {2017})},\ \Eprint {http://arxiv.org/abs/1705.05699} {arXiv:1705.05699
  [cond-mat.str-el]} \BibitemShut {NoStop}%
\bibitem [{\citenamefont {Li}\ \emph {et~al.}(2017{\natexlab{b}})\citenamefont
  {Li}, \citenamefont {Adroja}, \citenamefont {Bewley}, \citenamefont
  {Voneshen}, \citenamefont {Tsirlin}, \citenamefont {Gegenwart},\ and\
  \citenamefont {Zhang}}]{Zhang2017}%
  \BibitemOpen
  \bibfield  {author} {\bibinfo {author} {\bibfnamefont {Yuesheng}\
  \bibnamefont {Li}}, \bibinfo {author} {\bibfnamefont {Devashibhai}\
  \bibnamefont {Adroja}}, \bibinfo {author} {\bibfnamefont {Robert~I.}\
  \bibnamefont {Bewley}}, \bibinfo {author} {\bibfnamefont {David}\
  \bibnamefont {Voneshen}}, \bibinfo {author} {\bibfnamefont {Alexander~A.}\
  \bibnamefont {Tsirlin}}, \bibinfo {author} {\bibfnamefont {Philipp}\
  \bibnamefont {Gegenwart}}, \ and\ \bibinfo {author} {\bibfnamefont
  {Qingming}\ \bibnamefont {Zhang}},\ }\bibfield  {title} {\enquote {\bibinfo
  {title} {Crystalline electric-field randomness in the triangular lattice
  spin-liquid ${\mathrm{ybmggao}}_{4}$},}\ }\href {\doibase
  10.1103/PhysRevLett.118.107202} {\bibfield  {journal} {\bibinfo  {journal}
  {Phys. Rev. Lett.}\ }\textbf {\bibinfo {volume} {118}},\ \bibinfo {pages}
  {107202} (\bibinfo {year} {2017}{\natexlab{b}})}\BibitemShut {NoStop}%
\bibitem [{\citenamefont {{Shen}}\ \emph {et~al.}(2017)\citenamefont {{Shen}},
  \citenamefont {{Li}}, \citenamefont {{Walker}}, \citenamefont {{Steffens}},
  \citenamefont {{Boehm}}, \citenamefont {{Zhang}}, \citenamefont {{Shen}},
  \citenamefont {{Wo}}, \citenamefont {{Chen}},\ and\ \citenamefont
  {{Zhao}}}]{Zhao2017}%
  \BibitemOpen
  \bibfield  {author} {\bibinfo {author} {\bibfnamefont {Y.}~\bibnamefont
  {{Shen}}}, \bibinfo {author} {\bibfnamefont {Y.-D.}\ \bibnamefont {{Li}}},
  \bibinfo {author} {\bibfnamefont {H.~C.}\ \bibnamefont {{Walker}}}, \bibinfo
  {author} {\bibfnamefont {P.}~\bibnamefont {{Steffens}}}, \bibinfo {author}
  {\bibfnamefont {M.}~\bibnamefont {{Boehm}}}, \bibinfo {author} {\bibfnamefont
  {X.}~\bibnamefont {{Zhang}}}, \bibinfo {author} {\bibfnamefont
  {S.}~\bibnamefont {{Shen}}}, \bibinfo {author} {\bibfnamefont
  {H.}~\bibnamefont {{Wo}}}, \bibinfo {author} {\bibfnamefont {G.}~\bibnamefont
  {{Chen}}}, \ and\ \bibinfo {author} {\bibfnamefont {J.}~\bibnamefont
  {{Zhao}}},\ }\bibfield  {title} {\enquote {\bibinfo {title} {{Fractionalized
  excitations in the partially magnetized spin liquid candidate YbMgGaO4}},}\
  }\href@noop {} {\bibfield  {journal} {\bibinfo  {journal} {ArXiv e-prints}\ }
  (\bibinfo {year} {2017})},\ \Eprint {http://arxiv.org/abs/1708.06655}
  {arXiv:1708.06655 [cond-mat.str-el]} \BibitemShut {NoStop}%
\bibitem [{\citenamefont {{Zhang}}\ \emph {et~al.}(2017)\citenamefont
  {{Zhang}}, \citenamefont {{Mahmood}}, \citenamefont {{Daum}}, \citenamefont
  {{Dun}}, \citenamefont {{Paddison}}, \citenamefont {{Laurita}}, \citenamefont
  {{Hong}}, \citenamefont {{Zhou}}, \citenamefont {{Armitage}},\ and\
  \citenamefont {{Mourigal}}}]{Mourigal2017}%
  \BibitemOpen
  \bibfield  {author} {\bibinfo {author} {\bibfnamefont {X.}~\bibnamefont
  {{Zhang}}}, \bibinfo {author} {\bibfnamefont {F.}~\bibnamefont {{Mahmood}}},
  \bibinfo {author} {\bibfnamefont {M.}~\bibnamefont {{Daum}}}, \bibinfo
  {author} {\bibfnamefont {Z.}~\bibnamefont {{Dun}}}, \bibinfo {author}
  {\bibfnamefont {J.~A.~M.}\ \bibnamefont {{Paddison}}}, \bibinfo {author}
  {\bibfnamefont {N.~J.}\ \bibnamefont {{Laurita}}}, \bibinfo {author}
  {\bibfnamefont {T.}~\bibnamefont {{Hong}}}, \bibinfo {author} {\bibfnamefont
  {H.}~\bibnamefont {{Zhou}}}, \bibinfo {author} {\bibfnamefont {N.~P.}\
  \bibnamefont {{Armitage}}}, \ and\ \bibinfo {author} {\bibfnamefont
  {M.}~\bibnamefont {{Mourigal}}},\ }\bibfield  {title} {\enquote {\bibinfo
  {title} {{Hierarchy of exchange interactions in the triangular-lattice
  spin-liquid YbMgGaO$\_{4}$}},}\ }\href@noop {} {\bibfield  {journal}
  {\bibinfo  {journal} {ArXiv e-prints}\ } (\bibinfo {year} {2017})},\ \Eprint
  {http://arxiv.org/abs/1708.07503} {arXiv:1708.07503 [cond-mat.str-el]}
  \BibitemShut {NoStop}%
\bibitem [{\citenamefont {Ma}\ \emph {et~al.}(2018)\citenamefont {Ma},
  \citenamefont {Wang}, \citenamefont {Dong}, \citenamefont {Zhang},
  \citenamefont {Li}, \citenamefont {Zheng}, \citenamefont {Yu}, \citenamefont
  {Wang}, \citenamefont {Che}, \citenamefont {Ran}, \citenamefont {Bao},
  \citenamefont {Cai}, \citenamefont {Cermak}, \citenamefont {Schneidewind},
  \citenamefont {Yano}, \citenamefont {Gardner}, \citenamefont {Lu},
  \citenamefont {Yu}, \citenamefont {Liu}, \citenamefont {Li}, \citenamefont
  {Li},\ and\ \citenamefont {Wen}}]{Wen2017}%
  \BibitemOpen
  \bibfield  {author} {\bibinfo {author} {\bibfnamefont {Zhen}\ \bibnamefont
  {Ma}}, \bibinfo {author} {\bibfnamefont {Jinghui}\ \bibnamefont {Wang}},
  \bibinfo {author} {\bibfnamefont {Zhao-Yang}\ \bibnamefont {Dong}}, \bibinfo
  {author} {\bibfnamefont {Jun}\ \bibnamefont {Zhang}}, \bibinfo {author}
  {\bibfnamefont {Shichao}\ \bibnamefont {Li}}, \bibinfo {author}
  {\bibfnamefont {Shu-Han}\ \bibnamefont {Zheng}}, \bibinfo {author}
  {\bibfnamefont {Yunjie}\ \bibnamefont {Yu}}, \bibinfo {author} {\bibfnamefont
  {Wei}\ \bibnamefont {Wang}}, \bibinfo {author} {\bibfnamefont {Liqiang}\
  \bibnamefont {Che}}, \bibinfo {author} {\bibfnamefont {Kejing}\ \bibnamefont
  {Ran}}, \bibinfo {author} {\bibfnamefont {Song}\ \bibnamefont {Bao}},
  \bibinfo {author} {\bibfnamefont {Zhengwei}\ \bibnamefont {Cai}}, \bibinfo
  {author} {\bibfnamefont {P.}~\bibnamefont {Cermak}}, \bibinfo {author}
  {\bibfnamefont {A.}~\bibnamefont {Schneidewind}}, \bibinfo {author}
  {\bibfnamefont {S.}~\bibnamefont {Yano}}, \bibinfo {author} {\bibfnamefont
  {J.~S.}\ \bibnamefont {Gardner}}, \bibinfo {author} {\bibfnamefont {Xin}\
  \bibnamefont {Lu}}, \bibinfo {author} {\bibfnamefont {Shun-Li}\ \bibnamefont
  {Yu}}, \bibinfo {author} {\bibfnamefont {Jun-Ming}\ \bibnamefont {Liu}},
  \bibinfo {author} {\bibfnamefont {Shiyan}\ \bibnamefont {Li}}, \bibinfo
  {author} {\bibfnamefont {Jian-Xin}\ \bibnamefont {Li}}, \ and\ \bibinfo
  {author} {\bibfnamefont {Jinsheng}\ \bibnamefont {Wen}},\ }\bibfield  {title}
  {\enquote {\bibinfo {title} {Spin-glass ground state in a triangular-lattice
  compound ${\mathrm{ybzngao}}_{4}$},}\ }\href {\doibase
  10.1103/PhysRevLett.120.087201} {\bibfield  {journal} {\bibinfo  {journal}
  {Phys. Rev. Lett.}\ }\textbf {\bibinfo {volume} {120}},\ \bibinfo {pages}
  {087201} (\bibinfo {year} {2018})}\BibitemShut {NoStop}%
\bibitem [{\citenamefont {Li}\ \emph {et~al.}(2016{\natexlab{b}})\citenamefont
  {Li}, \citenamefont {Wang},\ and\ \citenamefont {Chen}}]{Chen2016}%
  \BibitemOpen
  \bibfield  {author} {\bibinfo {author} {\bibfnamefont {Yao-Dong}\
  \bibnamefont {Li}}, \bibinfo {author} {\bibfnamefont {Xiaoqun}\ \bibnamefont
  {Wang}}, \ and\ \bibinfo {author} {\bibfnamefont {Gang}\ \bibnamefont
  {Chen}},\ }\bibfield  {title} {\enquote {\bibinfo {title} {Anisotropic spin
  model of strong spin-orbit-coupled triangular antiferromagnets},}\ }\href
  {\doibase 10.1103/PhysRevB.94.035107} {\bibfield  {journal} {\bibinfo
  {journal} {Phys. Rev. B}\ }\textbf {\bibinfo {volume} {94}},\ \bibinfo
  {pages} {035107} (\bibinfo {year} {2016}{\natexlab{b}})}\BibitemShut
  {NoStop}%
\bibitem [{\citenamefont {Liu}\ \emph {et~al.}(2016)\citenamefont {Liu},
  \citenamefont {Yu},\ and\ \citenamefont {Wang}}]{Wang2016}%
  \BibitemOpen
  \bibfield  {author} {\bibinfo {author} {\bibfnamefont {Changle}\ \bibnamefont
  {Liu}}, \bibinfo {author} {\bibfnamefont {Rong}\ \bibnamefont {Yu}}, \ and\
  \bibinfo {author} {\bibfnamefont {Xiaoqun}\ \bibnamefont {Wang}},\ }\bibfield
   {title} {\enquote {\bibinfo {title} {Semiclassical ground-state phase
  diagram and $\text{multi-}q$ phase of a spin-orbit-coupled model on
  triangular lattice},}\ }\href {\doibase 10.1103/PhysRevB.94.174424}
  {\bibfield  {journal} {\bibinfo  {journal} {Phys. Rev. B}\ }\textbf {\bibinfo
  {volume} {94}},\ \bibinfo {pages} {174424} (\bibinfo {year}
  {2016})}\BibitemShut {NoStop}%
\bibitem [{\citenamefont {Li}\ \emph {et~al.}(2017{\natexlab{c}})\citenamefont
  {Li}, \citenamefont {Lu},\ and\ \citenamefont {Chen}}]{Chen2016b}%
  \BibitemOpen
  \bibfield  {author} {\bibinfo {author} {\bibfnamefont {Yao-Dong}\
  \bibnamefont {Li}}, \bibinfo {author} {\bibfnamefont {Yuan-Ming}\
  \bibnamefont {Lu}}, \ and\ \bibinfo {author} {\bibfnamefont {Gang}\
  \bibnamefont {Chen}},\ }\bibfield  {title} {\enquote {\bibinfo {title}
  {Spinon fermi surface $u(1)$ spin liquid in the spin-orbit-coupled
  triangular-lattice mott insulator ${\mathrm{ybmggao}}_{4}$},}\ }\href
  {\doibase 10.1103/PhysRevB.96.054445} {\bibfield  {journal} {\bibinfo
  {journal} {Phys. Rev. B}\ }\textbf {\bibinfo {volume} {96}},\ \bibinfo
  {pages} {054445} (\bibinfo {year} {2017}{\natexlab{c}})}\BibitemShut
  {NoStop}%
\bibitem [{\citenamefont {Li}\ and\ \citenamefont {Chen}(2017)}]{Chen2017}%
  \BibitemOpen
  \bibfield  {author} {\bibinfo {author} {\bibfnamefont {Yao-Dong}\
  \bibnamefont {Li}}\ and\ \bibinfo {author} {\bibfnamefont {Gang}\
  \bibnamefont {Chen}},\ }\bibfield  {title} {\enquote {\bibinfo {title}
  {Detecting spin fractionalization in a spinon fermi surface spin liquid},}\
  }\href {\doibase 10.1103/PhysRevB.96.075105} {\bibfield  {journal} {\bibinfo
  {journal} {Phys. Rev. B}\ }\textbf {\bibinfo {volume} {96}},\ \bibinfo
  {pages} {075105} (\bibinfo {year} {2017})}\BibitemShut {NoStop}%
\bibitem [{\citenamefont {Takemori}\ and\ \citenamefont
  {Kamimura}(1982)}]{Kamimura1982}%
  \BibitemOpen
  \bibfield  {author} {\bibinfo {author} {\bibfnamefont {T.}~\bibnamefont
  {Takemori}}\ and\ \bibinfo {author} {\bibfnamefont {H.}~\bibnamefont
  {Kamimura}},\ }\bibfield  {title} {\enquote {\bibinfo {title} {Effect of
  interstate interactions on specific heat and magnetic susceptibility in
  anderson-localized system},}\ }\href {\doibase
  http://dx.doi.org/10.1016/0038-1098(82)91229-7} {\bibfield  {journal}
  {\bibinfo  {journal} {Solid State Communications}\ }\textbf {\bibinfo
  {volume} {41}},\ \bibinfo {pages} {885 -- 888} (\bibinfo {year}
  {1982})}\BibitemShut {NoStop}%
\bibitem [{\citenamefont {Lee}\ and\ \citenamefont
  {Ramakrishnan}(1985)}]{Ramakrishnan1985}%
  \BibitemOpen
  \bibfield  {author} {\bibinfo {author} {\bibfnamefont {Patrick~A.}\
  \bibnamefont {Lee}}\ and\ \bibinfo {author} {\bibfnamefont {T.~V.}\
  \bibnamefont {Ramakrishnan}},\ }\bibfield  {title} {\enquote {\bibinfo
  {title} {Disordered electronic systems},}\ }\href {\doibase
  10.1103/RevModPhys.57.287} {\bibfield  {journal} {\bibinfo  {journal} {Rev.
  Mod. Phys.}\ }\textbf {\bibinfo {volume} {57}},\ \bibinfo {pages} {287--337}
  (\bibinfo {year} {1985})}\BibitemShut {NoStop}%
\bibitem [{\citenamefont {Bhatt}\ and\ \citenamefont {Lee}(1981)}]{Lee1981}%
  \BibitemOpen
  \bibfield  {author} {\bibinfo {author} {\bibfnamefont {R.~N.}\ \bibnamefont
  {Bhatt}}\ and\ \bibinfo {author} {\bibfnamefont {P.~A.}\ \bibnamefont
  {Lee}},\ }\bibfield  {title} {\enquote {\bibinfo {title} {A scaling method
  for low temperature behavior of random antiferromagnetic systems
  (invited)},}\ }\href {\doibase 10.1063/1.329684} {\bibfield  {journal}
  {\bibinfo  {journal} {Journal of Applied Physics}\ }\textbf {\bibinfo
  {volume} {52}},\ \bibinfo {pages} {1703--1707} (\bibinfo {year} {1981})},\
  \Eprint
  {http://arxiv.org/abs/http://aip.scitation.org/doi/pdf/10.1063/1.329684}
  {http://aip.scitation.org/doi/pdf/10.1063/1.329684} \BibitemShut {NoStop}%
\bibitem [{\citenamefont {Bhatt}\ and\ \citenamefont {Lee}(1982)}]{Lee1982}%
  \BibitemOpen
  \bibfield  {author} {\bibinfo {author} {\bibfnamefont {R.~N.}\ \bibnamefont
  {Bhatt}}\ and\ \bibinfo {author} {\bibfnamefont {P.~A.}\ \bibnamefont
  {Lee}},\ }\bibfield  {title} {\enquote {\bibinfo {title} {Scaling studies of
  highly disordered spin-1/2 antiferromagnetic systems},}\ }\href {\doibase
  10.1103/PhysRevLett.48.344} {\bibfield  {journal} {\bibinfo  {journal} {Phys.
  Rev. Lett.}\ }\textbf {\bibinfo {volume} {48}},\ \bibinfo {pages} {344--347}
  (\bibinfo {year} {1982})}\BibitemShut {NoStop}%
\bibitem [{\citenamefont {Paalanen}\ \emph {et~al.}(1988)\citenamefont
  {Paalanen}, \citenamefont {Graebner}, \citenamefont {Bhatt},\ and\
  \citenamefont {Sachdev}}]{Sachdev1988}%
  \BibitemOpen
  \bibfield  {author} {\bibinfo {author} {\bibfnamefont {M.~A.}\ \bibnamefont
  {Paalanen}}, \bibinfo {author} {\bibfnamefont {J.~E.}\ \bibnamefont
  {Graebner}}, \bibinfo {author} {\bibfnamefont {R.~N.}\ \bibnamefont {Bhatt}},
  \ and\ \bibinfo {author} {\bibfnamefont {S.}~\bibnamefont {Sachdev}},\
  }\bibfield  {title} {\enquote {\bibinfo {title} {Thermodynamic behavior near
  a metal-insulator transition},}\ }\href {\doibase 10.1103/PhysRevLett.61.597}
  {\bibfield  {journal} {\bibinfo  {journal} {Phys. Rev. Lett.}\ }\textbf
  {\bibinfo {volume} {61}},\ \bibinfo {pages} {597--600} (\bibinfo {year}
  {1988})}\BibitemShut {NoStop}%
\bibitem [{\citenamefont {Fisher}(1994)}]{Fisher1994}%
  \BibitemOpen
  \bibfield  {author} {\bibinfo {author} {\bibfnamefont {Daniel~S.}\
  \bibnamefont {Fisher}},\ }\bibfield  {title} {\enquote {\bibinfo {title}
  {Random antiferromagnetic quantum spin chains},}\ }\href {\doibase
  10.1103/PhysRevB.50.3799} {\bibfield  {journal} {\bibinfo  {journal} {Phys.
  Rev. B}\ }\textbf {\bibinfo {volume} {50}},\ \bibinfo {pages} {3799--3821}
  (\bibinfo {year} {1994})}\BibitemShut {NoStop}%
\bibitem [{\citenamefont {Lin}\ \emph {et~al.}(2003)\citenamefont {Lin},
  \citenamefont {M\'elin}, \citenamefont {Rieger},\ and\ \citenamefont
  {Igl\'oi}}]{Igloi2003}%
  \BibitemOpen
  \bibfield  {author} {\bibinfo {author} {\bibfnamefont {Y.-C.}\ \bibnamefont
  {Lin}}, \bibinfo {author} {\bibfnamefont {R.}~\bibnamefont {M\'elin}},
  \bibinfo {author} {\bibfnamefont {H.}~\bibnamefont {Rieger}}, \ and\ \bibinfo
  {author} {\bibfnamefont {F.}~\bibnamefont {Igl\'oi}},\ }\bibfield  {title}
  {\enquote {\bibinfo {title} {Low-energy fixed points of random heisenberg
  models},}\ }\href {\doibase 10.1103/PhysRevB.68.024424} {\bibfield  {journal}
  {\bibinfo  {journal} {Phys. Rev. B}\ }\textbf {\bibinfo {volume} {68}},\
  \bibinfo {pages} {024424} (\bibinfo {year} {2003})}\BibitemShut {NoStop}%
\bibitem [{\citenamefont {Motrunich}\ \emph {et~al.}(2000)\citenamefont
  {Motrunich}, \citenamefont {Mau}, \citenamefont {Huse},\ and\ \citenamefont
  {Fisher}}]{Fisher2000}%
  \BibitemOpen
  \bibfield  {author} {\bibinfo {author} {\bibfnamefont {Olexei}\ \bibnamefont
  {Motrunich}}, \bibinfo {author} {\bibfnamefont {Siun-Chuon}\ \bibnamefont
  {Mau}}, \bibinfo {author} {\bibfnamefont {David~A.}\ \bibnamefont {Huse}}, \
  and\ \bibinfo {author} {\bibfnamefont {Daniel~S.}\ \bibnamefont {Fisher}},\
  }\bibfield  {title} {\enquote {\bibinfo {title} {Infinite-randomness quantum
  ising critical fixed points},}\ }\href {\doibase 10.1103/PhysRevB.61.1160}
  {\bibfield  {journal} {\bibinfo  {journal} {Phys. Rev. B}\ }\textbf {\bibinfo
  {volume} {61}},\ \bibinfo {pages} {1160--1172} (\bibinfo {year}
  {2000})}\BibitemShut {NoStop}%
\bibitem [{\citenamefont {Laumann}\ \emph {et~al.}(2012)\citenamefont
  {Laumann}, \citenamefont {Huse}, \citenamefont {Ludwig}, \citenamefont
  {Refael}, \citenamefont {Trebst},\ and\ \citenamefont {Troyer}}]{Troyer2012}%
  \BibitemOpen
  \bibfield  {author} {\bibinfo {author} {\bibfnamefont {C.~R.}\ \bibnamefont
  {Laumann}}, \bibinfo {author} {\bibfnamefont {D.~A.}\ \bibnamefont {Huse}},
  \bibinfo {author} {\bibfnamefont {A.~W.~W.}\ \bibnamefont {Ludwig}}, \bibinfo
  {author} {\bibfnamefont {G.}~\bibnamefont {Refael}}, \bibinfo {author}
  {\bibfnamefont {S.}~\bibnamefont {Trebst}}, \ and\ \bibinfo {author}
  {\bibfnamefont {M.}~\bibnamefont {Troyer}},\ }\bibfield  {title} {\enquote
  {\bibinfo {title} {Strong-disorder renormalization for interacting
  non-abelian anyon systems in two dimensions},}\ }\href {\doibase
  10.1103/PhysRevB.85.224201} {\bibfield  {journal} {\bibinfo  {journal} {Phys.
  Rev. B}\ }\textbf {\bibinfo {volume} {85}},\ \bibinfo {pages} {224201}
  (\bibinfo {year} {2012})}\BibitemShut {NoStop}%
\bibitem [{\citenamefont {Sandvik}\ and\ \citenamefont
  {Veki{\'c}}(1995)}]{sandvik1995disorder}%
  \BibitemOpen
  \bibfield  {author} {\bibinfo {author} {\bibfnamefont {Anders~W}\
  \bibnamefont {Sandvik}}\ and\ \bibinfo {author} {\bibfnamefont {Marco}\
  \bibnamefont {Veki{\'c}}},\ }\bibfield  {title} {\enquote {\bibinfo {title}
  {Disorder induced phase transition in a two-dimensional random quantum
  antiferromagnet},}\ }\href@noop {} {\bibfield  {journal} {\bibinfo  {journal}
  {Physical review letters}\ }\textbf {\bibinfo {volume} {74}},\ \bibinfo
  {pages} {1226} (\bibinfo {year} {1995})}\BibitemShut {NoStop}%
\bibitem [{\citenamefont {Watanabe}\ \emph {et~al.}(2014)\citenamefont
  {Watanabe}, \citenamefont {Kawamura}, \citenamefont {Nakano},\ and\
  \citenamefont {Sakai}}]{Sakai2014}%
  \BibitemOpen
  \bibfield  {author} {\bibinfo {author} {\bibfnamefont {Ken}\ \bibnamefont
  {Watanabe}}, \bibinfo {author} {\bibfnamefont {Hikaru}\ \bibnamefont
  {Kawamura}}, \bibinfo {author} {\bibfnamefont {Hiroki}\ \bibnamefont
  {Nakano}}, \ and\ \bibinfo {author} {\bibfnamefont {Tôru}\ \bibnamefont
  {Sakai}},\ }\bibfield  {title} {\enquote {\bibinfo {title} {Quantum
  spin-liquid behavior in the spin-1/2 random heisenberg antiferromagnet on the
  triangular lattice},}\ }\href {\doibase 10.7566/JPSJ.83.034714} {\bibfield
  {journal} {\bibinfo  {journal} {Journal of the Physical Society of Japan}\
  }\textbf {\bibinfo {volume} {83}},\ \bibinfo {pages} {034714} (\bibinfo
  {year} {2014})},\ \Eprint
  {http://arxiv.org/abs/http://dx.doi.org/10.7566/JPSJ.83.034714}
  {http://dx.doi.org/10.7566/JPSJ.83.034714} \BibitemShut {NoStop}%
\bibitem [{\citenamefont {Kawamura}\ \emph {et~al.}(2014)\citenamefont
  {Kawamura}, \citenamefont {Watanabe},\ and\ \citenamefont
  {Shimokawa}}]{Shimokawa2014}%
  \BibitemOpen
  \bibfield  {author} {\bibinfo {author} {\bibfnamefont {Hikaru}\ \bibnamefont
  {Kawamura}}, \bibinfo {author} {\bibfnamefont {Ken}\ \bibnamefont
  {Watanabe}}, \ and\ \bibinfo {author} {\bibfnamefont {Tokuro}\ \bibnamefont
  {Shimokawa}},\ }\bibfield  {title} {\enquote {\bibinfo {title} {Quantum
  spin-liquid behavior in the spin-1/2 random-bond heisenberg antiferromagnet
  on the kagome lattice},}\ }\href {\doibase 10.7566/JPSJ.83.103704} {\bibfield
   {journal} {\bibinfo  {journal} {Journal of the Physical Society of Japan}\
  }\textbf {\bibinfo {volume} {83}},\ \bibinfo {pages} {103704} (\bibinfo
  {year} {2014})},\ \Eprint
  {http://arxiv.org/abs/http://dx.doi.org/10.7566/JPSJ.83.103704}
  {http://dx.doi.org/10.7566/JPSJ.83.103704} \BibitemShut {NoStop}%
\bibitem [{\citenamefont {Uematsu}\ and\ \citenamefont
  {Kawamura}(2017)}]{Kawamura2017}%
  \BibitemOpen
  \bibfield  {author} {\bibinfo {author} {\bibfnamefont {Kazuki}\ \bibnamefont
  {Uematsu}}\ and\ \bibinfo {author} {\bibfnamefont {Hikaru}\ \bibnamefont
  {Kawamura}},\ }\bibfield  {title} {\enquote {\bibinfo {title}
  {Randomness-induced quantum spin liquid behavior in the s = 1/2 random
  j1–j2 heisenberg antiferromagnet on the honeycomb lattice},}\ }\href
  {\doibase 10.7566/JPSJ.86.044704} {\bibfield  {journal} {\bibinfo  {journal}
  {Journal of the Physical Society of Japan}\ }\textbf {\bibinfo {volume}
  {86}},\ \bibinfo {pages} {044704} (\bibinfo {year} {2017})},\ \Eprint
  {http://arxiv.org/abs/https://doi.org/10.7566/JPSJ.86.044704}
  {https://doi.org/10.7566/JPSJ.86.044704} \BibitemShut {NoStop}%
\bibitem [{\citenamefont {Sandvik}(2002)}]{Sandvik2002}%
  \BibitemOpen
  \bibfield  {author} {\bibinfo {author} {\bibfnamefont {Anders~W.}\
  \bibnamefont {Sandvik}},\ }\bibfield  {title} {\enquote {\bibinfo {title}
  {Classical percolation transition in the diluted two-dimensional
  $s=\frac{1}{2}$ heisenberg antiferromagnet},}\ }\href {\doibase
  10.1103/PhysRevB.66.024418} {\bibfield  {journal} {\bibinfo  {journal} {Phys.
  Rev. B}\ }\textbf {\bibinfo {volume} {66}},\ \bibinfo {pages} {024418}
  (\bibinfo {year} {2002})}\BibitemShut {NoStop}%
\bibitem [{\citenamefont {Laflorencie}\ \emph {et~al.}(2006)\citenamefont
  {Laflorencie}, \citenamefont {Wessel}, \citenamefont {L\"auchli},\ and\
  \citenamefont {Rieger}}]{Rieger2006}%
  \BibitemOpen
  \bibfield  {author} {\bibinfo {author} {\bibfnamefont {Nicolas}\ \bibnamefont
  {Laflorencie}}, \bibinfo {author} {\bibfnamefont {Stefan}\ \bibnamefont
  {Wessel}}, \bibinfo {author} {\bibfnamefont {Andreas}\ \bibnamefont
  {L\"auchli}}, \ and\ \bibinfo {author} {\bibfnamefont {Heikog}\ \bibnamefont
  {Rieger}},\ }\bibfield  {title} {\enquote {\bibinfo {title} {Random-exchange
  quantum heisenberg antiferromagnets on a square lattice},}\ }\href {\doibase
  10.1103/PhysRevB.73.060403} {\bibfield  {journal} {\bibinfo  {journal} {Phys.
  Rev. B}\ }\textbf {\bibinfo {volume} {73}},\ \bibinfo {pages} {060403}
  (\bibinfo {year} {2006})}\BibitemShut {NoStop}%
\bibitem [{\citenamefont {Zhu}\ \emph {et~al.}(2017)\citenamefont {Zhu},
  \citenamefont {Maksimov}, \citenamefont {White},\ and\ \citenamefont
  {Chernyshev}}]{Chernyshev2017}%
  \BibitemOpen
  \bibfield  {author} {\bibinfo {author} {\bibfnamefont {Zhenyue}\ \bibnamefont
  {Zhu}}, \bibinfo {author} {\bibfnamefont {P.~A.}\ \bibnamefont {Maksimov}},
  \bibinfo {author} {\bibfnamefont {Steven~R.}\ \bibnamefont {White}}, \ and\
  \bibinfo {author} {\bibfnamefont {A.~L.}\ \bibnamefont {Chernyshev}},\
  }\bibfield  {title} {\enquote {\bibinfo {title} {Disorder-induced mimicry of
  a spin liquid in ${\mathrm{ybmggao}}_{4}$},}\ }\href {\doibase
  10.1103/PhysRevLett.119.157201} {\bibfield  {journal} {\bibinfo  {journal}
  {Phys. Rev. Lett.}\ }\textbf {\bibinfo {volume} {119}},\ \bibinfo {pages}
  {157201} (\bibinfo {year} {2017})}\BibitemShut {NoStop}%
\bibitem [{\citenamefont {Singh}(2010)}]{Singh2010}%
  \BibitemOpen
  \bibfield  {author} {\bibinfo {author} {\bibfnamefont {R.~R.~P.}\
  \bibnamefont {Singh}},\ }\bibfield  {title} {\enquote {\bibinfo {title}
  {Valence bond glass phase in dilute kagome antiferromagnets},}\ }\href
  {\doibase 10.1103/PhysRevLett.104.177203} {\bibfield  {journal} {\bibinfo
  {journal} {Phys. Rev. Lett.}\ }\textbf {\bibinfo {volume} {104}},\ \bibinfo
  {pages} {177203} (\bibinfo {year} {2010})}\BibitemShut {NoStop}%
\bibitem [{\citenamefont {Han}\ \emph {et~al.}(2016)\citenamefont {Han},
  \citenamefont {Norman}, \citenamefont {Wen}, \citenamefont
  {Rodriguez-Rivera}, \citenamefont {Helton}, \citenamefont {Broholm},\ and\
  \citenamefont {Lee}}]{Lee2016}%
  \BibitemOpen
  \bibfield  {author} {\bibinfo {author} {\bibfnamefont {Tian-Heng}\
  \bibnamefont {Han}}, \bibinfo {author} {\bibfnamefont {M.~R.}\ \bibnamefont
  {Norman}}, \bibinfo {author} {\bibfnamefont {J.-J.}\ \bibnamefont {Wen}},
  \bibinfo {author} {\bibfnamefont {Jose~A.}\ \bibnamefont {Rodriguez-Rivera}},
  \bibinfo {author} {\bibfnamefont {Joel~S.}\ \bibnamefont {Helton}}, \bibinfo
  {author} {\bibfnamefont {Collin}\ \bibnamefont {Broholm}}, \ and\ \bibinfo
  {author} {\bibfnamefont {Young~S.}\ \bibnamefont {Lee}},\ }\bibfield  {title}
  {\enquote {\bibinfo {title} {Correlated impurities and intrinsic spin-liquid
  physics in the kagome material herbertsmithite},}\ }\href {\doibase
  10.1103/PhysRevB.94.060409} {\bibfield  {journal} {\bibinfo  {journal} {Phys.
  Rev. B}\ }\textbf {\bibinfo {volume} {94}},\ \bibinfo {pages} {060409}
  (\bibinfo {year} {2016})}\BibitemShut {NoStop}%
\bibitem [{\citenamefont {Tarzia}\ and\ \citenamefont
  {Biroli}(2008)}]{Biroli2008}%
  \BibitemOpen
  \bibfield  {author} {\bibinfo {author} {\bibfnamefont {M.}~\bibnamefont
  {Tarzia}}\ and\ \bibinfo {author} {\bibfnamefont {G.}~\bibnamefont
  {Biroli}},\ }\bibfield  {title} {\enquote {\bibinfo {title} {The valence bond
  glass phase},}\ }\href {http://stacks.iop.org/0295-5075/82/i=6/a=67008}
  {\bibfield  {journal} {\bibinfo  {journal} {EPL (Europhysics Letters)}\
  }\textbf {\bibinfo {volume} {82}},\ \bibinfo {pages} {67008} (\bibinfo {year}
  {2008})}\BibitemShut {NoStop}%
\bibitem [{\citenamefont {Affleck}\ \emph {et~al.}(1988)\citenamefont
  {Affleck}, \citenamefont {Kennedy}, \citenamefont {Lieb},\ and\ \citenamefont
  {Tasaki}}]{Tasaki1988}%
  \BibitemOpen
  \bibfield  {author} {\bibinfo {author} {\bibfnamefont {Ian}\ \bibnamefont
  {Affleck}}, \bibinfo {author} {\bibfnamefont {Tom}\ \bibnamefont {Kennedy}},
  \bibinfo {author} {\bibfnamefont {Elliott~H.}\ \bibnamefont {Lieb}}, \ and\
  \bibinfo {author} {\bibfnamefont {Hal}\ \bibnamefont {Tasaki}},\ }\bibfield
  {title} {\enquote {\bibinfo {title} {Valence bond ground states in isotropic
  quantum antiferromagnets},}\ }\href
  {https://projecteuclid.org:443/euclid.cmp/1104161001} {\bibfield  {journal}
  {\bibinfo  {journal} {Comm. Math. Phys.}\ }\textbf {\bibinfo {volume}
  {115}},\ \bibinfo {pages} {477--528} (\bibinfo {year} {1988})}\BibitemShut
  {NoStop}%
\bibitem [{\citenamefont {Sandvik}(2007)}]{Sandvik2007}%
  \BibitemOpen
  \bibfield  {author} {\bibinfo {author} {\bibfnamefont {Anders~W.}\
  \bibnamefont {Sandvik}},\ }\bibfield  {title} {\enquote {\bibinfo {title}
  {Evidence for deconfined quantum criticality in a two-dimensional heisenberg
  model with four-spin interactions},}\ }\href {\doibase
  10.1103/PhysRevLett.98.227202} {\bibfield  {journal} {\bibinfo  {journal}
  {Phys. Rev. Lett.}\ }\textbf {\bibinfo {volume} {98}},\ \bibinfo {pages}
  {227202} (\bibinfo {year} {2007})}\BibitemShut {NoStop}%
\bibitem [{\citenamefont {Zhu}\ \emph {et~al.}(2013)\citenamefont {Zhu},
  \citenamefont {Huse},\ and\ \citenamefont {White}}]{White2013}%
  \BibitemOpen
  \bibfield  {author} {\bibinfo {author} {\bibfnamefont {Zhenyue}\ \bibnamefont
  {Zhu}}, \bibinfo {author} {\bibfnamefont {David~A.}\ \bibnamefont {Huse}}, \
  and\ \bibinfo {author} {\bibfnamefont {Steven~R.}\ \bibnamefont {White}},\
  }\bibfield  {title} {\enquote {\bibinfo {title} {Weak plaquette valence bond
  order in the $s\mathbf{=}1/2$ honeycomb
  ${J}_{1}\mathbf{\ensuremath{-}}{J}_{2}$ heisenberg model},}\ }\href {\doibase
  10.1103/PhysRevLett.110.127205} {\bibfield  {journal} {\bibinfo  {journal}
  {Phys. Rev. Lett.}\ }\textbf {\bibinfo {volume} {110}},\ \bibinfo {pages}
  {127205} (\bibinfo {year} {2013})}\BibitemShut {NoStop}%
\bibitem [{\citenamefont {Gong}\ \emph {et~al.}(2014)\citenamefont {Gong},
  \citenamefont {Zhu}, \citenamefont {Sheng}, \citenamefont {Motrunich},\ and\
  \citenamefont {Fisher}}]{Fisher2014}%
  \BibitemOpen
  \bibfield  {author} {\bibinfo {author} {\bibfnamefont {Shou-Shu}\
  \bibnamefont {Gong}}, \bibinfo {author} {\bibfnamefont {Wei}\ \bibnamefont
  {Zhu}}, \bibinfo {author} {\bibfnamefont {D.~N.}\ \bibnamefont {Sheng}},
  \bibinfo {author} {\bibfnamefont {Olexei~I.}\ \bibnamefont {Motrunich}}, \
  and\ \bibinfo {author} {\bibfnamefont {Matthew P.~A.}\ \bibnamefont
  {Fisher}},\ }\bibfield  {title} {\enquote {\bibinfo {title} {Plaquette
  ordered phase and quantum phase diagram in the spin-$\frac{1}{2}$
  ${J}_{1}\text{\ensuremath{-}}{J}_{2}$ square heisenberg model},}\ }\href
  {\doibase 10.1103/PhysRevLett.113.027201} {\bibfield  {journal} {\bibinfo
  {journal} {Phys. Rev. Lett.}\ }\textbf {\bibinfo {volume} {113}},\ \bibinfo
  {pages} {027201} (\bibinfo {year} {2014})}\BibitemShut {NoStop}%
\bibitem [{\citenamefont {Lee}\ \emph {et~al.}(2000)\citenamefont {Lee},
  \citenamefont {Broholm}, \citenamefont {Kim}, \citenamefont {Ratcliff},\ and\
  \citenamefont {Cheong}}]{Cheong2000}%
  \BibitemOpen
  \bibfield  {author} {\bibinfo {author} {\bibfnamefont {S.-H.}\ \bibnamefont
  {Lee}}, \bibinfo {author} {\bibfnamefont {C.}~\bibnamefont {Broholm}},
  \bibinfo {author} {\bibfnamefont {T.~H.}\ \bibnamefont {Kim}}, \bibinfo
  {author} {\bibfnamefont {W.}~\bibnamefont {Ratcliff}}, \ and\ \bibinfo
  {author} {\bibfnamefont {S-W.}\ \bibnamefont {Cheong}},\ }\bibfield  {title}
  {\enquote {\bibinfo {title} {Local spin resonance and spin-peierls-like phase
  transition in a geometrically frustrated antiferromagnet},}\ }\href {\doibase
  10.1103/PhysRevLett.84.3718} {\bibfield  {journal} {\bibinfo  {journal}
  {Phys. Rev. Lett.}\ }\textbf {\bibinfo {volume} {84}},\ \bibinfo {pages}
  {3718--3721} (\bibinfo {year} {2000})}\BibitemShut {NoStop}%
\bibitem [{\citenamefont {Radaelli}\ \emph {et~al.}(2002)\citenamefont
  {Radaelli}, \citenamefont {Horibe}, \citenamefont {Gutmann}, \citenamefont
  {Ishibashi}, \citenamefont {Chen}, \citenamefont {Ibberson}, \citenamefont
  {Koyama}, \citenamefont {Hor}, \citenamefont {Kiryukhin},\ and\ \citenamefont
  {Cheong}}]{Cheong2002}%
  \BibitemOpen
  \bibfield  {author} {\bibinfo {author} {\bibfnamefont {Paolo~G.}\
  \bibnamefont {Radaelli}}, \bibinfo {author} {\bibfnamefont {Y.}~\bibnamefont
  {Horibe}}, \bibinfo {author} {\bibfnamefont {Matthias~J.}\ \bibnamefont
  {Gutmann}}, \bibinfo {author} {\bibfnamefont {Hiroki}\ \bibnamefont
  {Ishibashi}}, \bibinfo {author} {\bibfnamefont {C.~H.}\ \bibnamefont {Chen}},
  \bibinfo {author} {\bibfnamefont {Richard~M.}\ \bibnamefont {Ibberson}},
  \bibinfo {author} {\bibfnamefont {Y.}~\bibnamefont {Koyama}}, \bibinfo
  {author} {\bibfnamefont {Yew-San}\ \bibnamefont {Hor}}, \bibinfo {author}
  {\bibfnamefont {Valery}\ \bibnamefont {Kiryukhin}}, \ and\ \bibinfo {author}
  {\bibfnamefont {Sang-Wook}\ \bibnamefont {Cheong}},\ }\bibfield  {title}
  {\enquote {\bibinfo {title} {Formation of isomorphic ir3+ and ir4+ octamers
  and spin dimerization in the spinel cuir2s4},}\ }\href
  {http://dx.doi.org/10.1038/416155a} {\bibfield  {journal} {\bibinfo
  {journal} {Nature}\ }\textbf {\bibinfo {volume} {416}},\ \bibinfo {pages}
  {155--158} (\bibinfo {year} {2002})}\BibitemShut {NoStop}%
\bibitem [{\citenamefont {Tsunetsugu}\ and\ \citenamefont
  {Motome}(2003)}]{Motome2003}%
  \BibitemOpen
  \bibfield  {author} {\bibinfo {author} {\bibfnamefont {Hirokazu}\
  \bibnamefont {Tsunetsugu}}\ and\ \bibinfo {author} {\bibfnamefont
  {Yukitoshi}\ \bibnamefont {Motome}},\ }\bibfield  {title} {\enquote {\bibinfo
  {title} {Magnetic transition and orbital degrees of freedom in vanadium
  spinels},}\ }\href {\doibase 10.1103/PhysRevB.68.060405} {\bibfield
  {journal} {\bibinfo  {journal} {Phys. Rev. B}\ }\textbf {\bibinfo {volume}
  {68}},\ \bibinfo {pages} {060405} (\bibinfo {year} {2003})}\BibitemShut
  {NoStop}%
\bibitem [{\citenamefont {Shimizu}\ \emph {et~al.}(2007)\citenamefont
  {Shimizu}, \citenamefont {Akimoto}, \citenamefont {Tsujii}, \citenamefont
  {Tajima},\ and\ \citenamefont {Kato}}]{Kato2007}%
  \BibitemOpen
  \bibfield  {author} {\bibinfo {author} {\bibfnamefont {Y.}~\bibnamefont
  {Shimizu}}, \bibinfo {author} {\bibfnamefont {H.}~\bibnamefont {Akimoto}},
  \bibinfo {author} {\bibfnamefont {H.}~\bibnamefont {Tsujii}}, \bibinfo
  {author} {\bibfnamefont {A.}~\bibnamefont {Tajima}}, \ and\ \bibinfo {author}
  {\bibfnamefont {R.}~\bibnamefont {Kato}},\ }\bibfield  {title} {\enquote
  {\bibinfo {title} {Mott transition in a valence-bond solid insulator with a
  triangular lattice},}\ }\href {\doibase 10.1103/PhysRevLett.99.256403}
  {\bibfield  {journal} {\bibinfo  {journal} {Phys. Rev. Lett.}\ }\textbf
  {\bibinfo {volume} {99}},\ \bibinfo {pages} {256403} (\bibinfo {year}
  {2007})}\BibitemShut {NoStop}%
\bibitem [{\citenamefont {Okamoto}\ \emph {et~al.}(2008)\citenamefont
  {Okamoto}, \citenamefont {Niitaka}, \citenamefont {Uchida}, \citenamefont
  {Waki}, \citenamefont {Takigawa}, \citenamefont {Nakatsu}, \citenamefont
  {Sekiyama}, \citenamefont {Suga}, \citenamefont {Arita},\ and\ \citenamefont
  {Takagi}}]{Takagi2008}%
  \BibitemOpen
  \bibfield  {author} {\bibinfo {author} {\bibfnamefont {Yoshihiko}\
  \bibnamefont {Okamoto}}, \bibinfo {author} {\bibfnamefont {Seiji}\
  \bibnamefont {Niitaka}}, \bibinfo {author} {\bibfnamefont {Masaya}\
  \bibnamefont {Uchida}}, \bibinfo {author} {\bibfnamefont {Takeshi}\
  \bibnamefont {Waki}}, \bibinfo {author} {\bibfnamefont {Masashi}\
  \bibnamefont {Takigawa}}, \bibinfo {author} {\bibfnamefont {Yoshitaka}\
  \bibnamefont {Nakatsu}}, \bibinfo {author} {\bibfnamefont {Akira}\
  \bibnamefont {Sekiyama}}, \bibinfo {author} {\bibfnamefont {Shigemasa}\
  \bibnamefont {Suga}}, \bibinfo {author} {\bibfnamefont {Ryotaro}\
  \bibnamefont {Arita}}, \ and\ \bibinfo {author} {\bibfnamefont {Hidenori}\
  \bibnamefont {Takagi}},\ }\bibfield  {title} {\enquote {\bibinfo {title}
  {Band jahn-teller instability and formation of valence bond solid in a
  mixed-valent spinel oxide ${\mathrm{lirh}}_{2}{\mathrm{o}}_{4}$},}\ }\href
  {\doibase 10.1103/PhysRevLett.101.086404} {\bibfield  {journal} {\bibinfo
  {journal} {Phys. Rev. Lett.}\ }\textbf {\bibinfo {volume} {101}},\ \bibinfo
  {pages} {086404} (\bibinfo {year} {2008})}\BibitemShut {NoStop}%
\bibitem [{\citenamefont {{Lieb}}\ \emph {et~al.}(1961)\citenamefont {{Lieb}},
  \citenamefont {{Schultz}},\ and\ \citenamefont {{Mattis}}}]{Mattis1961}%
  \BibitemOpen
  \bibfield  {author} {\bibinfo {author} {\bibfnamefont {E.}~\bibnamefont
  {{Lieb}}}, \bibinfo {author} {\bibfnamefont {T.}~\bibnamefont {{Schultz}}}, \
  and\ \bibinfo {author} {\bibfnamefont {D.}~\bibnamefont {{Mattis}}},\
  }\bibfield  {title} {\enquote {\bibinfo {title} {{Two soluble models of an
  antiferromagnetic chain}},}\ }\href {\doibase 10.1016/0003-4916(61)90115-4}
  {\bibfield  {journal} {\bibinfo  {journal} {Annals of Physics}\ }\textbf
  {\bibinfo {volume} {16}},\ \bibinfo {pages} {407--466} (\bibinfo {year}
  {1961})}\BibitemShut {NoStop}%
\bibitem [{\citenamefont {Oshikawa}(2000)}]{Oshikawa2000}%
  \BibitemOpen
  \bibfield  {author} {\bibinfo {author} {\bibfnamefont {Masaki}\ \bibnamefont
  {Oshikawa}},\ }\bibfield  {title} {\enquote {\bibinfo {title}
  {Commensurability, excitation gap, and topology in quantum many-particle
  systems on a periodic lattice},}\ }\href {\doibase
  10.1103/PhysRevLett.84.1535} {\bibfield  {journal} {\bibinfo  {journal}
  {Phys. Rev. Lett.}\ }\textbf {\bibinfo {volume} {84}},\ \bibinfo {pages}
  {1535--1538} (\bibinfo {year} {2000})}\BibitemShut {NoStop}%
\bibitem [{\citenamefont {Hastings}(2004)}]{Hastings2004}%
  \BibitemOpen
  \bibfield  {author} {\bibinfo {author} {\bibfnamefont {M.~B.}\ \bibnamefont
  {Hastings}},\ }\bibfield  {title} {\enquote {\bibinfo {title}
  {Lieb-schultz-mattis in higher dimensions},}\ }\href {\doibase
  10.1103/PhysRevB.69.104431} {\bibfield  {journal} {\bibinfo  {journal} {Phys.
  Rev. B}\ }\textbf {\bibinfo {volume} {69}},\ \bibinfo {pages} {104431}
  (\bibinfo {year} {2004})}\BibitemShut {NoStop}%
\bibitem [{\citenamefont {Senthil}\ and\ \citenamefont
  {Fisher}(2000)}]{SenthilFisher2000}%
  \BibitemOpen
  \bibfield  {author} {\bibinfo {author} {\bibfnamefont {T.}~\bibnamefont
  {Senthil}}\ and\ \bibinfo {author} {\bibfnamefont {Matthew P.~A.}\
  \bibnamefont {Fisher}},\ }\bibfield  {title} {\enquote {\bibinfo {title}
  {${Z}_{2}$ gauge theory of electron fractionalization in strongly correlated
  systems},}\ }\href {\doibase 10.1103/PhysRevB.62.7850} {\bibfield  {journal}
  {\bibinfo  {journal} {Phys. Rev. B}\ }\textbf {\bibinfo {volume} {62}},\
  \bibinfo {pages} {7850--7881} (\bibinfo {year} {2000})}\BibitemShut {NoStop}%
\bibitem [{\citenamefont {Jalabert}\ and\ \citenamefont
  {Sachdev}(1991)}]{Sachdev1991}%
  \BibitemOpen
  \bibfield  {author} {\bibinfo {author} {\bibfnamefont {Rodolfo~A.}\
  \bibnamefont {Jalabert}}\ and\ \bibinfo {author} {\bibfnamefont {Subir}\
  \bibnamefont {Sachdev}},\ }\bibfield  {title} {\enquote {\bibinfo {title}
  {Spontaneous alignment of frustrated bonds in an anisotropic,
  three-dimensional ising model},}\ }\href {\doibase 10.1103/PhysRevB.44.686}
  {\bibfield  {journal} {\bibinfo  {journal} {Phys. Rev. B}\ }\textbf {\bibinfo
  {volume} {44}},\ \bibinfo {pages} {686--690} (\bibinfo {year}
  {1991})}\BibitemShut {NoStop}%
\bibitem [{\citenamefont {{Sachdev}}\ and\ \citenamefont
  {{Vojta}}(2000)}]{Vojta1999}%
  \BibitemOpen
  \bibfield  {author} {\bibinfo {author} {\bibfnamefont {S.}~\bibnamefont
  {{Sachdev}}}\ and\ \bibinfo {author} {\bibfnamefont {M.}~\bibnamefont
  {{Vojta}}},\ }\bibfield  {title} {\enquote {\bibinfo {title} {{Translational
  symmetry breaking in two-dimensional antiferromagnets and
  superconductors}},}\ }\href@noop {} {\bibfield  {journal} {\bibinfo
  {journal} {eprint arXiv:cond-mat/9910231, Journal of the Physical Society of
  Japan, 69 Suppl B, 1}\ } (\bibinfo {year} {2000})},\ \Eprint
  {http://arxiv.org/abs/cond-mat/9910231} {cond-mat/9910231} \BibitemShut
  {NoStop}%
\bibitem [{\citenamefont {Hermele}\ and\ \citenamefont
  {Chen}(2016)}]{Chen2016c}%
  \BibitemOpen
  \bibfield  {author} {\bibinfo {author} {\bibfnamefont {Michael}\ \bibnamefont
  {Hermele}}\ and\ \bibinfo {author} {\bibfnamefont {Xie}\ \bibnamefont
  {Chen}},\ }\bibfield  {title} {\enquote {\bibinfo {title} {Flux-fusion
  anomaly test and bosonic topological crystalline insulators},}\ }\href
  {\doibase 10.1103/PhysRevX.6.041006} {\bibfield  {journal} {\bibinfo
  {journal} {Phys. Rev. X}\ }\textbf {\bibinfo {volume} {6}},\ \bibinfo {pages}
  {041006} (\bibinfo {year} {2016})}\BibitemShut {NoStop}%
\bibitem [{\citenamefont {{Qi}}\ \emph {et~al.}(2015)\citenamefont {{Qi}},
  \citenamefont {{Cheng}},\ and\ \citenamefont {{Fang}}}]{Fang2015}%
  \BibitemOpen
  \bibfield  {author} {\bibinfo {author} {\bibfnamefont {Y.}~\bibnamefont
  {{Qi}}}, \bibinfo {author} {\bibfnamefont {M.}~\bibnamefont {{Cheng}}}, \
  and\ \bibinfo {author} {\bibfnamefont {C.}~\bibnamefont {{Fang}}},\
  }\bibfield  {title} {\enquote {\bibinfo {title} {{Symmetry fractionalization
  of visons in $\mathbb Z\_2$ spin liquids}},}\ }\href@noop {} {\bibfield
  {journal} {\bibinfo  {journal} {ArXiv e-prints}\ } (\bibinfo {year}
  {2015})},\ \Eprint {http://arxiv.org/abs/1509.02927} {arXiv:1509.02927
  [cond-mat.str-el]} \BibitemShut {NoStop}%
\bibitem [{\citenamefont {Cheng}\ \emph {et~al.}(2016)\citenamefont {Cheng},
  \citenamefont {Zaletel}, \citenamefont {Barkeshli}, \citenamefont
  {Vishwanath},\ and\ \citenamefont {Bonderson}}]{Bonderson2016}%
  \BibitemOpen
  \bibfield  {author} {\bibinfo {author} {\bibfnamefont {Meng}\ \bibnamefont
  {Cheng}}, \bibinfo {author} {\bibfnamefont {Michael}\ \bibnamefont
  {Zaletel}}, \bibinfo {author} {\bibfnamefont {Maissam}\ \bibnamefont
  {Barkeshli}}, \bibinfo {author} {\bibfnamefont {Ashvin}\ \bibnamefont
  {Vishwanath}}, \ and\ \bibinfo {author} {\bibfnamefont {Parsa}\ \bibnamefont
  {Bonderson}},\ }\bibfield  {title} {\enquote {\bibinfo {title} {Translational
  symmetry and microscopic constraints on symmetry-enriched topological phases
  a view from the surface},}\ }\href {\doibase 10.1103/PhysRevX.6.041068}
  {\bibfield  {journal} {\bibinfo  {journal} {Phys. Rev. X}\ }\textbf {\bibinfo
  {volume} {6}},\ \bibinfo {pages} {041068} (\bibinfo {year}
  {2016})}\BibitemShut {NoStop}%
\bibitem [{\citenamefont {Hu}\ \emph {et~al.}(2015)\citenamefont {Hu},
  \citenamefont {Gong}, \citenamefont {Zhu},\ and\ \citenamefont
  {Sheng}}]{Sheng2015}%
  \BibitemOpen
  \bibfield  {author} {\bibinfo {author} {\bibfnamefont {Wen-Jun}\ \bibnamefont
  {Hu}}, \bibinfo {author} {\bibfnamefont {Shou-Shu}\ \bibnamefont {Gong}},
  \bibinfo {author} {\bibfnamefont {Wei}\ \bibnamefont {Zhu}}, \ and\ \bibinfo
  {author} {\bibfnamefont {D.~N.}\ \bibnamefont {Sheng}},\ }\bibfield  {title}
  {\enquote {\bibinfo {title} {Competing spin-liquid states in the spin-1/2
  heisenberg model on the triangular lattice},}\ }\href {\doibase
  10.1103/PhysRevB.92.140403} {\bibfield  {journal} {\bibinfo  {journal} {Phys.
  Rev. B}\ }\textbf {\bibinfo {volume} {92}},\ \bibinfo {pages} {140403}
  (\bibinfo {year} {2015})}\BibitemShut {NoStop}%
\bibitem [{\citenamefont {Zhu}\ and\ \citenamefont {White}(2015)}]{White2015}%
  \BibitemOpen
  \bibfield  {author} {\bibinfo {author} {\bibfnamefont {Zhenyue}\ \bibnamefont
  {Zhu}}\ and\ \bibinfo {author} {\bibfnamefont {Steven~R.}\ \bibnamefont
  {White}},\ }\bibfield  {title} {\enquote {\bibinfo {title} {Spin liquid phase
  of the s=1/2 j1-j2 heisenberg model on the triangular lattice},}\ }\href
  {\doibase 10.1103/PhysRevB.92.041105} {\bibfield  {journal} {\bibinfo
  {journal} {Phys. Rev. B}\ }\textbf {\bibinfo {volume} {92}},\ \bibinfo
  {pages} {041105} (\bibinfo {year} {2015})}\BibitemShut {NoStop}%
\bibitem [{\citenamefont {Imry}\ and\ \citenamefont {Ma}(1975)}]{Ma1975}%
  \BibitemOpen
  \bibfield  {author} {\bibinfo {author} {\bibfnamefont {Yoseph}\ \bibnamefont
  {Imry}}\ and\ \bibinfo {author} {\bibfnamefont {Shang-keng}\ \bibnamefont
  {Ma}},\ }\bibfield  {title} {\enquote {\bibinfo {title} {Random-field
  instability of the ordered state of continuous symmetry},}\ }\href {\doibase
  10.1103/PhysRevLett.35.1399} {\bibfield  {journal} {\bibinfo  {journal}
  {Phys. Rev. Lett.}\ }\textbf {\bibinfo {volume} {35}},\ \bibinfo {pages}
  {1399--1401} (\bibinfo {year} {1975})}\BibitemShut {NoStop}%
\bibitem [{\citenamefont {Binder}(1983)}]{Binder1983}%
  \BibitemOpen
  \bibfield  {author} {\bibinfo {author} {\bibfnamefont {K.}~\bibnamefont
  {Binder}},\ }\bibfield  {title} {\enquote {\bibinfo {title} {Random-field
  induced interface widths in ising systems},}\ }\href {\doibase
  10.1007/BF01470045} {\bibfield  {journal} {\bibinfo  {journal} {Zeitschrift
  f{\"u}r Physik B Condensed Matter}\ }\textbf {\bibinfo {volume} {50}},\
  \bibinfo {pages} {343--352} (\bibinfo {year} {1983})}\BibitemShut {NoStop}%
\bibitem [{\citenamefont {Nattermann}\ and\ \citenamefont
  {Villain}(1988)}]{nattermann1988random}%
  \BibitemOpen
  \bibfield  {author} {\bibinfo {author} {\bibfnamefont {T}~\bibnamefont
  {Nattermann}}\ and\ \bibinfo {author} {\bibfnamefont {J}~\bibnamefont
  {Villain}},\ }\bibfield  {title} {\enquote {\bibinfo {title} {Random-field
  ising systems: A survey of current theoretical views},}\ }\href@noop {}
  {\bibfield  {journal} {\bibinfo  {journal} {Phase Transitions: A
  Multinational Journal}\ }\textbf {\bibinfo {volume} {11}},\ \bibinfo {pages}
  {5--51} (\bibinfo {year} {1988})}\BibitemShut {NoStop}%
\bibitem [{\citenamefont {Hyman}\ \emph {et~al.}(1996)\citenamefont {Hyman},
  \citenamefont {Yang}, \citenamefont {Bhatt},\ and\ \citenamefont
  {Girvin}}]{Girvin1996}%
  \BibitemOpen
  \bibfield  {author} {\bibinfo {author} {\bibfnamefont {R.~A.}\ \bibnamefont
  {Hyman}}, \bibinfo {author} {\bibfnamefont {Kun}\ \bibnamefont {Yang}},
  \bibinfo {author} {\bibfnamefont {R.~N.}\ \bibnamefont {Bhatt}}, \ and\
  \bibinfo {author} {\bibfnamefont {S.~M.}\ \bibnamefont {Girvin}},\ }\bibfield
   {title} {\enquote {\bibinfo {title} {Random bonds and topological stability
  in gapped quantum spin chains},}\ }\href {\doibase
  10.1103/PhysRevLett.76.839} {\bibfield  {journal} {\bibinfo  {journal} {Ref.
  21 in Phys. Rev. Lett.}\ }\textbf {\bibinfo {volume} {76}},\ \bibinfo {pages}
  {839--842} (\bibinfo {year} {1996})}\BibitemShut {NoStop}%
\bibitem [{\citenamefont {Yang}\ \emph {et~al.}(1996)\citenamefont {Yang},
  \citenamefont {Hyman}, \citenamefont {Bhatt},\ and\ \citenamefont
  {Girvin}}]{yang1996effects}%
  \BibitemOpen
  \bibfield  {author} {\bibinfo {author} {\bibfnamefont {Kun}\ \bibnamefont
  {Yang}}, \bibinfo {author} {\bibfnamefont {RA}~\bibnamefont {Hyman}},
  \bibinfo {author} {\bibfnamefont {RN}~\bibnamefont {Bhatt}}, \ and\ \bibinfo
  {author} {\bibfnamefont {SM}~\bibnamefont {Girvin}},\ }\bibfield  {title}
  {\enquote {\bibinfo {title} {Effects of randomness in gapped
  antiferromagnetic quantum spin chains},}\ }\href@noop {} {\bibfield
  {journal} {\bibinfo  {journal} {Journal of applied physics}\ }\textbf
  {\bibinfo {volume} {79}},\ \bibinfo {pages} {5096--5098} (\bibinfo {year}
  {1996})}\BibitemShut {NoStop}%
\bibitem [{\citenamefont {Lavar{\'e}lo}\ and\ \citenamefont
  {Roux}(2013)}]{lavarelo2013localization}%
  \BibitemOpen
  \bibfield  {author} {\bibinfo {author} {\bibfnamefont {Arthur}\ \bibnamefont
  {Lavar{\'e}lo}}\ and\ \bibinfo {author} {\bibfnamefont {Guillaume}\
  \bibnamefont {Roux}},\ }\bibfield  {title} {\enquote {\bibinfo {title}
  {Localization of spinons in random majumdar-ghosh chains},}\ }\href@noop {}
  {\bibfield  {journal} {\bibinfo  {journal} {Physical review letters}\
  }\textbf {\bibinfo {volume} {110}},\ \bibinfo {pages} {087204} (\bibinfo
  {year} {2013})}\BibitemShut {NoStop}%
\bibitem [{\citenamefont {Shu}\ \emph {et~al.}(2016)\citenamefont {Shu},
  \citenamefont {Yao}, \citenamefont {Ke}, \citenamefont {Lin},\ and\
  \citenamefont {Sandvik}}]{shu2016properties}%
  \BibitemOpen
  \bibfield  {author} {\bibinfo {author} {\bibfnamefont {Yu-Rong}\ \bibnamefont
  {Shu}}, \bibinfo {author} {\bibfnamefont {Dao-Xin}\ \bibnamefont {Yao}},
  \bibinfo {author} {\bibfnamefont {Chih-Wei}\ \bibnamefont {Ke}}, \bibinfo
  {author} {\bibfnamefont {Yu-Cheng}\ \bibnamefont {Lin}}, \ and\ \bibinfo
  {author} {\bibfnamefont {Anders~W}\ \bibnamefont {Sandvik}},\ }\bibfield
  {title} {\enquote {\bibinfo {title} {Properties of the random-singlet phase:
  From the disordered heisenberg chain to an amorphous valence-bond solid},}\
  }\href@noop {} {\bibfield  {journal} {\bibinfo  {journal} {Physical Review
  B}\ }\textbf {\bibinfo {volume} {94}},\ \bibinfo {pages} {174442} (\bibinfo
  {year} {2016})}\BibitemShut {NoStop}%
\bibitem [{\citenamefont {Bursill}\ \emph {et~al.}(1995)\citenamefont
  {Bursill}, \citenamefont {Gehring}, \citenamefont {Farnell}, \citenamefont
  {Parkinson}, \citenamefont {Xiang},\ and\ \citenamefont {Zeng}}]{Zeng1995}%
  \BibitemOpen
  \bibfield  {author} {\bibinfo {author} {\bibfnamefont {R}~\bibnamefont
  {Bursill}}, \bibinfo {author} {\bibfnamefont {G~A}\ \bibnamefont {Gehring}},
  \bibinfo {author} {\bibfnamefont {D~J~J}\ \bibnamefont {Farnell}}, \bibinfo
  {author} {\bibfnamefont {J~B}\ \bibnamefont {Parkinson}}, \bibinfo {author}
  {\bibfnamefont {Tao}\ \bibnamefont {Xiang}}, \ and\ \bibinfo {author}
  {\bibfnamefont {Chen}\ \bibnamefont {Zeng}},\ }\bibfield  {title} {\enquote
  {\bibinfo {title} {Numerical and approximate analytical results for the
  frustrated spin- 1/2 quantum spin chain},}\ }\href
  {http://stacks.iop.org/0953-8984/7/i=45/a=016} {\bibfield  {journal}
  {\bibinfo  {journal} {Journal of Physics: Condensed Matter}\ }\textbf
  {\bibinfo {volume} {7}},\ \bibinfo {pages} {8605} (\bibinfo {year}
  {1995})}\BibitemShut {NoStop}%
\bibitem [{\citenamefont {Deschner}\ and\ \citenamefont
  {S\o{}rensen}(2013)}]{Sorensen2013}%
  \BibitemOpen
  \bibfield  {author} {\bibinfo {author} {\bibfnamefont {Andreas}\ \bibnamefont
  {Deschner}}\ and\ \bibinfo {author} {\bibfnamefont {Erik~S.}\ \bibnamefont
  {S\o{}rensen}},\ }\bibfield  {title} {\enquote {\bibinfo {title}
  {Incommensurability effects in odd length ${J}_{1}$-${J}_{2}$ quantum spin
  chains: On-site magnetization and entanglement},}\ }\href {\doibase
  10.1103/PhysRevB.87.094415} {\bibfield  {journal} {\bibinfo  {journal} {Phys.
  Rev. B}\ }\textbf {\bibinfo {volume} {87}},\ \bibinfo {pages} {094415}
  (\bibinfo {year} {2013})}\BibitemShut {NoStop}%
\bibitem [{\citenamefont {Dasgupta}\ and\ \citenamefont {Ma}(1980)}]{Ma1980}%
  \BibitemOpen
  \bibfield  {author} {\bibinfo {author} {\bibfnamefont {Chandan}\ \bibnamefont
  {Dasgupta}}\ and\ \bibinfo {author} {\bibfnamefont {Shang-keng}\ \bibnamefont
  {Ma}},\ }\bibfield  {title} {\enquote {\bibinfo {title} {Low-temperature
  properties of the random heisenberg antiferromagnetic chain},}\ }\href
  {\doibase 10.1103/PhysRevB.22.1305} {\bibfield  {journal} {\bibinfo
  {journal} {Phys. Rev. B}\ }\textbf {\bibinfo {volume} {22}},\ \bibinfo
  {pages} {1305--1319} (\bibinfo {year} {1980})}\BibitemShut {NoStop}%
\bibitem [{\citenamefont {Ma}\ \emph {et~al.}(1979)\citenamefont {Ma},
  \citenamefont {Dasgupta},\ and\ \citenamefont {Hu}}]{Hu1979}%
  \BibitemOpen
  \bibfield  {author} {\bibinfo {author} {\bibfnamefont {Shang-keng}\
  \bibnamefont {Ma}}, \bibinfo {author} {\bibfnamefont {Chandan}\ \bibnamefont
  {Dasgupta}}, \ and\ \bibinfo {author} {\bibfnamefont {Chin-kun}\ \bibnamefont
  {Hu}},\ }\bibfield  {title} {\enquote {\bibinfo {title} {Random
  antiferromagnetic chain},}\ }\href {\doibase 10.1103/PhysRevLett.43.1434}
  {\bibfield  {journal} {\bibinfo  {journal} {Phys. Rev. Lett.}\ }\textbf
  {\bibinfo {volume} {43}},\ \bibinfo {pages} {1434--1437} (\bibinfo {year}
  {1979})}\BibitemShut {NoStop}%
\bibitem [{\citenamefont {Westerberg}\ \emph {et~al.}(1995)\citenamefont
  {Westerberg}, \citenamefont {Furusaki}, \citenamefont {Sigrist},\ and\
  \citenamefont {Lee}}]{Lee1995}%
  \BibitemOpen
  \bibfield  {author} {\bibinfo {author} {\bibfnamefont {E.}~\bibnamefont
  {Westerberg}}, \bibinfo {author} {\bibfnamefont {A.}~\bibnamefont
  {Furusaki}}, \bibinfo {author} {\bibfnamefont {M.}~\bibnamefont {Sigrist}}, \
  and\ \bibinfo {author} {\bibfnamefont {P.~A.}\ \bibnamefont {Lee}},\
  }\bibfield  {title} {\enquote {\bibinfo {title} {Random quantum spin chains:
  A real-space renormalization group study},}\ }\href {\doibase
  10.1103/PhysRevLett.75.4302} {\bibfield  {journal} {\bibinfo  {journal}
  {Phys. Rev. Lett.}\ }\textbf {\bibinfo {volume} {75}},\ \bibinfo {pages}
  {4302--4305} (\bibinfo {year} {1995})}\BibitemShut {NoStop}%
\bibitem [{\citenamefont {Westerberg}\ \emph {et~al.}(1997)\citenamefont
  {Westerberg}, \citenamefont {Furusaki}, \citenamefont {Sigrist},\ and\
  \citenamefont {Lee}}]{Lee1997}%
  \BibitemOpen
  \bibfield  {author} {\bibinfo {author} {\bibfnamefont {E.}~\bibnamefont
  {Westerberg}}, \bibinfo {author} {\bibfnamefont {A.}~\bibnamefont
  {Furusaki}}, \bibinfo {author} {\bibfnamefont {M.}~\bibnamefont {Sigrist}}, \
  and\ \bibinfo {author} {\bibfnamefont {P.~A.}\ \bibnamefont {Lee}},\
  }\bibfield  {title} {\enquote {\bibinfo {title} {Low-energy fixed points of
  random quantum spin chains},}\ }\href {\doibase 10.1103/PhysRevB.55.12578}
  {\bibfield  {journal} {\bibinfo  {journal} {Phys. Rev. B}\ }\textbf {\bibinfo
  {volume} {55}},\ \bibinfo {pages} {12578--12593} (\bibinfo {year}
  {1997})}\BibitemShut {NoStop}%
\bibitem [{\citenamefont {Azuma}\ \emph {et~al.}(1997)\citenamefont {Azuma},
  \citenamefont {Fujishiro}, \citenamefont {Takano}, \citenamefont {Nohara},\
  and\ \citenamefont {Takagi}}]{Takagi1997}%
  \BibitemOpen
  \bibfield  {author} {\bibinfo {author} {\bibfnamefont {M.}~\bibnamefont
  {Azuma}}, \bibinfo {author} {\bibfnamefont {Y.}~\bibnamefont {Fujishiro}},
  \bibinfo {author} {\bibfnamefont {M.}~\bibnamefont {Takano}}, \bibinfo
  {author} {\bibfnamefont {M.}~\bibnamefont {Nohara}}, \ and\ \bibinfo {author}
  {\bibfnamefont {H.}~\bibnamefont {Takagi}},\ }\bibfield  {title} {\enquote
  {\bibinfo {title} {Switching of the gapped singlet spin-liquid state to an
  antiferromagnetically ordered state in
  sr(${\mathrm{cu}}_{1\mathrm{\ensuremath{-}}\mathrm{x}}$${\mathrm{zn}}_{\mathrm{x}}$${)}_{2}$${\mathrm{o}}_{3}$},}\
  }\href {\doibase 10.1103/PhysRevB.55.R8658} {\bibfield  {journal} {\bibinfo
  {journal} {Phys. Rev. B}\ }\textbf {\bibinfo {volume} {55}},\ \bibinfo
  {pages} {R8658--R8661} (\bibinfo {year} {1997})}\BibitemShut {NoStop}%
\bibitem [{\citenamefont {Bobroff}\ \emph {et~al.}(2009)\citenamefont
  {Bobroff}, \citenamefont {Laflorencie}, \citenamefont {Alexander},
  \citenamefont {Mahajan}, \citenamefont {Koteswararao},\ and\ \citenamefont
  {Mendels}}]{Mendels2009}%
  \BibitemOpen
  \bibfield  {author} {\bibinfo {author} {\bibfnamefont {J.}~\bibnamefont
  {Bobroff}}, \bibinfo {author} {\bibfnamefont {N.}~\bibnamefont
  {Laflorencie}}, \bibinfo {author} {\bibfnamefont {L.~K.}\ \bibnamefont
  {Alexander}}, \bibinfo {author} {\bibfnamefont {A.~V.}\ \bibnamefont
  {Mahajan}}, \bibinfo {author} {\bibfnamefont {B.}~\bibnamefont
  {Koteswararao}}, \ and\ \bibinfo {author} {\bibfnamefont {P.}~\bibnamefont
  {Mendels}},\ }\bibfield  {title} {\enquote {\bibinfo {title}
  {Impurity-induced magnetic order in low-dimensional spin-gapped materials},}\
  }\href {\doibase 10.1103/PhysRevLett.103.047201} {\bibfield  {journal}
  {\bibinfo  {journal} {Phys. Rev. Lett.}\ }\textbf {\bibinfo {volume} {103}},\
  \bibinfo {pages} {047201} (\bibinfo {year} {2009})}\BibitemShut {NoStop}%
\bibitem [{\citenamefont {Levin}\ and\ \citenamefont
  {Senthil}(2004)}]{Senthil2004}%
  \BibitemOpen
  \bibfield  {author} {\bibinfo {author} {\bibfnamefont {Michael}\ \bibnamefont
  {Levin}}\ and\ \bibinfo {author} {\bibfnamefont {T.}~\bibnamefont
  {Senthil}},\ }\bibfield  {title} {\enquote {\bibinfo {title} {Deconfined
  quantum criticality and n\'eel order via dimer disorder},}\ }\href {\doibase
  10.1103/PhysRevB.70.220403} {\bibfield  {journal} {\bibinfo  {journal} {Phys.
  Rev. B}\ }\textbf {\bibinfo {volume} {70}},\ \bibinfo {pages} {220403}
  (\bibinfo {year} {2004})}\BibitemShut {NoStop}%
\bibitem [{\citenamefont {Gingras}\ and\ \citenamefont
  {Huse}(1996)}]{gingras1996topological}%
  \BibitemOpen
  \bibfield  {author} {\bibinfo {author} {\bibfnamefont {Michel~JP}\
  \bibnamefont {Gingras}}\ and\ \bibinfo {author} {\bibfnamefont {David~A}\
  \bibnamefont {Huse}},\ }\bibfield  {title} {\enquote {\bibinfo {title}
  {Topological defects in the random-field xy model and the pinned vortex
  lattice to vortex glass transition in type-ii superconductors},}\ }\href@noop
  {} {\bibfield  {journal} {\bibinfo  {journal} {Physical Review B}\ }\textbf
  {\bibinfo {volume} {53}},\ \bibinfo {pages} {15193} (\bibinfo {year}
  {1996})}\BibitemShut {NoStop}%
\bibitem [{\citenamefont {Beach}(2011)}]{beachtalk}%
  \BibitemOpen
  \bibfield  {author} {\bibinfo {author} {\bibfnamefont {K.}~\bibnamefont
  {Beach}},\ }\bibfield  {title} {\enquote {\bibinfo {title} {Random singlet
  phases beyond one spatial dimension},}\ }in\ \href@noop {} {\emph {\bibinfo
  {booktitle} {Talk at ICTP Trieste}}}\ (\bibinfo {year} {2011})\BibitemShut
  {NoStop}%
\bibitem [{\citenamefont {{Liu}}\ \emph {et~al.}(2018)\citenamefont {{Liu}},
  \citenamefont {{Shao}}, \citenamefont {{Lin}}, \citenamefont {{Guo}},\ and\
  \citenamefont {{Sandvik}}}]{forthcomingAnders}%
  \BibitemOpen
  \bibfield  {author} {\bibinfo {author} {\bibfnamefont {L.}~\bibnamefont
  {{Liu}}}, \bibinfo {author} {\bibfnamefont {H.}~\bibnamefont {{Shao}}},
  \bibinfo {author} {\bibfnamefont {Y.-C.}\ \bibnamefont {{Lin}}}, \bibinfo
  {author} {\bibfnamefont {W.}~\bibnamefont {{Guo}}}, \ and\ \bibinfo {author}
  {\bibfnamefont {A.~W.}\ \bibnamefont {{Sandvik}}},\ }\bibfield  {title}
  {\enquote {\bibinfo {title} {{Random-Singlet Phase in Disordered
  Two-Dimensional Quantum Magnets}},}\ }\href@noop {} {\bibfield  {journal}
  {\bibinfo  {journal} {ArXiv e-prints}\ } (\bibinfo {year} {2018})},\ \Eprint
  {http://arxiv.org/abs/1804.06108} {arXiv:1804.06108 [cond-mat.str-el]}
  \BibitemShut {NoStop}%
\bibitem [{\citenamefont {Nie}\ \emph {et~al.}(2014)\citenamefont {Nie},
  \citenamefont {Tarjus},\ and\ \citenamefont {Kivelson}}]{Kivelson2014}%
  \BibitemOpen
  \bibfield  {author} {\bibinfo {author} {\bibfnamefont {Laimei}\ \bibnamefont
  {Nie}}, \bibinfo {author} {\bibfnamefont {Gilles}\ \bibnamefont {Tarjus}}, \
  and\ \bibinfo {author} {\bibfnamefont {Steven~Allan}\ \bibnamefont
  {Kivelson}},\ }\bibfield  {title} {\enquote {\bibinfo {title} {Quenched
  disorder and vestigial nematicity in the pseudogap regime of the cuprates},}\
  }\href {\doibase 10.1073/pnas.1406019111} {\bibfield  {journal} {\bibinfo
  {journal} {Proceedings of the National Academy of Sciences}\ }\textbf
  {\bibinfo {volume} {111}},\ \bibinfo {pages} {7980--7985} (\bibinfo {year}
  {2014})},\ \Eprint
  {http://arxiv.org/abs/http://www.pnas.org/content/111/22/7980.full.pdf}
  {http://www.pnas.org/content/111/22/7980.full.pdf} \BibitemShut {NoStop}%
\bibitem [{\citenamefont {Moessner}\ and\ \citenamefont
  {Sondhi}(2001)}]{Sondhi2001}%
  \BibitemOpen
  \bibfield  {author} {\bibinfo {author} {\bibfnamefont {R.}~\bibnamefont
  {Moessner}}\ and\ \bibinfo {author} {\bibfnamefont {S.~L.}\ \bibnamefont
  {Sondhi}},\ }\bibfield  {title} {\enquote {\bibinfo {title} {Ising models of
  quantum frustration},}\ }\href {\doibase 10.1103/PhysRevB.63.224401}
  {\bibfield  {journal} {\bibinfo  {journal} {Phys. Rev. B}\ }\textbf {\bibinfo
  {volume} {63}},\ \bibinfo {pages} {224401} (\bibinfo {year}
  {2001})}\BibitemShut {NoStop}%
\bibitem [{\citenamefont {Misguich}\ and\ \citenamefont
  {Mila}(2008)}]{Mila2008}%
  \BibitemOpen
  \bibfield  {author} {\bibinfo {author} {\bibfnamefont {Gr\'egoire}\
  \bibnamefont {Misguich}}\ and\ \bibinfo {author} {\bibfnamefont
  {Fr\'ed\'eric}\ \bibnamefont {Mila}},\ }\bibfield  {title} {\enquote
  {\bibinfo {title} {Quantum dimer model on the triangular lattice:
  Semiclassical and variational approaches to vison dispersion and
  condensation},}\ }\href {\doibase 10.1103/PhysRevB.77.134421} {\bibfield
  {journal} {\bibinfo  {journal} {Phys. Rev. B}\ }\textbf {\bibinfo {volume}
  {77}},\ \bibinfo {pages} {134421} (\bibinfo {year} {2008})}\BibitemShut
  {NoStop}%
\bibitem [{\citenamefont {Slagle}\ and\ \citenamefont {Xu}(2014)}]{Xu2014}%
  \BibitemOpen
  \bibfield  {author} {\bibinfo {author} {\bibfnamefont {Kevin}\ \bibnamefont
  {Slagle}}\ and\ \bibinfo {author} {\bibfnamefont {Cenke}\ \bibnamefont
  {Xu}},\ }\bibfield  {title} {\enquote {\bibinfo {title} {Quantum phase
  transition between the ${Z}_{2}$ spin liquid and valence bond crystals on a
  triangular lattice},}\ }\href {\doibase 10.1103/PhysRevB.89.104418}
  {\bibfield  {journal} {\bibinfo  {journal} {Phys. Rev. B}\ }\textbf {\bibinfo
  {volume} {89}},\ \bibinfo {pages} {104418} (\bibinfo {year}
  {2014})}\BibitemShut {NoStop}%
\bibitem [{\citenamefont {Lammert}\ \emph {et~al.}(1995)\citenamefont
  {Lammert}, \citenamefont {Rokhsar},\ and\ \citenamefont {Toner}}]{Toner1995}%
  \BibitemOpen
  \bibfield  {author} {\bibinfo {author} {\bibfnamefont {Paul~E.}\ \bibnamefont
  {Lammert}}, \bibinfo {author} {\bibfnamefont {Daniel~S.}\ \bibnamefont
  {Rokhsar}}, \ and\ \bibinfo {author} {\bibfnamefont {John}\ \bibnamefont
  {Toner}},\ }\bibfield  {title} {\enquote {\bibinfo {title} {Topology and
  nematic ordering. i. a gauge theory},}\ }\href {\doibase
  10.1103/PhysRevE.52.1778} {\bibfield  {journal} {\bibinfo  {journal} {Phys.
  Rev. E}\ }\textbf {\bibinfo {volume} {52}},\ \bibinfo {pages} {1778--1800}
  (\bibinfo {year} {1995})}\BibitemShut {NoStop}%
\bibitem [{\citenamefont {Zeng}\ \emph {et~al.}(1999)\citenamefont {Zeng},
  \citenamefont {Leath},\ and\ \citenamefont {Fisher}}]{Fisher1999}%
  \BibitemOpen
  \bibfield  {author} {\bibinfo {author} {\bibfnamefont {Chen}\ \bibnamefont
  {Zeng}}, \bibinfo {author} {\bibfnamefont {P.~L.}\ \bibnamefont {Leath}}, \
  and\ \bibinfo {author} {\bibfnamefont {Daniel~S.}\ \bibnamefont {Fisher}},\
  }\bibfield  {title} {\enquote {\bibinfo {title} {Absence of two-dimensional
  bragg glasses},}\ }\href {\doibase 10.1103/PhysRevLett.82.1935} {\bibfield
  {journal} {\bibinfo  {journal} {Phys. Rev. Lett.}\ }\textbf {\bibinfo
  {volume} {82}},\ \bibinfo {pages} {1935--1938} (\bibinfo {year}
  {1999})}\BibitemShut {NoStop}%
\bibitem [{\citenamefont {Middleton}(2000)}]{Middleton2000}%
  \BibitemOpen
  \bibfield  {author} {\bibinfo {author} {\bibfnamefont {A.~Alan}\ \bibnamefont
  {Middleton}},\ }\bibfield  {title} {\enquote {\bibinfo {title}
  {Disorder-induced topological defects in a $d=2$ elastic medium at zero
  temperature},}\ }\href {\doibase 10.1103/PhysRevB.61.14787} {\bibfield
  {journal} {\bibinfo  {journal} {Phys. Rev. B}\ }\textbf {\bibinfo {volume}
  {61}},\ \bibinfo {pages} {14787--14790} (\bibinfo {year} {2000})}\BibitemShut
  {NoStop}%
\bibitem [{\citenamefont {Cardy}\ and\ \citenamefont
  {Ostlund}(1982)}]{Ostlund1982}%
  \BibitemOpen
  \bibfield  {author} {\bibinfo {author} {\bibfnamefont {John~L.}\ \bibnamefont
  {Cardy}}\ and\ \bibinfo {author} {\bibfnamefont {S.}~\bibnamefont
  {Ostlund}},\ }\bibfield  {title} {\enquote {\bibinfo {title} {Random
  symmetry-breaking fields and the $\mathrm{XY}$ model},}\ }\href {\doibase
  10.1103/PhysRevB.25.6899} {\bibfield  {journal} {\bibinfo  {journal} {Phys.
  Rev. B}\ }\textbf {\bibinfo {volume} {25}},\ \bibinfo {pages} {6899--6909}
  (\bibinfo {year} {1982})}\BibitemShut {NoStop}%
\bibitem [{\citenamefont {Giamarchi}\ and\ \citenamefont
  {Le~Doussal}(1995)}]{LeDoussal1995}%
  \BibitemOpen
  \bibfield  {author} {\bibinfo {author} {\bibfnamefont {Thierry}\ \bibnamefont
  {Giamarchi}}\ and\ \bibinfo {author} {\bibfnamefont {Pierre}\ \bibnamefont
  {Le~Doussal}},\ }\bibfield  {title} {\enquote {\bibinfo {title} {Elastic
  theory of flux lattices in the presence of weak disorder},}\ }\href {\doibase
  10.1103/PhysRevB.52.1242} {\bibfield  {journal} {\bibinfo  {journal} {Phys.
  Rev. B}\ }\textbf {\bibinfo {volume} {52}},\ \bibinfo {pages} {1242--1270}
  (\bibinfo {year} {1995})}\BibitemShut {NoStop}%
\bibitem [{\citenamefont {Bray}\ and\ \citenamefont
  {Moore}(1984)}]{bray1984lower}%
  \BibitemOpen
  \bibfield  {author} {\bibinfo {author} {\bibfnamefont {AJ}~\bibnamefont
  {Bray}}\ and\ \bibinfo {author} {\bibfnamefont {MA}~\bibnamefont {Moore}},\
  }\bibfield  {title} {\enquote {\bibinfo {title} {Lower critical dimension of
  ising spin glasses: a numerical study},}\ }\href@noop {} {\bibfield
  {journal} {\bibinfo  {journal} {Journal of Physics C: Solid State Physics}\
  }\textbf {\bibinfo {volume} {17}},\ \bibinfo {pages} {L463} (\bibinfo {year}
  {1984})}\BibitemShut {NoStop}%
\bibitem [{\citenamefont {Bray}\ and\ \citenamefont
  {Moore}(1987)}]{bray1987chaotic}%
  \BibitemOpen
  \bibfield  {author} {\bibinfo {author} {\bibfnamefont {AJ}~\bibnamefont
  {Bray}}\ and\ \bibinfo {author} {\bibfnamefont {MA}~\bibnamefont {Moore}},\
  }\bibfield  {title} {\enquote {\bibinfo {title} {Chaotic nature of the
  spin-glass phase},}\ }\href@noop {} {\bibfield  {journal} {\bibinfo
  {journal} {Physical review letters}\ }\textbf {\bibinfo {volume} {58}},\
  \bibinfo {pages} {57} (\bibinfo {year} {1987})}\BibitemShut {NoStop}%
\bibitem [{\citenamefont {Middleton}(2001)}]{middleton2001energetics}%
  \BibitemOpen
  \bibfield  {author} {\bibinfo {author} {\bibfnamefont {A~Alan}\ \bibnamefont
  {Middleton}},\ }\bibfield  {title} {\enquote {\bibinfo {title} {Energetics
  and geometry of excitations in random systems},}\ }\href@noop {} {\bibfield
  {journal} {\bibinfo  {journal} {Physical Review B}\ }\textbf {\bibinfo
  {volume} {63}},\ \bibinfo {pages} {060202} (\bibinfo {year}
  {2001})}\BibitemShut {NoStop}%
\bibitem [{\citenamefont {Hartmann}\ and\ \citenamefont
  {Young}(2002)}]{hartmann2002large}%
  \BibitemOpen
  \bibfield  {author} {\bibinfo {author} {\bibfnamefont {AK}~\bibnamefont
  {Hartmann}}\ and\ \bibinfo {author} {\bibfnamefont {AP}~\bibnamefont
  {Young}},\ }\bibfield  {title} {\enquote {\bibinfo {title} {Large-scale
  low-energy excitations in the two-dimensional ising spin glass},}\
  }\href@noop {} {\bibfield  {journal} {\bibinfo  {journal} {Physical Review
  B}\ }\textbf {\bibinfo {volume} {66}},\ \bibinfo {pages} {094419} (\bibinfo
  {year} {2002})}\BibitemShut {NoStop}%
\bibitem [{\citenamefont {Fisher}\ and\ \citenamefont {Huse}(1986)}]{Huse1986}%
  \BibitemOpen
  \bibfield  {author} {\bibinfo {author} {\bibfnamefont {Daniel~S.}\
  \bibnamefont {Fisher}}\ and\ \bibinfo {author} {\bibfnamefont {David~A.}\
  \bibnamefont {Huse}},\ }\bibfield  {title} {\enquote {\bibinfo {title}
  {Ordered phase of short-range ising spin-glasses},}\ }\href {\doibase
  10.1103/PhysRevLett.56.1601} {\bibfield  {journal} {\bibinfo  {journal}
  {Phys. Rev. Lett.}\ }\textbf {\bibinfo {volume} {56}},\ \bibinfo {pages}
  {1601--1604} (\bibinfo {year} {1986})}\BibitemShut {NoStop}%
\bibitem [{\citenamefont {Fisher}\ and\ \citenamefont
  {Huse}(1988)}]{fisher1988equilibrium}%
  \BibitemOpen
  \bibfield  {author} {\bibinfo {author} {\bibfnamefont {Daniel~S}\
  \bibnamefont {Fisher}}\ and\ \bibinfo {author} {\bibfnamefont {David~A}\
  \bibnamefont {Huse}},\ }\bibfield  {title} {\enquote {\bibinfo {title}
  {Equilibrium behavior of the spin-glass ordered phase},}\ }\href@noop {}
  {\bibfield  {journal} {\bibinfo  {journal} {Physical Review B}\ }\textbf
  {\bibinfo {volume} {38}},\ \bibinfo {pages} {386} (\bibinfo {year}
  {1988})}\BibitemShut {NoStop}%
\bibitem [{\citenamefont {Jayaprakash}\ and\ \citenamefont
  {Kirkpatrick}(1980)}]{jayaprakash1980random}%
  \BibitemOpen
  \bibfield  {author} {\bibinfo {author} {\bibfnamefont {C}~\bibnamefont
  {Jayaprakash}}\ and\ \bibinfo {author} {\bibfnamefont {S}~\bibnamefont
  {Kirkpatrick}},\ }\bibfield  {title} {\enquote {\bibinfo {title} {Random
  anisotropy models in the ising limit},}\ }\href@noop {} {\bibfield  {journal}
  {\bibinfo  {journal} {Physical Review B}\ }\textbf {\bibinfo {volume} {21}},\
  \bibinfo {pages} {4072} (\bibinfo {year} {1980})}\BibitemShut {NoStop}%
\bibitem [{\citenamefont {Liers}\ \emph {et~al.}(2007)\citenamefont {Liers},
  \citenamefont {Lukic}, \citenamefont {Marinari}, \citenamefont {Pelissetto},\
  and\ \citenamefont {Vicari}}]{vicari2007zerotemperature}%
  \BibitemOpen
  \bibfield  {author} {\bibinfo {author} {\bibfnamefont {Frauke}\ \bibnamefont
  {Liers}}, \bibinfo {author} {\bibfnamefont {Jovanka}\ \bibnamefont {Lukic}},
  \bibinfo {author} {\bibfnamefont {Enzo}\ \bibnamefont {Marinari}}, \bibinfo
  {author} {\bibfnamefont {Andrea}\ \bibnamefont {Pelissetto}}, \ and\ \bibinfo
  {author} {\bibfnamefont {Ettore}\ \bibnamefont {Vicari}},\ }\bibfield
  {title} {\enquote {\bibinfo {title} {Zero-temperature behavior of the
  random-anisotropy model in the strong-anisotropy limit},}\ }\href {\doibase
  10.1103/PhysRevB.76.174423} {\bibfield  {journal} {\bibinfo  {journal} {Phys.
  Rev. B}\ }\textbf {\bibinfo {volume} {76}},\ \bibinfo {pages} {174423}
  (\bibinfo {year} {2007})}\BibitemShut {NoStop}%
\bibitem [{\citenamefont {Oshikawa}\ \emph {et~al.}(1997)\citenamefont
  {Oshikawa}, \citenamefont {Yamanaka},\ and\ \citenamefont
  {Affleck}}]{Affleck1997}%
  \BibitemOpen
  \bibfield  {author} {\bibinfo {author} {\bibfnamefont {Masaki}\ \bibnamefont
  {Oshikawa}}, \bibinfo {author} {\bibfnamefont {Masanori}\ \bibnamefont
  {Yamanaka}}, \ and\ \bibinfo {author} {\bibfnamefont {Ian}\ \bibnamefont
  {Affleck}},\ }\bibfield  {title} {\enquote {\bibinfo {title} {Magnetization
  plateaus in spin chains: ``haldane gap'' for half-integer spins},}\ }\href
  {\doibase 10.1103/PhysRevLett.78.1984} {\bibfield  {journal} {\bibinfo
  {journal} {Phys. Rev. Lett.}\ }\textbf {\bibinfo {volume} {78}},\ \bibinfo
  {pages} {1984--1987} (\bibinfo {year} {1997})}\BibitemShut {NoStop}%
\bibitem [{\citenamefont {Fu}\ and\ \citenamefont
  {Kane}(2012)}]{FuKaneTopology2012}%
  \BibitemOpen
  \bibfield  {author} {\bibinfo {author} {\bibfnamefont {Liang}\ \bibnamefont
  {Fu}}\ and\ \bibinfo {author} {\bibfnamefont {C.~L.}\ \bibnamefont {Kane}},\
  }\bibfield  {title} {\enquote {\bibinfo {title} {Topology, delocalization via
  average symmetry and the symplectic anderson transition},}\ }\href {\doibase
  10.1103/PhysRevLett.109.246605} {\bibfield  {journal} {\bibinfo  {journal}
  {Phys. Rev. Lett.}\ }\textbf {\bibinfo {volume} {109}},\ \bibinfo {pages}
  {246605} (\bibinfo {year} {2012})}\BibitemShut {NoStop}%
\bibitem [{\citenamefont {Fulga}\ \emph {et~al.}(2014)\citenamefont {Fulga},
  \citenamefont {van Heck}, \citenamefont {Edge},\ and\ \citenamefont
  {Akhmerov}}]{FulgavanHeckStatistical}%
  \BibitemOpen
  \bibfield  {author} {\bibinfo {author} {\bibfnamefont {I.~C.}\ \bibnamefont
  {Fulga}}, \bibinfo {author} {\bibfnamefont {B.}~\bibnamefont {van Heck}},
  \bibinfo {author} {\bibfnamefont {J.~M.}\ \bibnamefont {Edge}}, \ and\
  \bibinfo {author} {\bibfnamefont {A.~R.}\ \bibnamefont {Akhmerov}},\
  }\bibfield  {title} {\enquote {\bibinfo {title} {Statistical topological
  insulators},}\ }\href {\doibase 10.1103/PhysRevB.89.155424} {\bibfield
  {journal} {\bibinfo  {journal} {Phys. Rev. B}\ }\textbf {\bibinfo {volume}
  {89}},\ \bibinfo {pages} {155424} (\bibinfo {year} {2014})}\BibitemShut
  {NoStop}%
\bibitem [{\citenamefont {{Hastings}}(2010)}]{Hastings2010}%
  \BibitemOpen
  \bibfield  {author} {\bibinfo {author} {\bibfnamefont {M.~B.}\ \bibnamefont
  {{Hastings}}},\ }\bibfield  {title} {\enquote {\bibinfo {title}
  {{Quasi-adiabatic Continuation for Disordered Systems: Applications to
  Correlations, Lieb-Schultz-Mattis, and Hall Conductance}},}\ }\href@noop {}
  {\bibfield  {journal} {\bibinfo  {journal} {ArXiv e-prints}\ } (\bibinfo
  {year} {2010})},\ \Eprint {http://arxiv.org/abs/1001.5280} {arXiv:1001.5280
  [math-ph]} \BibitemShut {NoStop}%
\bibitem [{\citenamefont {Oshikawa}(2003)}]{Oshikawa2003}%
  \BibitemOpen
  \bibfield  {author} {\bibinfo {author} {\bibfnamefont {Masaki}\ \bibnamefont
  {Oshikawa}},\ }\bibfield  {title} {\enquote {\bibinfo {title} {Insulator,
  conductor, and commensurability: A topological approach},}\ }\href {\doibase
  10.1103/PhysRevLett.90.236401} {\bibfield  {journal} {\bibinfo  {journal}
  {Phys. Rev. Lett.}\ }\textbf {\bibinfo {volume} {90}},\ \bibinfo {pages}
  {236401} (\bibinfo {year} {2003})}\BibitemShut {NoStop}%
\bibitem [{\citenamefont {Kimchi}\ \emph {et~al.}(2013)\citenamefont {Kimchi},
  \citenamefont {Parameswaran}, \citenamefont {Turner}, \citenamefont {Wang},\
  and\ \citenamefont {Vishwanath}}]{Vishwanath2013}%
  \BibitemOpen
  \bibfield  {author} {\bibinfo {author} {\bibfnamefont {Itamar}\ \bibnamefont
  {Kimchi}}, \bibinfo {author} {\bibfnamefont {S.~A.}\ \bibnamefont
  {Parameswaran}}, \bibinfo {author} {\bibfnamefont {Ari~M.}\ \bibnamefont
  {Turner}}, \bibinfo {author} {\bibfnamefont {Fa}~\bibnamefont {Wang}}, \ and\
  \bibinfo {author} {\bibfnamefont {Ashvin}\ \bibnamefont {Vishwanath}},\
  }\bibfield  {title} {\enquote {\bibinfo {title} {Featureless and
  nonfractionalized mott insulators on the honeycomb lattice at 1/2 site
  filling},}\ }\href {\doibase 10.1073/pnas.1307245110} {\bibfield  {journal}
  {\bibinfo  {journal} {Proceedings of the National Academy of Sciences}\
  }\textbf {\bibinfo {volume} {110}},\ \bibinfo {pages} {16378--16383}
  (\bibinfo {year} {2013})},\ \Eprint
  {http://arxiv.org/abs/http://www.pnas.org/content/110/41/16378.full.pdf}
  {http://www.pnas.org/content/110/41/16378.full.pdf} \BibitemShut {NoStop}%
\bibitem [{\citenamefont {Kim}\ \emph {et~al.}(2016)\citenamefont {Kim},
  \citenamefont {Lee}, \citenamefont {Jiang}, \citenamefont {Ware},
  \citenamefont {Jian}, \citenamefont {Zaletel}, \citenamefont {Han},\ and\
  \citenamefont {Ran}}]{Ran2016}%
  \BibitemOpen
  \bibfield  {author} {\bibinfo {author} {\bibfnamefont {Panjin}\ \bibnamefont
  {Kim}}, \bibinfo {author} {\bibfnamefont {Hyunyong}\ \bibnamefont {Lee}},
  \bibinfo {author} {\bibfnamefont {Shenghan}\ \bibnamefont {Jiang}}, \bibinfo
  {author} {\bibfnamefont {Brayden}\ \bibnamefont {Ware}}, \bibinfo {author}
  {\bibfnamefont {Chao-Ming}\ \bibnamefont {Jian}}, \bibinfo {author}
  {\bibfnamefont {Michael}\ \bibnamefont {Zaletel}}, \bibinfo {author}
  {\bibfnamefont {Jung~Hoon}\ \bibnamefont {Han}}, \ and\ \bibinfo {author}
  {\bibfnamefont {Ying}\ \bibnamefont {Ran}},\ }\bibfield  {title} {\enquote
  {\bibinfo {title} {Featureless quantum insulator on the honeycomb lattice},}\
  }\href {\doibase 10.1103/PhysRevB.94.064432} {\bibfield  {journal} {\bibinfo
  {journal} {Phys. Rev. B}\ }\textbf {\bibinfo {volume} {94}},\ \bibinfo
  {pages} {064432} (\bibinfo {year} {2016})}\BibitemShut {NoStop}%
\bibitem [{\citenamefont {Aizenman}\ \emph {et~al.}(2001)\citenamefont
  {Aizenman}, \citenamefont {Goldstein},\ and\ \citenamefont
  {Lebowitz}}]{aizenman2001bounded}%
  \BibitemOpen
  \bibfield  {author} {\bibinfo {author} {\bibfnamefont {M}~\bibnamefont
  {Aizenman}}, \bibinfo {author} {\bibfnamefont {S}~\bibnamefont {Goldstein}},
  \ and\ \bibinfo {author} {\bibfnamefont {JL}~\bibnamefont {Lebowitz}},\
  }\bibfield  {title} {\enquote {\bibinfo {title} {Bounded fluctuations and
  translation symmetry breaking in one-dimensional particle systems},}\
  }\href@noop {} {\bibfield  {journal} {\bibinfo  {journal} {Journal of
  Statistical Physics}\ }\textbf {\bibinfo {volume} {103}},\ \bibinfo {pages}
  {601--618} (\bibinfo {year} {2001})}\BibitemShut {NoStop}%
\bibitem [{\citenamefont {Aizenman}\ and\ \citenamefont
  {Nachtergaele}(1994)}]{aizenman1994geometric}%
  \BibitemOpen
  \bibfield  {author} {\bibinfo {author} {\bibfnamefont {Michael}\ \bibnamefont
  {Aizenman}}\ and\ \bibinfo {author} {\bibfnamefont {Bruno}\ \bibnamefont
  {Nachtergaele}},\ }\bibfield  {title} {\enquote {\bibinfo {title} {Geometric
  aspects of quantum spin states},}\ }\href@noop {} {\bibfield  {journal}
  {\bibinfo  {journal} {Communications in Mathematical Physics}\ }\textbf
  {\bibinfo {volume} {164}},\ \bibinfo {pages} {17--63} (\bibinfo {year}
  {1994})}\BibitemShut {NoStop}%
\bibitem [{\citenamefont {Affleck}\ and\ \citenamefont
  {Lieb}(1986)}]{affleck1986proof}%
  \BibitemOpen
  \bibfield  {author} {\bibinfo {author} {\bibfnamefont {Ian}\ \bibnamefont
  {Affleck}}\ and\ \bibinfo {author} {\bibfnamefont {Elliott~H}\ \bibnamefont
  {Lieb}},\ }\bibfield  {title} {\enquote {\bibinfo {title} {A proof of part of
  haldane?s conjecture on spin chains},}\ }in\ \href@noop {} {\emph {\bibinfo
  {booktitle} {Condensed Matter Physics and Exactly Soluble Models}}}\
  (\bibinfo  {publisher} {Springer},\ \bibinfo {year} {1986})\ pp.\ \bibinfo
  {pages} {235--247}\BibitemShut {NoStop}%
\bibitem [{for(2017)}]{forthcoming3dglass}%
  \BibitemOpen
  \href@noop {} {\bibfield  {journal} {\bibinfo  {journal} {Nahum et al,
  unpublished}\ } (\bibinfo {year} {2017})}\BibitemShut {NoStop}%
\bibitem [{\citenamefont {{Tasaki}}(2017)}]{Tasaki2017}%
  \BibitemOpen
  \bibfield  {author} {\bibinfo {author} {\bibfnamefont {H.}~\bibnamefont
  {{Tasaki}}},\ }\bibfield  {title} {\enquote {\bibinfo {title}
  {{Lieb-Schultz-Mattis theorem with a local twist for general one-dimensional
  quantum systems}},}\ }\href@noop {} {\bibfield  {journal} {\bibinfo
  {journal} {ArXiv e-prints}\ } (\bibinfo {year} {2017})},\ \Eprint
  {http://arxiv.org/abs/1708.05186} {arXiv:1708.05186 [cond-mat.stat-mech]}
  \BibitemShut {NoStop}%
\bibitem [{\citenamefont {Bravyi}\ \emph {et~al.}(2006)\citenamefont {Bravyi},
  \citenamefont {Hastings},\ and\ \citenamefont
  {Verstraete}}]{BravyiHastingsVerstraete}%
  \BibitemOpen
  \bibfield  {author} {\bibinfo {author} {\bibfnamefont {S.}~\bibnamefont
  {Bravyi}}, \bibinfo {author} {\bibfnamefont {M.~B.}\ \bibnamefont
  {Hastings}}, \ and\ \bibinfo {author} {\bibfnamefont {F.}~\bibnamefont
  {Verstraete}},\ }\bibfield  {title} {\enquote {\bibinfo {title}
  {Lieb-robinson bounds and the generation of correlations and topological
  quantum order},}\ }\href {\doibase 10.1103/PhysRevLett.97.050401} {\bibfield
  {journal} {\bibinfo  {journal} {Phys. Rev. Lett.}\ }\textbf {\bibinfo
  {volume} {97}},\ \bibinfo {pages} {050401} (\bibinfo {year}
  {2006})}\BibitemShut {NoStop}%
\bibitem [{\citenamefont {Haah}(2016)}]{haah2016invariant}%
  \BibitemOpen
  \bibfield  {author} {\bibinfo {author} {\bibfnamefont {Jeongwan}\
  \bibnamefont {Haah}},\ }\bibfield  {title} {\enquote {\bibinfo {title} {An
  invariant of topologically ordered states under local unitary
  transformations},}\ }\href@noop {} {\bibfield  {journal} {\bibinfo  {journal}
  {Communications in Mathematical Physics}\ }\textbf {\bibinfo {volume}
  {342}},\ \bibinfo {pages} {771--801} (\bibinfo {year} {2016})}\BibitemShut
  {NoStop}%
\bibitem [{\citenamefont {Refael}\ and\ \citenamefont
  {Moore}(2009)}]{Moore2009}%
  \BibitemOpen
  \bibfield  {author} {\bibinfo {author} {\bibfnamefont {G}~\bibnamefont
  {Refael}}\ and\ \bibinfo {author} {\bibfnamefont {J~E}\ \bibnamefont
  {Moore}},\ }\bibfield  {title} {\enquote {\bibinfo {title} {Criticality and
  entanglement in random quantum systems},}\ }\href
  {http://stacks.iop.org/1751-8121/42/i=50/a=504010} {\bibfield  {journal}
  {\bibinfo  {journal} {Journal of Physics A: Mathematical and Theoretical}\
  }\textbf {\bibinfo {volume} {42}},\ \bibinfo {pages} {504010} (\bibinfo
  {year} {2009})}\BibitemShut {NoStop}%
\bibitem [{\citenamefont {Jackeli}\ and\ \citenamefont
  {Khaliullin}(2009)}]{Khaliullin2009}%
  \BibitemOpen
  \bibfield  {author} {\bibinfo {author} {\bibfnamefont {G.}~\bibnamefont
  {Jackeli}}\ and\ \bibinfo {author} {\bibfnamefont {G.}~\bibnamefont
  {Khaliullin}},\ }\bibfield  {title} {\enquote {\bibinfo {title} {Mott
  insulators in the strong spin-orbit coupling limit: From heisenberg to a
  quantum compass and kitaev models},}\ }\href {\doibase
  10.1103/PhysRevLett.102.017205} {\bibfield  {journal} {\bibinfo  {journal}
  {Phys. Rev. Lett.}\ }\textbf {\bibinfo {volume} {102}},\ \bibinfo {pages}
  {017205} (\bibinfo {year} {2009})}\BibitemShut {NoStop}%
\bibitem [{\citenamefont {Kimchi}\ \emph {et~al.}(2014)\citenamefont {Kimchi},
  \citenamefont {Analytis},\ and\ \citenamefont {Vishwanath}}]{Vishwanath2014}%
  \BibitemOpen
  \bibfield  {author} {\bibinfo {author} {\bibfnamefont {Itamar}\ \bibnamefont
  {Kimchi}}, \bibinfo {author} {\bibfnamefont {James~G.}\ \bibnamefont
  {Analytis}}, \ and\ \bibinfo {author} {\bibfnamefont {Ashvin}\ \bibnamefont
  {Vishwanath}},\ }\bibfield  {title} {\enquote {\bibinfo {title}
  {Three-dimensional quantum spin liquids in models of harmonic-honeycomb
  iridates and phase diagram in an infinite-$d$ approximation},}\ }\href
  {\doibase 10.1103/PhysRevB.90.205126} {\bibfield  {journal} {\bibinfo
  {journal} {Phys. Rev. B}\ }\textbf {\bibinfo {volume} {90}},\ \bibinfo
  {pages} {205126} (\bibinfo {year} {2014})}\BibitemShut {NoStop}%
\bibitem [{\citenamefont {w.~Anderson}\ \emph {et~al.}(1972)\citenamefont
  {w.~Anderson}, \citenamefont {Halperin},\ and\ \citenamefont
  {c.~M.~Varma}}]{Varma1972}%
  \BibitemOpen
  \bibfield  {author} {\bibinfo {author} {\bibfnamefont {P.}~\bibnamefont
  {w.~Anderson}}, \bibinfo {author} {\bibfnamefont {B.~I.}\ \bibnamefont
  {Halperin}}, \ and\ \bibinfo {author} {\bibnamefont {c.~M.~Varma}},\
  }\bibfield  {title} {\enquote {\bibinfo {title} {Anomalous low-temperature
  thermal properties of glasses and spin glasses},}\ }\href {\doibase
  10.1080/14786437208229210} {\bibfield  {journal} {\bibinfo  {journal}
  {Philosophical Magazine}\ }\textbf {\bibinfo {volume} {25}},\ \bibinfo
  {pages} {1--9} (\bibinfo {year} {1972})},\ \Eprint
  {http://arxiv.org/abs/http://dx.doi.org/10.1080/14786437208229210}
  {http://dx.doi.org/10.1080/14786437208229210} \BibitemShut {NoStop}%
\bibitem [{\citenamefont {Pohl}\ \emph {et~al.}(2002)\citenamefont {Pohl},
  \citenamefont {Liu},\ and\ \citenamefont {Thompson}}]{Thompson2002}%
  \BibitemOpen
  \bibfield  {author} {\bibinfo {author} {\bibfnamefont {Robert~O.}\
  \bibnamefont {Pohl}}, \bibinfo {author} {\bibfnamefont {Xiao}\ \bibnamefont
  {Liu}}, \ and\ \bibinfo {author} {\bibfnamefont {EunJoo}\ \bibnamefont
  {Thompson}},\ }\bibfield  {title} {\enquote {\bibinfo {title}
  {Low-temperature thermal conductivity and acoustic attenuation in amorphous
  solids},}\ }\href {\doibase 10.1103/RevModPhys.74.991} {\bibfield  {journal}
  {\bibinfo  {journal} {Rev. Mod. Phys.}\ }\textbf {\bibinfo {volume} {74}},\
  \bibinfo {pages} {991--1013} (\bibinfo {year} {2002})}\BibitemShut {NoStop}%
\bibitem [{\citenamefont {Kimchi}\ \emph {et~al.}(2018)\citenamefont {Kimchi},
  \citenamefont {Sheckelton}, \citenamefont {McQueen},\ and\ \citenamefont
  {Lee}}]{Scaling2018}%
  \BibitemOpen
  \bibfield  {author} {\bibinfo {author} {\bibfnamefont {I.}~\bibnamefont
  {Kimchi}}, \bibinfo {author} {\bibfnamefont {J.}~\bibnamefont {Sheckelton}},
  \bibinfo {author} {\bibfnamefont {T.}~\bibnamefont {McQueen}}, \ and\
  \bibinfo {author} {\bibfnamefont {P.}~\bibnamefont {Lee}},\ }\href@noop {}
  {\bibfield  {journal} {\bibinfo  {journal} {ArXiv e-prints}\ } (\bibinfo
  {year} {2018})},\ \Eprint {http://arxiv.org/abs/1803.00013} {arXiv:1803.00013
  [cond-mat.str-el]} \BibitemShut {NoStop}%
\bibitem [{\citenamefont {Furukawa}\ \emph {et~al.}(2015)\citenamefont
  {Furukawa}, \citenamefont {Miyagawa}, \citenamefont {Itou}, \citenamefont
  {Ito}, \citenamefont {Taniguchi}, \citenamefont {Saito}, \citenamefont
  {Iguchi}, \citenamefont {Sasaki},\ and\ \citenamefont {Kanoda}}]{Kanoda2015}%
  \BibitemOpen
  \bibfield  {author} {\bibinfo {author} {\bibfnamefont {T.}~\bibnamefont
  {Furukawa}}, \bibinfo {author} {\bibfnamefont {K.}~\bibnamefont {Miyagawa}},
  \bibinfo {author} {\bibfnamefont {T.}~\bibnamefont {Itou}}, \bibinfo {author}
  {\bibfnamefont {M.}~\bibnamefont {Ito}}, \bibinfo {author} {\bibfnamefont
  {H.}~\bibnamefont {Taniguchi}}, \bibinfo {author} {\bibfnamefont
  {M.}~\bibnamefont {Saito}}, \bibinfo {author} {\bibfnamefont
  {S.}~\bibnamefont {Iguchi}}, \bibinfo {author} {\bibfnamefont
  {T.}~\bibnamefont {Sasaki}}, \ and\ \bibinfo {author} {\bibfnamefont
  {K.}~\bibnamefont {Kanoda}},\ }\bibfield  {title} {\enquote {\bibinfo {title}
  {Quantum spin liquid emerging from antiferromagnetic order by introducing
  disorder},}\ }\href {\doibase 10.1103/PhysRevLett.115.077001} {\bibfield
  {journal} {\bibinfo  {journal} {Phys. Rev. Lett.}\ }\textbf {\bibinfo
  {volume} {115}},\ \bibinfo {pages} {077001} (\bibinfo {year}
  {2015})}\BibitemShut {NoStop}%
\bibitem [{\citenamefont {Kitagawa}\ \emph {et~al.}(2018)\citenamefont
  {Kitagawa}, \citenamefont {Takayama}, \citenamefont {Matsumoto},
  \citenamefont {Kato}, \citenamefont {Takano}, \citenamefont {Kishimoto},
  \citenamefont {Bette}, \citenamefont {Dinnebier}, \citenamefont {Jackeli},\
  and\ \citenamefont {Takagi}}]{Takagi2018}%
  \BibitemOpen
  \bibfield  {author} {\bibinfo {author} {\bibfnamefont {K.}~\bibnamefont
  {Kitagawa}}, \bibinfo {author} {\bibfnamefont {T.}~\bibnamefont {Takayama}},
  \bibinfo {author} {\bibfnamefont {Y.}~\bibnamefont {Matsumoto}}, \bibinfo
  {author} {\bibfnamefont {A.}~\bibnamefont {Kato}}, \bibinfo {author}
  {\bibfnamefont {R.}~\bibnamefont {Takano}}, \bibinfo {author} {\bibfnamefont
  {Y.}~\bibnamefont {Kishimoto}}, \bibinfo {author} {\bibfnamefont
  {S.}~\bibnamefont {Bette}}, \bibinfo {author} {\bibfnamefont
  {R.}~\bibnamefont {Dinnebier}}, \bibinfo {author} {\bibfnamefont
  {G.}~\bibnamefont {Jackeli}}, \ and\ \bibinfo {author} {\bibfnamefont
  {H.}~\bibnamefont {Takagi}},\ }\bibfield  {title} {\enquote {\bibinfo {title}
  {A spin-orbital-entangled quantum liquid on a honeycomb lattice},}\ }\href
  {http://dx.doi.org/10.1038/nature25482} {\bibfield  {journal} {\bibinfo
  {journal} {Nature}\ }\textbf {\bibinfo {volume} {554}},\ \bibinfo {pages}
  {341} (\bibinfo {year} {2018})},\ \bibinfo {note}
  {http://rdcu.be/GYFa}\BibitemShut {NoStop}%
\bibitem [{\citenamefont {Sheckelton}\ \emph {et~al.}(2012)\citenamefont
  {Sheckelton}, \citenamefont {Neilson}, \citenamefont {Soltan},\ and\
  \citenamefont {McQueen}}]{McQueen2012}%
  \BibitemOpen
  \bibfield  {author} {\bibinfo {author} {\bibfnamefont {J.~P.}\ \bibnamefont
  {Sheckelton}}, \bibinfo {author} {\bibfnamefont {J.~R.}\ \bibnamefont
  {Neilson}}, \bibinfo {author} {\bibfnamefont {D.~G.}\ \bibnamefont {Soltan}},
  \ and\ \bibinfo {author} {\bibfnamefont {T.~M.}\ \bibnamefont {McQueen}},\
  }\bibfield  {title} {\enquote {\bibinfo {title} {Possible valence-bond
  condensation in the frustrated cluster magnet lizn2mo3o8},}\ }\href
  {http://dx.doi.org/10.1038/nmat3329} {\bibfield  {journal} {\bibinfo
  {journal} {Nature Materials}\ }\textbf {\bibinfo {volume} {11}},\ \bibinfo
  {pages} {493} (\bibinfo {year} {2012})}\BibitemShut {NoStop}%
\bibitem [{\citenamefont {Sheckelton}\ \emph {et~al.}(2014)\citenamefont
  {Sheckelton}, \citenamefont {Foronda}, \citenamefont {Pan}, \citenamefont
  {Moir}, \citenamefont {McDonald}, \citenamefont {Lancaster}, \citenamefont
  {Baker}, \citenamefont {Armitage}, \citenamefont {Imai}, \citenamefont
  {Blundell},\ and\ \citenamefont {McQueen}}]{McQueen2014}%
  \BibitemOpen
  \bibfield  {author} {\bibinfo {author} {\bibfnamefont {J.~P.}\ \bibnamefont
  {Sheckelton}}, \bibinfo {author} {\bibfnamefont {F.~R.}\ \bibnamefont
  {Foronda}}, \bibinfo {author} {\bibfnamefont {LiDong}\ \bibnamefont {Pan}},
  \bibinfo {author} {\bibfnamefont {C.}~\bibnamefont {Moir}}, \bibinfo {author}
  {\bibfnamefont {R.~D.}\ \bibnamefont {McDonald}}, \bibinfo {author}
  {\bibfnamefont {T.}~\bibnamefont {Lancaster}}, \bibinfo {author}
  {\bibfnamefont {P.~J.}\ \bibnamefont {Baker}}, \bibinfo {author}
  {\bibfnamefont {N.~P.}\ \bibnamefont {Armitage}}, \bibinfo {author}
  {\bibfnamefont {T.}~\bibnamefont {Imai}}, \bibinfo {author} {\bibfnamefont
  {S.~J.}\ \bibnamefont {Blundell}}, \ and\ \bibinfo {author} {\bibfnamefont
  {T.~M.}\ \bibnamefont {McQueen}},\ }\bibfield  {title} {\enquote {\bibinfo
  {title} {Local magnetism and spin correlations in the geometrically
  frustrated cluster magnet
  ${\text{lizn}}_{2}{\text{mo}}_{3}{\text{o}}_{8}$},}\ }\href {\doibase
  10.1103/PhysRevB.89.064407} {\bibfield  {journal} {\bibinfo  {journal} {Phys.
  Rev. B}\ }\textbf {\bibinfo {volume} {89}},\ \bibinfo {pages} {064407}
  (\bibinfo {year} {2014})}\BibitemShut {NoStop}%
\bibitem [{\citenamefont {Mourigal}\ \emph {et~al.}(2014)\citenamefont
  {Mourigal}, \citenamefont {Fuhrman}, \citenamefont {Sheckelton},
  \citenamefont {Wartelle}, \citenamefont {Rodriguez-Rivera}, \citenamefont
  {Abernathy}, \citenamefont {McQueen},\ and\ \citenamefont
  {Broholm}}]{Broholm2014}%
  \BibitemOpen
  \bibfield  {author} {\bibinfo {author} {\bibfnamefont {M.}~\bibnamefont
  {Mourigal}}, \bibinfo {author} {\bibfnamefont {W.~T.}\ \bibnamefont
  {Fuhrman}}, \bibinfo {author} {\bibfnamefont {J.~P.}\ \bibnamefont
  {Sheckelton}}, \bibinfo {author} {\bibfnamefont {A.}~\bibnamefont
  {Wartelle}}, \bibinfo {author} {\bibfnamefont {J.~A.}\ \bibnamefont
  {Rodriguez-Rivera}}, \bibinfo {author} {\bibfnamefont {D.~L.}\ \bibnamefont
  {Abernathy}}, \bibinfo {author} {\bibfnamefont {T.~M.}\ \bibnamefont
  {McQueen}}, \ and\ \bibinfo {author} {\bibfnamefont {C.~L.}\ \bibnamefont
  {Broholm}},\ }\bibfield  {title} {\enquote {\bibinfo {title} {Molecular
  quantum magnetism in
  ${\mathrm{lizn}}_{2}{\mathrm{mo}}_{3}{\mathrm{o}}_{8}$},}\ }\href {\doibase
  10.1103/PhysRevLett.112.027202} {\bibfield  {journal} {\bibinfo  {journal}
  {Phys. Rev. Lett.}\ }\textbf {\bibinfo {volume} {112}},\ \bibinfo {pages}
  {027202} (\bibinfo {year} {2014})}\BibitemShut {NoStop}%
\bibitem [{\citenamefont {Helton}\ \emph {et~al.}(2010)\citenamefont {Helton},
  \citenamefont {Matan}, \citenamefont {Shores}, \citenamefont {Nytko},
  \citenamefont {Bartlett}, \citenamefont {Qiu}, \citenamefont {Nocera},\ and\
  \citenamefont {Lee}}]{Lee2010}%
  \BibitemOpen
  \bibfield  {author} {\bibinfo {author} {\bibfnamefont {J.~S.}\ \bibnamefont
  {Helton}}, \bibinfo {author} {\bibfnamefont {K.}~\bibnamefont {Matan}},
  \bibinfo {author} {\bibfnamefont {M.~P.}\ \bibnamefont {Shores}}, \bibinfo
  {author} {\bibfnamefont {E.~A.}\ \bibnamefont {Nytko}}, \bibinfo {author}
  {\bibfnamefont {B.~M.}\ \bibnamefont {Bartlett}}, \bibinfo {author}
  {\bibfnamefont {Y.}~\bibnamefont {Qiu}}, \bibinfo {author} {\bibfnamefont
  {D.~G.}\ \bibnamefont {Nocera}}, \ and\ \bibinfo {author} {\bibfnamefont
  {Y.~S.}\ \bibnamefont {Lee}},\ }\bibfield  {title} {\enquote {\bibinfo
  {title} {Dynamic scaling in the susceptibility of the spin-$\frac{1}{2}$
  kagome lattice antiferromagnet herbertsmithite},}\ }\href {\doibase
  10.1103/PhysRevLett.104.147201} {\bibfield  {journal} {\bibinfo  {journal}
  {Phys. Rev. Lett.}\ }\textbf {\bibinfo {volume} {104}},\ \bibinfo {pages}
  {147201} (\bibinfo {year} {2010})}\BibitemShut {NoStop}%
\bibitem [{\citenamefont {Ribak}\ \emph {et~al.}(2017)\citenamefont {Ribak},
  \citenamefont {Silber}, \citenamefont {Baines}, \citenamefont {Chashka},
  \citenamefont {Salman}, \citenamefont {Dagan},\ and\ \citenamefont
  {Kanigel}}]{Kanigel2017}%
  \BibitemOpen
  \bibfield  {author} {\bibinfo {author} {\bibfnamefont {A.}~\bibnamefont
  {Ribak}}, \bibinfo {author} {\bibfnamefont {I.}~\bibnamefont {Silber}},
  \bibinfo {author} {\bibfnamefont {C.}~\bibnamefont {Baines}}, \bibinfo
  {author} {\bibfnamefont {K.}~\bibnamefont {Chashka}}, \bibinfo {author}
  {\bibfnamefont {Z.}~\bibnamefont {Salman}}, \bibinfo {author} {\bibfnamefont
  {Y.}~\bibnamefont {Dagan}}, \ and\ \bibinfo {author} {\bibfnamefont
  {A.}~\bibnamefont {Kanigel}},\ }\bibfield  {title} {\enquote {\bibinfo
  {title} {Gapless excitations in the ground state of
  $1t\text{\ensuremath{-}}{\mathrm{tas}}_{2}$},}\ }\href {\doibase
  10.1103/PhysRevB.96.195131} {\bibfield  {journal} {\bibinfo  {journal} {Phys.
  Rev. B}\ }\textbf {\bibinfo {volume} {96}},\ \bibinfo {pages} {195131}
  (\bibinfo {year} {2017})}\BibitemShut {NoStop}%
\bibitem [{for(2018)}]{forthcomingDagan}%
  \BibitemOpen
  \href@noop {} {\bibfield  {journal} {\bibinfo  {journal} {Y. Dagan and I.
  Silber, unpublished}\ } (\bibinfo {year} {2018})}\BibitemShut {NoStop}%
\bibitem [{\citenamefont {de~Vries}\ \emph {et~al.}(2010)\citenamefont
  {de~Vries}, \citenamefont {Mclaughlin},\ and\ \citenamefont {Bos}}]{Bos2010}%
  \BibitemOpen
  \bibfield  {author} {\bibinfo {author} {\bibfnamefont {M.~A.}\ \bibnamefont
  {de~Vries}}, \bibinfo {author} {\bibfnamefont {A.~C.}\ \bibnamefont
  {Mclaughlin}}, \ and\ \bibinfo {author} {\bibfnamefont {J.-W.~G.}\
  \bibnamefont {Bos}},\ }\bibfield  {title} {\enquote {\bibinfo {title}
  {Valence bond glass on an fcc lattice in the double perovskite
  ${\mathrm{ba}}_{2}{\mathrm{ymoo}}_{6}$},}\ }\href {\doibase
  10.1103/PhysRevLett.104.177202} {\bibfield  {journal} {\bibinfo  {journal}
  {Phys. Rev. Lett.}\ }\textbf {\bibinfo {volume} {104}},\ \bibinfo {pages}
  {177202} (\bibinfo {year} {2010})}\BibitemShut {NoStop}%
\bibitem [{\citenamefont {de~Vries}\ \emph {et~al.}(2013)\citenamefont
  {de~Vries}, \citenamefont {Piatek}, \citenamefont {Misek}, \citenamefont
  {Lord}, \citenamefont {Rønnow},\ and\ \citenamefont {Bos}}]{Bos2013}%
  \BibitemOpen
  \bibfield  {author} {\bibinfo {author} {\bibfnamefont {M~A}\ \bibnamefont
  {de~Vries}}, \bibinfo {author} {\bibfnamefont {J~O}\ \bibnamefont {Piatek}},
  \bibinfo {author} {\bibfnamefont {M}~\bibnamefont {Misek}}, \bibinfo {author}
  {\bibfnamefont {J~S}\ \bibnamefont {Lord}}, \bibinfo {author} {\bibfnamefont
  {H~M}\ \bibnamefont {Rønnow}}, \ and\ \bibinfo {author} {\bibfnamefont
  {J-W~G}\ \bibnamefont {Bos}},\ }\bibfield  {title} {\enquote {\bibinfo
  {title} {Low-temperature spin dynamics of a valence bond glass in ba 2 ymoo
  6},}\ }\href {http://stacks.iop.org/1367-2630/15/i=4/a=043024} {\bibfield
  {journal} {\bibinfo  {journal} {New Journal of Physics}\ }\textbf {\bibinfo
  {volume} {15}},\ \bibinfo {pages} {043024} (\bibinfo {year}
  {2013})}\BibitemShut {NoStop}%
\bibitem [{\citenamefont {Romh\'anyi}\ \emph {et~al.}(2017)\citenamefont
  {Romh\'anyi}, \citenamefont {Balents},\ and\ \citenamefont
  {Jackeli}}]{Jackeli2017}%
  \BibitemOpen
  \bibfield  {author} {\bibinfo {author} {\bibfnamefont {Judit}\ \bibnamefont
  {Romh\'anyi}}, \bibinfo {author} {\bibfnamefont {Leon}\ \bibnamefont
  {Balents}}, \ and\ \bibinfo {author} {\bibfnamefont {George}\ \bibnamefont
  {Jackeli}},\ }\bibfield  {title} {\enquote {\bibinfo {title} {Spin-orbit
  dimers and noncollinear phases in ${d}^{1}$ cubic double perovskites},}\
  }\href {\doibase 10.1103/PhysRevLett.118.217202} {\bibfield  {journal}
  {\bibinfo  {journal} {Phys. Rev. Lett.}\ }\textbf {\bibinfo {volume} {118}},\
  \bibinfo {pages} {217202} (\bibinfo {year} {2017})}\BibitemShut {NoStop}%
\bibitem [{\citenamefont {Aharen}\ \emph {et~al.}(2010)\citenamefont {Aharen},
  \citenamefont {Greedan}, \citenamefont {Bridges}, \citenamefont {Aczel},
  \citenamefont {Rodriguez}, \citenamefont {MacDougall}, \citenamefont {Luke},
  \citenamefont {Imai}, \citenamefont {Michaelis}, \citenamefont {Kroeker},
  \citenamefont {Zhou}, \citenamefont {Wiebe},\ and\ \citenamefont
  {Cranswick}}]{Cranswick2010}%
  \BibitemOpen
  \bibfield  {author} {\bibinfo {author} {\bibfnamefont {Tomoko}\ \bibnamefont
  {Aharen}}, \bibinfo {author} {\bibfnamefont {John~E.}\ \bibnamefont
  {Greedan}}, \bibinfo {author} {\bibfnamefont {Craig~A.}\ \bibnamefont
  {Bridges}}, \bibinfo {author} {\bibfnamefont {Adam~A.}\ \bibnamefont
  {Aczel}}, \bibinfo {author} {\bibfnamefont {Jose}\ \bibnamefont {Rodriguez}},
  \bibinfo {author} {\bibfnamefont {Greg}\ \bibnamefont {MacDougall}}, \bibinfo
  {author} {\bibfnamefont {Graeme~M.}\ \bibnamefont {Luke}}, \bibinfo {author}
  {\bibfnamefont {Takashi}\ \bibnamefont {Imai}}, \bibinfo {author}
  {\bibfnamefont {Vladimir~K.}\ \bibnamefont {Michaelis}}, \bibinfo {author}
  {\bibfnamefont {Scott}\ \bibnamefont {Kroeker}}, \bibinfo {author}
  {\bibfnamefont {Haidong}\ \bibnamefont {Zhou}}, \bibinfo {author}
  {\bibfnamefont {Chris~R.}\ \bibnamefont {Wiebe}}, \ and\ \bibinfo {author}
  {\bibfnamefont {Lachlan M.~D.}\ \bibnamefont {Cranswick}},\ }\bibfield
  {title} {\enquote {\bibinfo {title} {Magnetic properties of the geometrically
  frustrated $s=\frac{1}{2}$ antiferromagnets,
  ${\text{la}}_{2}{\text{limoo}}_{6}$ and ${\text{ba}}_{2}{\text{ymoo}}_{6}$,
  with the b-site ordered double perovskite structure: Evidence for a
  collective spin-singlet ground state},}\ }\href {\doibase
  10.1103/PhysRevB.81.224409} {\bibfield  {journal} {\bibinfo  {journal} {Phys.
  Rev. B}\ }\textbf {\bibinfo {volume} {81}},\ \bibinfo {pages} {224409}
  (\bibinfo {year} {2010})}\BibitemShut {NoStop}%
\bibitem [{\citenamefont {Kardar}(2007)}]{mehran_book}%
  \BibitemOpen
  \bibfield  {author} {\bibinfo {author} {\bibfnamefont {Mehran}\ \bibnamefont
  {Kardar}},\ }\href@noop {} {\emph {\bibinfo {title} {Statistical physics of
  fields}}}\ (\bibinfo  {publisher} {Cambridge University Press},\ \bibinfo
  {year} {2007})\BibitemShut {NoStop}%
\bibitem [{\citenamefont {Evertz}(2003)}]{evertz2003loop}%
  \BibitemOpen
  \bibfield  {author} {\bibinfo {author} {\bibfnamefont {Hans~Gerd}\
  \bibnamefont {Evertz}},\ }\bibfield  {title} {\enquote {\bibinfo {title} {The
  loop algorithm},}\ }\href@noop {} {\bibfield  {journal} {\bibinfo  {journal}
  {Advances in Physics}\ }\textbf {\bibinfo {volume} {52}},\ \bibinfo {pages}
  {1--66} (\bibinfo {year} {2003})}\BibitemShut {NoStop}%
\bibitem [{\citenamefont {Huse}\ \emph {et~al.}(1985)\citenamefont {Huse},
  \citenamefont {Henley},\ and\ \citenamefont
  {Fisher}}]{HuseHenleyFisher1985respond}%
  \BibitemOpen
  \bibfield  {author} {\bibinfo {author} {\bibfnamefont {D.~A.}\ \bibnamefont
  {Huse}}, \bibinfo {author} {\bibfnamefont {C.~L.}\ \bibnamefont {Henley}}, \
  and\ \bibinfo {author} {\bibfnamefont {D.~S.}\ \bibnamefont {Fisher}},\
  }\bibfield  {title} {\enquote {\bibinfo {title} {respond},}\ }\href {\doibase
  10.1103/PhysRevLett.55.2924} {\bibfield  {journal} {\bibinfo  {journal}
  {Phys. Rev. Lett.}\ }\textbf {\bibinfo {volume} {55}},\ \bibinfo {pages}
  {2924--2924} (\bibinfo {year} {1985})}\BibitemShut {NoStop}%
\bibitem [{\citenamefont {Kardar}\ and\ \citenamefont
  {Nelson}(1985)}]{kardar1985commensurate}%
  \BibitemOpen
  \bibfield  {author} {\bibinfo {author} {\bibfnamefont {Mehran}\ \bibnamefont
  {Kardar}}\ and\ \bibinfo {author} {\bibfnamefont {David~R}\ \bibnamefont
  {Nelson}},\ }\bibfield  {title} {\enquote {\bibinfo {title}
  {Commensurate-incommensurate transitions with quenched random impurities},}\
  }\href@noop {} {\bibfield  {journal} {\bibinfo  {journal} {Physical review
  letters}\ }\textbf {\bibinfo {volume} {55}},\ \bibinfo {pages} {1157}
  (\bibinfo {year} {1985})}\BibitemShut {NoStop}%
\bibitem [{\citenamefont {Kardar}\ \emph {et~al.}(1986)\citenamefont {Kardar},
  \citenamefont {Parisi},\ and\ \citenamefont {Zhang}}]{kpz}%
  \BibitemOpen
  \bibfield  {author} {\bibinfo {author} {\bibfnamefont {M.}~\bibnamefont
  {Kardar}}, \bibinfo {author} {\bibfnamefont {G.}~\bibnamefont {Parisi}}, \
  and\ \bibinfo {author} {\bibfnamefont {Y-C.}\ \bibnamefont {Zhang}},\
  }\bibfield  {title} {\enquote {\bibinfo {title} {Dynamic scaling of growing
  interfaces},}\ }\href {\doibase 10.1103/PhysRevLett.56.889} {\bibfield
  {journal} {\bibinfo  {journal} {Phys. Rev. Lett.}\ }\textbf {\bibinfo
  {volume} {56}},\ \bibinfo {pages} {889--892} (\bibinfo {year}
  {1986})}\BibitemShut {NoStop}%
\bibitem [{\citenamefont {Nguyen}\ \emph {et~al.}(1985)\citenamefont {Nguyen},
  \citenamefont {Spivak},\ and\ \citenamefont {Shklovskii}}]{nguyen1985tunnel}%
  \BibitemOpen
  \bibfield  {author} {\bibinfo {author} {\bibfnamefont {VL}~\bibnamefont
  {Nguyen}}, \bibinfo {author} {\bibfnamefont {BZ}~\bibnamefont {Spivak}}, \
  and\ \bibinfo {author} {\bibfnamefont {BI}~\bibnamefont {Shklovskii}},\
  }\bibfield  {title} {\enquote {\bibinfo {title} {Tunnel hopping in disordered
  systems},}\ }\href@noop {} {\bibfield  {journal} {\bibinfo  {journal} {Sov.
  Phys. JETP}\ }\textbf {\bibinfo {volume} {62}},\ \bibinfo {pages}
  {1021--1029} (\bibinfo {year} {1985})}\BibitemShut {NoStop}%
\bibitem [{\citenamefont {Medina}\ \emph {et~al.}(1989)\citenamefont {Medina},
  \citenamefont {Kardar}, \citenamefont {Shapir},\ and\ \citenamefont
  {Wang}}]{medina1989interference}%
  \BibitemOpen
  \bibfield  {author} {\bibinfo {author} {\bibfnamefont {Ernesto}\ \bibnamefont
  {Medina}}, \bibinfo {author} {\bibfnamefont {Mehran}\ \bibnamefont {Kardar}},
  \bibinfo {author} {\bibfnamefont {Yonathan}\ \bibnamefont {Shapir}}, \ and\
  \bibinfo {author} {\bibfnamefont {Xiang~Rong}\ \bibnamefont {Wang}},\
  }\bibfield  {title} {\enquote {\bibinfo {title} {Interference of directed
  paths in disordered systems},}\ }\href@noop {} {\bibfield  {journal}
  {\bibinfo  {journal} {Physical review letters}\ }\textbf {\bibinfo {volume}
  {62}},\ \bibinfo {pages} {941} (\bibinfo {year} {1989})}\BibitemShut
  {NoStop}%
\bibitem [{\citenamefont {Wang}\ \emph {et~al.}(1990)\citenamefont {Wang},
  \citenamefont {Shapir}, \citenamefont {Medina},\ and\ \citenamefont
  {Kardar}}]{WangExact}%
  \BibitemOpen
  \bibfield  {author} {\bibinfo {author} {\bibfnamefont {Xiang~Rong}\
  \bibnamefont {Wang}}, \bibinfo {author} {\bibfnamefont {Yonathan}\
  \bibnamefont {Shapir}}, \bibinfo {author} {\bibfnamefont {Ernesto}\
  \bibnamefont {Medina}}, \ and\ \bibinfo {author} {\bibfnamefont {Mehran}\
  \bibnamefont {Kardar}},\ }\bibfield  {title} {\enquote {\bibinfo {title}
  {Exact-enumeration approach to tunneling in disordered systems},}\ }\href
  {\doibase 10.1103/PhysRevB.42.4559} {\bibfield  {journal} {\bibinfo
  {journal} {Phys. Rev. B}\ }\textbf {\bibinfo {volume} {42}},\ \bibinfo
  {pages} {4559--4562} (\bibinfo {year} {1990})}\BibitemShut {NoStop}%
\bibitem [{\citenamefont {Medina}\ and\ \citenamefont
  {Kardar}(1992)}]{medina1992quantum}%
  \BibitemOpen
  \bibfield  {author} {\bibinfo {author} {\bibfnamefont {Ernesto}\ \bibnamefont
  {Medina}}\ and\ \bibinfo {author} {\bibfnamefont {Mehran}\ \bibnamefont
  {Kardar}},\ }\bibfield  {title} {\enquote {\bibinfo {title} {Quantum
  interference effects for strongly localized electrons},}\ }\href@noop {}
  {\bibfield  {journal} {\bibinfo  {journal} {Physical Review B}\ }\textbf
  {\bibinfo {volume} {46}},\ \bibinfo {pages} {9984} (\bibinfo {year}
  {1992})}\BibitemShut {NoStop}%
\bibitem [{\citenamefont {Roux}\ and\ \citenamefont
  {Coniglio}(1994)}]{roux1994interference}%
  \BibitemOpen
  \bibfield  {author} {\bibinfo {author} {\bibfnamefont {S}~\bibnamefont
  {Roux}}\ and\ \bibinfo {author} {\bibfnamefont {Antonio}\ \bibnamefont
  {Coniglio}},\ }\bibfield  {title} {\enquote {\bibinfo {title} {Interference
  of directed paths},}\ }\href@noop {} {\bibfield  {journal} {\bibinfo
  {journal} {Journal of Physics A: Mathematical and General}\ }\textbf
  {\bibinfo {volume} {27}},\ \bibinfo {pages} {5467} (\bibinfo {year}
  {1994})}\BibitemShut {NoStop}%
\bibitem [{\citenamefont {Lien~Nguyen}\ and\ \citenamefont
  {Gamietea}(1996)}]{NguyenCrossover}%
  \BibitemOpen
  \bibfield  {author} {\bibinfo {author} {\bibfnamefont {V.}~\bibnamefont
  {Lien~Nguyen}}\ and\ \bibinfo {author} {\bibfnamefont {Arturo~D.}\
  \bibnamefont {Gamietea}},\ }\bibfield  {title} {\enquote {\bibinfo {title}
  {Crossover in tunneling hops in systems of strongly localized electrons},}\
  }\href {\doibase 10.1103/PhysRevB.53.7932} {\bibfield  {journal} {\bibinfo
  {journal} {Phys. Rev. B}\ }\textbf {\bibinfo {volume} {53}},\ \bibinfo
  {pages} {7932--7936} (\bibinfo {year} {1996})}\BibitemShut {NoStop}%
\bibitem [{\citenamefont {Kim}\ and\ \citenamefont
  {Huse}(2011)}]{husesigntransition}%
  \BibitemOpen
  \bibfield  {author} {\bibinfo {author} {\bibfnamefont {Hyungwon}\
  \bibnamefont {Kim}}\ and\ \bibinfo {author} {\bibfnamefont {David~A}\
  \bibnamefont {Huse}},\ }\bibfield  {title} {\enquote {\bibinfo {title}
  {Interfering directed paths and the sign phase transition},}\ }\href@noop {}
  {\bibfield  {journal} {\bibinfo  {journal} {Physical Review B}\ }\textbf
  {\bibinfo {volume} {83}},\ \bibinfo {pages} {052405} (\bibinfo {year}
  {2011})}\BibitemShut {NoStop}%
\bibitem [{\citenamefont {Baldwin}\ \emph {et~al.}(2017)\citenamefont
  {Baldwin}, \citenamefont {Laumann},\ and\ \citenamefont
  {Spivak}}]{laumannsigntransition}%
  \BibitemOpen
  \bibfield  {author} {\bibinfo {author} {\bibfnamefont {CL}~\bibnamefont
  {Baldwin}}, \bibinfo {author} {\bibfnamefont {CR}~\bibnamefont {Laumann}}, \
  and\ \bibinfo {author} {\bibfnamefont {B}~\bibnamefont {Spivak}},\ }\bibfield
   {title} {\enquote {\bibinfo {title} {The sign phase transition in the
  problem of interfering directed paths},}\ }\href@noop {} {\bibfield
  {journal} {\bibinfo  {journal} {arXiv preprint arXiv:1709.03516}\ } (\bibinfo
  {year} {2017})}\BibitemShut {NoStop}%
\bibitem [{\citenamefont {Kim}\ \emph {et~al.}(1991)\citenamefont {Kim},
  \citenamefont {Bray},\ and\ \citenamefont {Moore}}]{kim1991finite}%
  \BibitemOpen
  \bibfield  {author} {\bibinfo {author} {\bibfnamefont {JM}~\bibnamefont
  {Kim}}, \bibinfo {author} {\bibfnamefont {AJ}~\bibnamefont {Bray}}, \ and\
  \bibinfo {author} {\bibfnamefont {MA}~\bibnamefont {Moore}},\ }\bibfield
  {title} {\enquote {\bibinfo {title} {Finite-temperature directed polymers in
  a random potential},}\ }\href@noop {} {\bibfield  {journal} {\bibinfo
  {journal} {Physical Review A}\ }\textbf {\bibinfo {volume} {44}},\ \bibinfo
  {pages} {R4782} (\bibinfo {year} {1991})}\BibitemShut {NoStop}%
\bibitem [{\citenamefont {Chubukov}\ \emph {et~al.}(1994)\citenamefont
  {Chubukov}, \citenamefont {Senthil},\ and\ \citenamefont
  {Sachdev}}]{Sachdev1994}%
  \BibitemOpen
  \bibfield  {author} {\bibinfo {author} {\bibfnamefont {Andrey~V.}\
  \bibnamefont {Chubukov}}, \bibinfo {author} {\bibfnamefont {T.}~\bibnamefont
  {Senthil}}, \ and\ \bibinfo {author} {\bibfnamefont {Subir}\ \bibnamefont
  {Sachdev}},\ }\bibfield  {title} {\enquote {\bibinfo {title} {Universal
  magnetic properties of frustrated quantum antiferromagnets in two
  dimensions},}\ }\href {\doibase 10.1103/PhysRevLett.72.2089} {\bibfield
  {journal} {\bibinfo  {journal} {Phys. Rev. Lett.}\ }\textbf {\bibinfo
  {volume} {72}},\ \bibinfo {pages} {2089--2092} (\bibinfo {year}
  {1994})}\BibitemShut {NoStop}%
\bibitem [{\citenamefont {Kawamura}\ and\ \citenamefont
  {Miyashita}(1985)}]{Miyashita1985}%
  \BibitemOpen
  \bibfield  {author} {\bibinfo {author} {\bibfnamefont {Hikaru}\ \bibnamefont
  {Kawamura}}\ and\ \bibinfo {author} {\bibfnamefont {Seiji}\ \bibnamefont
  {Miyashita}},\ }\bibfield  {title} {\enquote {\bibinfo {title} {Phase
  transition of the heisenberg antiferromagnet on the triangular lattice in a
  magnetic field},}\ }\href {\doibase 10.1143/JPSJ.54.4530} {\bibfield
  {journal} {\bibinfo  {journal} {Journal of the Physical Society of Japan}\
  }\textbf {\bibinfo {volume} {54}},\ \bibinfo {pages} {4530--4538} (\bibinfo
  {year} {1985})},\ \Eprint
  {http://arxiv.org/abs/http://dx.doi.org/10.1143/JPSJ.54.4530}
  {http://dx.doi.org/10.1143/JPSJ.54.4530} \BibitemShut {NoStop}%
\bibitem [{\citenamefont {Edmonds}(1965)}]{Edmonds1965}%
  \BibitemOpen
  \bibfield  {author} {\bibinfo {author} {\bibfnamefont {Jack}\ \bibnamefont
  {Edmonds}},\ }\bibfield  {title} {\enquote {\bibinfo {title} {Paths, trees,
  and flowers},}\ }\href@noop {} {\bibfield  {journal} {\bibinfo  {journal}
  {ACM Comput. Surv. 17, 449-467}\ } (\bibinfo {year} {1965})}\BibitemShut
  {NoStop}%
\bibitem [{\citenamefont {Galil}(1986)}]{Galil1986}%
  \BibitemOpen
  \bibfield  {author} {\bibinfo {author} {\bibfnamefont {Zvi}\ \bibnamefont
  {Galil}},\ }\bibfield  {title} {\enquote {\bibinfo {title} {Efficient
  algorithms for finding maximum matching in graphs},}\ }\href {\doibase
  10.1145/6462.6502} {\bibfield  {journal} {\bibinfo  {journal} {ACM Comput.
  Surv.}\ }\textbf {\bibinfo {volume} {18}},\ \bibinfo {pages} {23--38}
  (\bibinfo {year} {1986})}\BibitemShut {NoStop}%
\bibitem [{\citenamefont {{Li}}\ \emph {et~al.}(2016)\citenamefont {{Li}},
  \citenamefont {{Shen}}, \citenamefont {{Li}}, \citenamefont {{Zhao}},\ and\
  \citenamefont {{Chen}}}]{Chen2016a}%
  \BibitemOpen
  \bibfield  {author} {\bibinfo {author} {\bibfnamefont {Y.-D.}\ \bibnamefont
  {{Li}}}, \bibinfo {author} {\bibfnamefont {Y.}~\bibnamefont {{Shen}}},
  \bibinfo {author} {\bibfnamefont {Y.}~\bibnamefont {{Li}}}, \bibinfo {author}
  {\bibfnamefont {J.}~\bibnamefont {{Zhao}}}, \ and\ \bibinfo {author}
  {\bibfnamefont {G.}~\bibnamefont {{Chen}}},\ }\bibfield  {title} {\enquote
  {\bibinfo {title} {{The effect of spin-orbit coupling on the effective-spin
  correlation in YbMgGaO4}},}\ }\href@noop {} {\bibfield  {journal} {\bibinfo
  {journal} {ArXiv e-prints}\ } (\bibinfo {year} {2016})},\ \Eprint
  {http://arxiv.org/abs/1608.06445} {arXiv:1608.06445 [cond-mat.str-el]}
  \BibitemShut {NoStop}%
\bibitem [{\citenamefont {Chaloupka}\ \emph {et~al.}(2010)\citenamefont
  {Chaloupka}, \citenamefont {Jackeli},\ and\ \citenamefont
  {Khaliullin}}]{Khaliullin2010}%
  \BibitemOpen
  \bibfield  {author} {\bibinfo {author} {\bibfnamefont {Jiri}\ \bibnamefont
  {Chaloupka}}, \bibinfo {author} {\bibfnamefont {George}\ \bibnamefont
  {Jackeli}}, \ and\ \bibinfo {author} {\bibfnamefont {Giniyat}\ \bibnamefont
  {Khaliullin}},\ }\bibfield  {title} {\enquote {\bibinfo {title}
  {Kitaev-heisenberg model on a honeycomb lattice: Possible exotic phases in
  iridium oxides ${A}_{2}{\mathrm{iro}}_{3}$},}\ }\href {\doibase
  10.1103/PhysRevLett.105.027204} {\bibfield  {journal} {\bibinfo  {journal}
  {Phys. Rev. Lett.}\ }\textbf {\bibinfo {volume} {105}},\ \bibinfo {pages}
  {027204} (\bibinfo {year} {2010})}\BibitemShut {NoStop}%
\bibitem [{\citenamefont {Kimchi}\ and\ \citenamefont
  {Vishwanath}(2014)}]{Vishwanath2014a}%
  \BibitemOpen
  \bibfield  {author} {\bibinfo {author} {\bibfnamefont {Itamar}\ \bibnamefont
  {Kimchi}}\ and\ \bibinfo {author} {\bibfnamefont {Ashvin}\ \bibnamefont
  {Vishwanath}},\ }\bibfield  {title} {\enquote {\bibinfo {title}
  {Kitaev-heisenberg models for iridates on the triangular, hyperkagome,
  kagome, fcc, and pyrochlore lattices},}\ }\href {\doibase
  10.1103/PhysRevB.89.014414} {\bibfield  {journal} {\bibinfo  {journal} {Phys.
  Rev. B}\ }\textbf {\bibinfo {volume} {89}},\ \bibinfo {pages} {014414}
  (\bibinfo {year} {2014})}\BibitemShut {NoStop}%
\bibitem [{\citenamefont {Rousochatzakis}\ \emph {et~al.}(2016)\citenamefont
  {Rousochatzakis}, \citenamefont {R\"ossler}, \citenamefont {van~den Brink},\
  and\ \citenamefont {Daghofer}}]{Daghofer2016}%
  \BibitemOpen
  \bibfield  {author} {\bibinfo {author} {\bibfnamefont {Ioannis}\ \bibnamefont
  {Rousochatzakis}}, \bibinfo {author} {\bibfnamefont {Ulrich~K.}\ \bibnamefont
  {R\"ossler}}, \bibinfo {author} {\bibfnamefont {Jeroen}\ \bibnamefont
  {van~den Brink}}, \ and\ \bibinfo {author} {\bibfnamefont {Maria}\
  \bibnamefont {Daghofer}},\ }\bibfield  {title} {\enquote {\bibinfo {title}
  {Kitaev anisotropy induces mesoscopic ${\mathbb{z}}_{2}$ vortex crystals in
  frustrated hexagonal antiferromagnets},}\ }\href {\doibase
  10.1103/PhysRevB.93.104417} {\bibfield  {journal} {\bibinfo  {journal} {Phys.
  Rev. B}\ }\textbf {\bibinfo {volume} {93}},\ \bibinfo {pages} {104417}
  (\bibinfo {year} {2016})}\BibitemShut {NoStop}%
\end{thebibliography}%

\end{document}